\documentclass[aps,pra,twocolumn,superscriptaddress,nofootinbib,longbibliography]{revtex4-2}
\usepackage{amsmath,amssymb,amsfonts,bm,braket}
\usepackage{graphicx}
\usepackage{booktabs}
\usepackage{hyperref}
\usepackage{xcolor}
\usepackage{mathtools}
\hypersetup{colorlinks=true,linkcolor=blue,citecolor=blue,urlcolor=blue}

\begin{document}

\title{Bright-mode parity synthesis for bosonic state transfer through a single ancilla}

\author{Pradip Laha}
\email{plaha@uni-mainz.de}
\affiliation{Institute of Physics, Johannes Gutenberg-Universit\"at Mainz, Staudingerweg 7, 55128 Mainz, Germany}
\author{Peter van Loock}
\email{loock@uni-mainz.de}
\affiliation{Institute of Physics, Johannes Gutenberg-Universit\"at Mainz, Staudingerweg 7, 55128 Mainz, Germany}

\begin{abstract}
Bosonic modes provide hardware-efficient quantum memories and logical registers, including highly non-Gaussian encoded states, but transferring finite-dimensional bosonic states through a restricted ancilla interface requires identifying which collective mode is actually controlled. We study a restricted setting in which two oscillators couple with opposite signs to a single two-level ancilla. In the normal-mode basis, the symmetric mode is dark, while transfer between the physical modes is equivalent to synthesizing parity on the antisymmetric bright mode. This reduction gives exact finite-sum transfer formulas for Fock states, Fock-state qubits, and finite Fock superpositions, and explains why resonant single-ancilla transfer is recurrence-limited beyond the single-photon sector. We then show that detuned Jaynes--Cummings evolution provides a native two-parameter route to high-fidelity finite-cutoff parity synthesis, with residual ancilla excitation, calibration sensitivity, and a minimal Markovian noise estimate quantified separately. Bosonic-code examples illustrate how transfer sensitivity is governed by photon-number support and residual bright-mode phase errors. The result provides a practical benchmark and organizing principle for constrained ancilla-mediated bosonic transfer when direct exchange is unavailable or undesirable.
\end{abstract}

\maketitle

\section{Introduction}
\label{sec:introduction}

Practical quantum technologies rely heavily on the ability to store, protect, process, and move quantum information without losing coherence.  Bosonic modes are especially attractive for these tasks because the photon-number ladder of a single harmonic oscillator provides a large, in principle infinite, Hilbert space within one physical degree of freedom~\cite{Braunstein2005,Weedbrook_RMP_2012,Serafini_2017,Albert2018}. In cavity and circuit quantum electrodynamics, this structure has enabled precise Fock-state control~\cite{Haroche2006,Li_PRL_2024}, stabilized cat states and autonomous bosonic protection~\cite{Cochrane_PRA_1999,Leghtas2013,Leghtas2015,Mirrahimi2014}, binomial encodings~\cite{Michael2016,Chen_PRR_2021,chang_2025}, Gottesman–Kitaev–Preskill (GKP) encoding and error correction~\cite{GKP2001,Baragiola_PRL_2019,Noh_IEEE_2019,CampagneIbarcq2020,Noh_PRA_2020,Noh_PRXQ_2022,Juliette_2024,Conrad_2024}, and logical gates on oscillator-encoded qubits~\cite{Eickbusch2022}.  These developments are part of a broader program in which rotation-symmetric bosonic codes (RSBCs)~\cite{Grimsmo2020,xu_2024}---including cat and binomial codes---as well as grid-state encodings such as GKP codes provide hardware-efficient quantum error correction resources~\cite{Albert2018,Cai2021,Ma2021,Albert2025,Grimsmo2020,xu_2024}.  Recent progress with concatenated bosonic cat qubits~\cite{Putterman2025} and hybrid cat--transmon architectures~\cite{Hann2025} further highlights the practical importance of controlling encoded oscillator states.

Once quantum information is stored in a bosonic mode, it must often be transferred between memories, processing modes, communication channels, or code blocks.  For non-Gaussian encoded states, such transfer is not certified by moving energy or preserving Gaussian moments.  In addition to photon-number populations, a useful transfer protocol must also preserve the relative phases across the finite Fock support that carries the encoded information.

Quantum state transfer is a broad problem, with foundational results in engineered spin networks~\cite{Christandl_PRL_2004} and modern realizations spanning modular superconducting circuits, bosonic memories, and multimode quantum channels~\cite{Axline2018,Kurpiers2018,CampagneIbarcq2018,XuEtAl2023NJP,HeZhang2025PRL,XiangEtAl2024NatCommun}. In the bosonic setting, the existing approaches fall into rather different control paradigms.  The most direct route is a beam splitter interaction, generated either by direct exchange or by parametrically activated couplings~\cite{Tian2008,Basilewitsch2022}.  Such interactions coherently swap two modes when a high-quality mode--mode coupling is available, and they are now a powerful tool for operations on bosonic-code states~\cite{Chapman2023,Lu2023}.  Another route is to convert stationary bosonic excitations into itinerant photons or intermediate transducer modes, enabling state transfer and entanglement generation between remote microwave memories and modular nodes~\cite{Axline2018,Kurpiers2018,CampagneIbarcq2018}. Interference-based protocols provide a related strategy: dark-mode or impedance-matching effects can suppress unwanted intermediate population and improve transfer through imperfect transducers or optomechanical channels~\cite{WangClerk2012,LauClerk2019}.  Measurement, error detection, engineered dissipation, and exchange-free interactions provide further ways to protect or mediate the transfer of multiphoton cavity states, including bosonic code states~\cite{Burkhart2021,Zhou2024}.  Strongly coupled and multimode bosonic settings provide another route to coherent state exchange and entanglement preparation~\cite{XuEtAl2023NJP,HeZhang2025PRL}.  More broadly, quantum state transfer has been studied in interacting multiple-excitation systems and in settings where, transport can be enhanced by avoiding chaotic dynamics~\cite{YueEtAl2024PRB,XiangEtAl2024NatCommun}.

\begin{figure*}[t]
\centering
\includegraphics[width=0.98\textwidth]{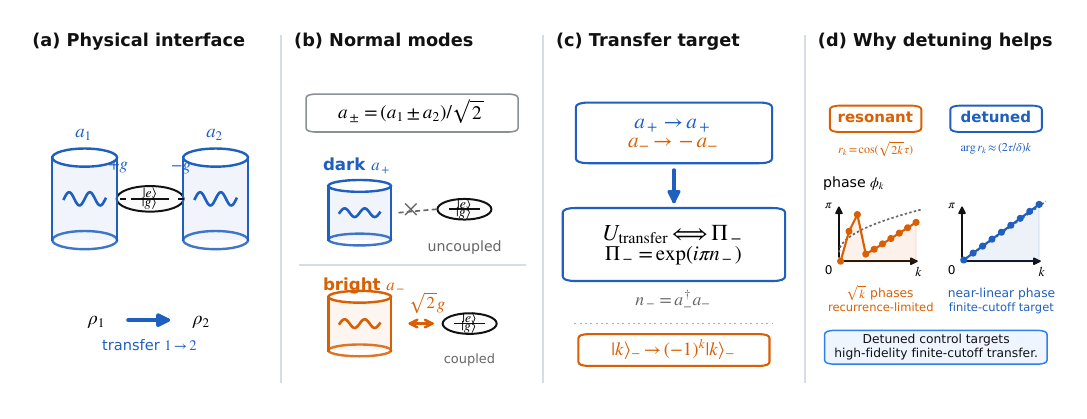}
\vspace{-3ex}
\caption{
Bright-mode parity as the transfer target. Two physical bosonic modes coupled with opposite signs to a single two-level ancilla reduce to a dark symmetric mode \(a_+\) and a bright antisymmetric mode \(a_-\).  State transfer \(1\!\to\!2\) is therefore equivalent to leaving \(a_+\) unchanged while applying parity to \(a_-\). Resonant dynamics approximate this parity only through recurrence of \(\sqrt{k}\)-dependent JC phases, whereas detuned native control produces an approximately photon-number-linear phase and directly targets high-fidelity finite-cutoff transfer.
}
\label{fig:schematic}
\end{figure*}

A complementary line of work concerns controllability and finite Fock-space phase engineering, including adiabatic transfer protocols, number-selective oscillator phase control, and broader pulse-design or bosonic network controllability frameworks~\cite{Bergmann1998,Larson2005,Kumar2016,Premaratne2017,Heeres2015,Krastanov2015,Eickbusch2022,Landgraf2024,PalaoKosloff2003,Khaneja2005,JinJing2026PRA,laha_PRR_2024}. Adiabatic and STIRAP-inspired protocols use dark-state structure to move excitations while suppressing population of intermediate states~\cite{Bergmann1998,Larson2005,Kumar2016,Premaratne2017}. In oscillator platforms, SNAP-type photon-number-selective phase gates and dispersive control provide a direct route to structured phase patterns across a finite Fock space~\cite{Heeres2015,Krastanov2015,Eickbusch2022,Landgraf2024}. More general pulse-design methods and bosonic network controllability results place such operations in a broader optimal-control framework, where, state transfer can appear as one application of a programmable bosonic control toolkit~\cite{Brif_NJP_2010,Koch_EPJ_2022}.

These transfer approaches are powerful when the required Hamiltonians or control capabilities are available.
They do not, however, answer what operation is being synthesized when the available interface is not a beam splitter, not a tunable network Hamiltonian, and not a broadly programmable control interface, but a single ancilla coupled to one collective mode.  In such a setting, a numerical fidelity maximum is not self-explanatory.  It may hide which normal mode is actually controlled, which photon-number phases must be synthesized, and why an evolution that works in the single-photon sector fails for larger oscillator cutoffs.  We therefore study the canonical antisymmetric single-ancilla interface, in which two bosonic modes couple with equal magnitude and opposite sign through the Jaynes–Cummings (JC) interaction to a two-level ancilla. Our aim is not to construct the fastest possible swap, but to identify the finite-cutoff operation that this restricted interface must synthesize in order to transfer an arbitrary bosonic state.

Figure~\ref{fig:schematic} summarizes the physical setting and the main reduction.  The structure becomes transparent after transforming to symmetric and antisymmetric normal modes.  The symmetric mode is exactly dark, while the ancilla couples only to the antisymmetric bright mode.  Consequently, physical-mode transfer is equivalent to leaving the dark mode unchanged and applying parity to the bright mode.  
This SWAP–parity relation is a known consequence of the permutation–parity exchange at a balanced beam splitter and has also been stated explicitly as a parity operator in the antisymmetric mode~\cite{Campos_PRA_2005,Karaev2026}. The constrained single-ancilla problem considered Here, is different: the required relative normal-mode phase is not directly available as a beam splitter Hamiltonian.
It must be synthesized through the photon-number-dependent return amplitudes of the bright-mode JC dynamics.  This turns bosonic transfer into a finite-cutoff phase-synthesis problem.

This viewpoint leads to a sequence of exact reductions and benchmarks.  First, we derive the exact bright-mode-parity reduction that applies to arbitrary finite superpositions, not only to Fock states.  Second, we obtain closed finite-sum transfer formulas for Fock states, vacuum--Fock-state qubits, and general finite Fock superpositions.  Third, we show that resonant single-ancilla transfer is recurrence-limited: it is exact in the single-photon sector but cannot realize perfect finite-time transfer on any cutoff containing the two-photon sector, although arbitrarily accurate recurrences exist on every fixed finite cutoff.  Fourth, we show that detuned JC evolution provides a native two-parameter route to finite-cutoff bright-mode parity synthesis, with coherent process fidelity, ancilla leakage, and calibration sensitivity quantified separately.  Finally, we give a minimal Markovian open-system estimate, which identifies oscillator loss as the dominant dissipative limitation over the rates considered.

The framework is complementary to an ideal beam splitter swap.  A beam splitter Hamiltonian supplies the required relative normal-mode phase and remains the natural closed-system benchmark when high-quality exchange is available.  The setting addressed Here, is the restricted-interface case, where, direct exchange is absent, undesirable, or replaced by ancilla-mediated control.  In that case, the parity formulation gives a concrete target for resonant, detuned, dispersive, or photon-number-selective implementations and a transparent way to diagnose why a given implementation succeeds or fails.


\section{Single-ancilla interface and  bright-mode parity target}
\label{sec:model}

We consider the closed-system setting in which two bosonic oscillators are coupled to a single two-level ancilla with equal magnitude and opposite sign, as illustrated in Fig.~\ref{fig:schematic}.  The opposite sign is the physically important feature, because it selects the collective coordinate addressed by the ancilla.  The two oscillator modes are degenerate in the rotating interaction frame used below; Sec.~\ref{sec:native} introduces a detuning between the ancilla and the bright mode.  This model is distinct from a direct beam splitter exchange Hamiltonian, so the first task is to identify the transfer operation that the restricted ancilla-mediated coupling must synthesize.

Let \(a_1\) and \(a_2\) (\(a_1^\dagger \) and \(a_2^\dagger\)) be the annihilation (creation) operators of the two oscillators, and let \(\sigma_-\) (\(\sigma_+\) ) be the ancilla lowering (raising) operator.  In the rotating interaction frame, the resonant antisymmetric coupling is governed by the interaction Hamiltonian
\begin{equation}
H_{\rm int} = ig\left[ (a_1-a_2)\sigma_+ - (a_1^\dagger-a_2^\dagger)\sigma_- \right],
\label{eq:H0}
\end{equation}
where, \(g\) is the physical-mode coupling magnitude and we set \(\hbar=1\) throughout the work.  The overall JC phase in Eq.~\eqref{eq:H0} is a convention: by rephasing the ancilla one may equivalently write the coupling as \(g[(a_1-a_2)\sigma_+ + (a_1^\dagger-a_2^\dagger)\sigma_-]\).  This changes only the phase convention of the excited-ancilla branch.

The equal-and-opposite signs in Eq.~\eqref{eq:H0} suggest introducing the normal modes
\begin{equation}
a_+ = \left(a_1+a_2\right)/\sqrt{2}, \qquad
a_- = \left(a_1-a_2\right)/\sqrt{2} .
\label{eq:normal-modes}
\end{equation}
This transformation identifies the collective coordinate actually coupled to the ancilla.  Substituting Eq.~\eqref{eq:normal-modes} into Eq.~\eqref{eq:H0} gives
\begin{equation}
H_{\rm int} = i\sqrt{2}g \left( a_-\sigma_+ - a_-^\dagger\sigma_- \right),
\label{eq:Hbright}
\end{equation}
making explicit the bright/dark decomposition already implied by the difference coordinate in Eq.~\eqref{eq:H0}: the ancilla couples only to the antisymmetric mode \(a_-\),
while the symmetric mode 
\(a_+\)is an exact dark mode.
This separation is exact within Eq.~\eqref{eq:H0}, uses no dispersive or perturbative approximation, and is the reason the transfer problem admits closed finite-sum expressions. 

The normal-mode basis also identifies the exact transfer target.  Although our benchmarks focus on one-way transfer from oscillator 1 to oscillator 2 with oscillator 2 initially in vacuum, the natural target is the bare two-mode swap \(S_{12}\), defined by
\begin{equation}
S_{12}^\dagger a_1 S_{12}=a_2,
\qquad
S_{12}^\dagger a_2 S_{12}=a_1 .
\label{eq:bare-swap}
\end{equation}
Using $ a_1=(a_+ + a_-)/\sqrt{2}$ and $a_2=(a_+ - a_-)/\sqrt{2}$, this swap is equivalent to 
\begin{equation}
a_+\rightarrow a_+, \qquad a_-\rightarrow -a_- .
\label{eq:normal-mode-swap}
\end{equation}
The unitary that implements this sign flip is the bright-mode parity operator
\begin{equation}
\Pi_-=\exp(i\pi a_-^\dagger a_-).
\label{eq:bright-parity}
\end{equation}
Therefore
\begin{equation}
\Pi_-^\dagger a_+\Pi_-=a_+, \qquad \Pi_-^\dagger a_-\Pi_-= -a_-,
\label{eq:parity-action-normal}
\end{equation}
and hence
\begin{equation}
\Pi_-^\dagger a_1\Pi_-=a_2, \qquad \Pi_-^\dagger a_2\Pi_-= a_1 .
\label{eq:parity-action-physical}
\end{equation}
Thus, in the antisymmetric interface, the transfer task reduces exactly to bright-mode parity synthesis. This is the two-mode normal-mode version of the known SWAP–parity equivalence: a balanced mode transformation exchanges permutation and parity observables, and the SWAP can be written as parity in the antisymmetric mode~\cite{Campos_PRA_2005,Karaev2026}.

Equivalently, Eq.~\eqref{eq:parity-action-physical} implies that each Fock state is transferred as $\Pi_- |n\rangle_1|0\rangle_2 = |0\rangle_1|n\rangle_2$. By linearity, for any finite oscillator state
\(|\psi\rangle=\sum_{n=0}^{N}c_n|n\rangle\),
\begin{equation}
\Pi_-\left(|\psi\rangle_1|0\rangle_2\right)
=
|0\rangle_1|\psi\rangle_2 .
\label{eq:state-transfer-target}
\end{equation}
This is the sense in which the parity target is not restricted to Fock states. It is the exact transfer operation on the full finite-dimensional subspace.

The antisymmetric relative sign in Eq.~\eqref{eq:H0} is chosen so that bright-mode parity equals the bare swap.  If the ancilla instead coupled to the symmetric mode, bright-mode parity would implement the swap followed by a known local parity on the output oscillator.  This distinction is invisible in fixed Fock state transfer probabilities but matters for superpositions across photon number.  Appendix~\ref{app:phase-conventions} gives the explicit relation between the two conventions.

The same equivalence can be seen at the number state level.  A number state localized in one physical oscillator is distributed over many bright-mode occupations
\begin{equation}
\ket{n}_1 \ket{0}_2 = \frac{1}{2^{n/2}} \sum_{k=0}^{n} \sqrt{\binom{n}{k}}\, \ket{n-k}_+ \ket{k}_- .
\label{eq:fock-normal-expansion}
\end{equation}
The desired transferred state,
\begin{equation}
\ket{0}_1 \ket{n}_2 = \frac{1}{2^{n/2}} \sum_{k=0}^{n} (-1)^k \sqrt{\binom{n}{k}}\, \ket{n-k}_+ \ket{k}_- ,
\label{eq:target-normal-expansion}
\end{equation}
has the same dark-mode weights but an alternating sign in the bright-mode component.
Equations~\eqref{eq:fock-normal-expansion} and \eqref{eq:target-normal-expansion} show that transfer is a photon-number phase synthesis problem: every bright component $\ket{k}_-$ must acquire the parity phase $(-1)^k$.

For an input cutoff \(N\), no bright-mode occupation larger than \(N\) can appear in the normal-mode expansion of \(|\psi\rangle_1|0\rangle_2\).  Transfer on this cutoff therefore requires the controlled bright-mode dynamics to implement
\begin{equation}
|k\rangle_-|g\rangle
\longrightarrow
(-1)^k |k\rangle_-|g\rangle,
\qquad
0\leqslant k\leqslant N,
\label{eq:finite-cutoff-parity-condition}
\end{equation}
up to known oscillator-frame phases and an overall ancilla phase independent of \(k\).  The ancilla must return to a state that is independent of the bright-mode occupation; otherwise it carries which-\(k\) information and dephases superpositions across the finite Fock support.


Equation~\eqref{eq:finite-cutoff-parity-condition} is the finite-cutoff control target.  
While the underlying SWAP–parity identity is known~\cite{Campos_PRA_2005,Karaev2026}, Eq.~\eqref{eq:finite-cutoff-parity-condition} makes explicit the dynamical synthesis problem specific to the present restricted single-ancilla interface:  the ancilla-mediated bright-mode dynamics must realize the parity phase \((-1)^k\) for all  \(0\leqslant k\leqslant N\) while returning the ancilla unentangled.
The following section shows that static resonant evolution reaches this target exactly only in the single-photon sector and otherwise becomes a recurrence problem; Sec.~\ref{sec:native} then shows how detuning provides a more direct route to the same parity pattern.

\section{Static resonant transfer: exact amplitudes and recurrence limits}
\label{sec:static}

The purpose of this section is to compute what the static resonant Hamiltonian $H_{\rm int}$ actually produces.  The calculation separates three ingredients: the kinematic transformation between physical and normal modes, the independent resonant JC evolution of each bright-mode component, and the reduced state fidelities obtained after tracing over oscillator 1 and the ancilla.  This gives exact finite-sum expressions for the state and process benchmarks used below.
The selected states considered below are deliberately non-Gaussian probes of the transfer channel.  They test whether the protocol preserves finite Fock-space amplitudes and relative phases, rather than only Gaussian mode moments.

\subsection{Physical-basis transition amplitudes}
\label{subsec:transition-amplitudes}

We first separate the kinematic basis change from the resonant JC dynamics.  Consider a fixed total joint oscillator photon number \(m\). Now, a physical state with \(s\) photons in the second mode and \(m-s\) photons in the first mode can be expanded as
\begin{equation}
|m-s\rangle_1 |s\rangle_2 = \sum_{k=0}^{m} U^{(m)}_{sk} |m-k\rangle_+ |k\rangle_- .
\label{eq:physical-normal-expansion}
\end{equation}
The coefficients \(U^{(m)}_{sk}\) are the fixed-\(m\) matrix elements of the balanced physical-to-normal-mode basis transformation, equivalent to a \(50{:}50\) beam splitter rotation:
\begin{equation}
U^{(m)}_{sk} = \frac{1}{2^{m/2}} \sqrt{ \frac{(m-k)!\,k!}{(m-s)!\,s!} } \sum_r (-1)^{k-r} \binom{m-s}{r} \binom{s}{k-r},
\label{eq:Ucoeff}
\end{equation}
where
\begin{equation}
\max(0,\,k-s)\leqslant r\leqslant \min(k,\,m-s).
\label{eq:Ucoeff-range}
\end{equation}
These coefficients are purely kinematic: the index \(k\) is the bright-mode occupation and \(m-k\) is the dark-mode occupation.  All dynamical time dependence enters only through the bright-mode JC factors introduced below. The derivation and orthogonality properties of \(U^{(m)}\) are given in Appendix~\ref{app:transition-amplitudes}.

Throughout this work, we use the scaled time
$\tau=gt$. For $H_{\rm int}$ in Eq.~\eqref{eq:Hbright}, the bright mode evolves
independently in each doublet
\(\{|k,g\rangle_-,|k-1,e\rangle_-\}\).  
In this ordered basis,
\[
H_{\rm int}^{(k)} = g
\begin{pmatrix}
0 & -i\sqrt{2k}\\
i\sqrt{2k} & 0
\end{pmatrix}
= g\sqrt{2k}\,\sigma_y .
\]
The time-evolution operator in this doublet is
\[
U_k(\tau)=\exp\left(-iH_{\rm int}^{(k)}t\right) = \cos(\sqrt{2k}\tau)I - i\sin(\sqrt{2k}\tau)\sigma_y .
\]
For \(k\geqslant 1\),
\begin{equation*}
U_k(\tau) |k,g\rangle_- =
\cos(\sqrt{2k}\,\tau)|k,g\rangle_- + \sin(\sqrt{2k}\,\tau)|k-1,e\rangle_-\, .
\label{eq:resonant-doublet-evolution}
\end{equation*}
Therefore, the ground-ancilla return factor is
\begin{equation}
r_k^{\rm stat}(\tau) = \cos\!\left(\sqrt{2k}\,\tau\right),
\label{eq:static-return-factor}
\end{equation}
and, with the phase convention of Eq.~\eqref{eq:H0}, the excited-ancilla factor is
\begin{equation}
e_k^{\rm stat}(\tau) = \sin\!\left(\sqrt{2k}\,\tau\right).
\label{eq:static-excited-factor}
\end{equation}
We take \(r_0^{\rm stat}=1\) and \(e_0^{\rm stat}=0\), since the zero-bright-photon component does not couple to the ancilla.  The key point is that all resonant bright-mode dynamics are contained in \(r_k^{\rm stat}\) and \(e_k^{\rm stat}\); the remaining sums only transform between normal and physical-mode bases. A different JC phase convention would only rephase the
excited-ancilla branch, provided the same convention is used throughout.

Starting from \(|m\rangle_1|0\rangle_2|g\rangle\), the amplitude to end with \(s\) photons in mode 2, \(m-s\) photons in mode 1, and the ancilla in \(|g\rangle\) is
\begin{equation}
G_{m,s}^{\rm stat}(\tau) = \sum_{k=0}^{m} U^{(m)}_{sk} U^{(m)}_{0k} r_k^{\rm stat}(\tau), \quad 0\leqslant s\leqslant m .
\label{eq:Gms}
\end{equation}
The amplitude to end with \(s\) photons in mode 2, \(m-1-s\) photons in mode 1, and one excitation stored in the ancilla is
\begin{equation}
E_{m,s}^{\rm stat}(\tau) = \sum_{k=1}^{m} U^{(m-1)}_{s,k-1} U^{(m)}_{0k} e_k^{\rm stat}(\tau), \quad 0\leqslant s\leqslant m-1 .
\label{eq:Ems}
\end{equation}
The \(G\) branch leaves the ancilla in the ground state; the \(E\) branch leaves one excitation in the ancilla and therefore one fewer photon in the oscillator pair.  These branch amplitudes are kept separately because the target-mode state is obtained after tracing over oscillator 1 and the ancilla; only paths with the same oscillator-1 photon number and the same final ancilla state can interfere.


\subsection{Fock-state transfer}
\label{subsec:fock-transfer}

For a Fock input \(|n\rangle_1|0\rangle_2\ket{g}\), the target is \(|0\rangle_1|n\rangle_2\ket{g}\).  The transfer amplitude is the special case \(G_{n,n}\), which gives
\begin{equation}
A^{\rm F,stat}_n(\tau) = G_{n,n}^{\rm stat}(\tau) = \frac{1}{2^n} \sum_{k=0}^{n} (-1)^k \binom{n}{k} \,r_k^{\rm stat}(\tau).
\label{eq:fock-amplitude}
\end{equation}
The corresponding Fock-state transfer fidelity is
\begin{equation}
F^{\rm F,stat}_n(\tau) = \left| A^{\rm F,stat}_n(\tau)
\right|^2 .
\label{eq:fock-fidelity}
\end{equation}
For the static resonant protocol, this simply becomes
\begin{equation}
F^{\rm F,stat}_n(\tau) = \left| \frac{1}{2^n} \sum_{k=0}^{n} (-1)^k \binom{n}{k} \cos\!\left(\sqrt{2k}\,\tau\right)
\right|^2 .
\label{eq:fock-fidelity-full}
\end{equation}
Equation~\eqref{eq:fock-amplitude} is the simplest manifestation of the parity requirement.  The initial physical Fock-state samples all bright occupations \(k=0,\ldots,n\) with binomial weights.  Perfect transfer requires all sampled bright components to acquire the parity phase \((-1)^k\).  The resonant protocol supplies instead the real return factors \(\cos(\sqrt{2k}\tau)\), so higher Fock states require simultaneous alignment of several square-root Rabi frequencies. 

A Fock state therefore tests population transfer in one photon-number sector, but it does not by itself test phase coherence between different photon-number components.

\subsection{Fock-state qubit transfer}
\label{subsec:fock-qubit-transfer}

We next consider the vacuum--\(n\)-photon Fock-state qubit
\begin{equation}
|\psi^{\rm FQ}_n\rangle = \left(|0\rangle+|n\rangle\right)/\sqrt{2},
\qquad n\geqslant 1 .
\label{eq:fock-qubit}
\end{equation}
This state tests more than population transfer.  The \(n\)-photon component must be delivered to the target oscillator, but it must also retain the correct phase relative to the vacuum component.  The fidelity is therefore sensitive both to the Fock-state-transfer amplitude and to residual branches that leave the target oscillator in vacuum after tracing out the unobserved degrees of freedom.

Besides the Fock-state-transfer amplitude \(A^{\rm F,stat}_{n}\), define the ground-ancilla return amplitude of the \(n\)-photon branch into the target-vacuum sector,
\begin{equation}
R_n^{\rm stat}(\tau) = \frac{1}{2^n} \sum_{k=0}^{n} \binom{n}{k} \,r_k^{\rm stat}(\tau) =
\frac{1}{2^n}
\sum_{k=0}^{n}
\binom{n}{k}
\cos\!\left(\sqrt{2k}\tau\right).
\label{eq:Rn}
\end{equation}
The corresponding target-vacuum branch with the ancilla excited is fixed by the derivative of \(R_n\).  Indeed,
\begin{equation}
\frac{dR_n^{\rm stat}}{d\tau}
=
-\frac{1}{2^n}
\sum_{k=1}^{n}
\binom{n}{k}
\sqrt{2k}
\sin\!\left(\sqrt{2k}\tau\right)
=
-\sqrt{n}\,Q_n^{\rm stat},
\label{eq:Q-derivative-relation}
\end{equation}
where
\begin{align}
Q_n^{\rm stat}(\tau) &= \frac{1}{2^n} \sum_{k=1}^{n} \binom{n}{k} \sqrt{\frac{2k}{n}} e_k^{\rm stat}(\tau) \nonumber\\
&= \frac{1}{2^n} \sum_{k=1}^{n} \binom{n}{k} \sqrt{\frac{2k}{n}} \sin\!\left(\sqrt{2k}\tau\right).
\label{eq:Qn}
\end{align}
Equivalently, $G_{n,0}^{\rm stat} = R_n^{\rm stat}$ and  $E_{n,0}^{\rm stat} = Q_n^{\rm stat}$. 
The total output-vacuum contribution of the \(n\)-photon branch is
\begin{equation}
P_{n,{\rm stat}}^{(0)}(\tau) = \left|R_n^{\rm stat}(\tau)\right|^2 + \left|Q_n^{\rm stat}(\tau)\right|^2 .
\label{eq:fock-qubit-vacuum-sector}
\end{equation}
Here, \(R_n^2\) is the ground-ancilla target-vacuum contribution, while \(Q_n^2\) is the excited-ancilla target-vacuum contribution.

Tracing over the residual oscillator and the ancilla gives the exact Fock-state qubit transfer fidelity
\begin{equation}
F_n^{\rm FQ,stat}(\tau) = \frac{1}{4} \left[ \left| 1+ A_n^{\rm F,stat}(\tau) \right|^2 + P_{n,{\rm stat}}^{(0)}(\tau) \right].
\label{eq:fock-qubit-fidelity}
\end{equation}
For the static resonant protocol \(A_n^{\rm F,stat}\), \(R_n^{\rm stat}\), and \(Q_n^{\rm stat}\) are real, so Eq.~\eqref{eq:fock-qubit-fidelity} reduces to the real-valued expression written below.  
\begin{align}
F_n^{\rm FQ,stat}(\tau) =& \frac{1}{4} \Bigg\{ \left[ 1+ \frac{1}{2^n} \sum_{k=0}^{n} (-1)^k \binom{n}{k} \cos\!\left(\sqrt{2k}\,\tau\right) \right]^2
\nonumber\\
&+
\left[ \frac{1}{2^n} \sum_{k=0}^{n} \binom{n}{k} \cos\!\left(\sqrt{2k}\,\tau\right) \right]^2 \nonumber\\
&+ \frac{1}{n} \left[ \frac{1}{2^n} \sum_{k=1}^{n} \binom{n}{k} \sqrt{2k}\, \sin\!\left(\sqrt{2k}\,\tau\right) \right]^2 \Bigg\}.
\label{eq:fock-qubit-fidelity-expanded}
\end{align}
The first term in Eq.~\eqref{eq:fock-qubit-fidelity} is the coherent overlap between the stationary vacuum branch and the transferred \(n\)-photon branch.  The second term is an incoherent contribution from branches in which the \(n\)-photon input leaves the target oscillator in vacuum after the residual oscillator and ancilla are traced out.

This form makes explicit why the Fock-state qubit is a stricter benchmark than a Fock-state: the fidelity depends on successful delivery of the \(n\)-photon component and on preserving its phase coherence with the vacuum.

\subsection{Finite Fock superposition states}
\label{subsec:finite-superpositions}

The static transition amplitudes \(G_{m,s}^{\rm stat}\) and \(E_{m,s}^{\rm stat}\) also determine the transfer fidelity of any finite Fock superposition
\begin{equation}
|\psi\rangle = \sum_{m=0}^{N} c_m|m\rangle
\label{eq:finite-input-state}
\end{equation}
initially placed in mode 1, with mode 2 in vacuum and the ancilla in \(|g\rangle\).  The target-mode fidelity is computed after tracing over oscillator 1 and the
ancilla.  Therefore, amplitudes add coherently only when they leave the same unobserved labels:
the same oscillator 1 photon number and the same final ancilla state.  Grouping paths by that label gives
\begin{align}
F_\psi^{\rm stat}(\tau) &= \sum_{p=0}^{N} \left| \sum_{s=0}^{N-p} c_s^*\, c_{p+s} \,G_{p+s,s}^{\rm stat}(\tau) \right|^2 \nonumber\\
&\quad+ \sum_{p=0}^{N-1} \left|
\sum_{s=0}^{N-p-1} c_s^* \,c_{p+s+1}\, E_{p+s+1,s}^{\rm stat}(\tau) \right|^2 .
\label{eq:static-general-state-fidelity}
\end{align}
The first sum collects ground-ancilla branches; the second collects excited-ancilla branches.  Thus, Eq.~\eqref{eq:static-general-state-fidelity} is the bookkeeping rule that decides which coherences survive the trace over the residual oscillator and ancilla.

A useful selected-state benchmark is the canonical finite superposition state
\begin{equation}
|\Phi_N\rangle = \frac{1}{\sqrt{N+1}} \sum_{m=0}^{N} |m\rangle .
\label{eq:canonical-superposition}
\end{equation}
This is not a substitute for the full cutoff process fidelity.  It is useful because it samples all photon numbers in the cutoff with equal amplitudes and therefore tests many relative phases at once.

The transfer fidelity of \(|\Phi_N\rangle\) can be written in a compact form. The coherent no-residual transfer amplitude,
\begin{equation}
A_{\Phi_N}^{\rm stat}(\tau) = \frac{1}{N+1} \sum_{m=0}^{N} A_m^{\rm F,stat}(\tau),
\label{eq:static-canonical-amplitude}
\end{equation}
where, \(A_m^{\rm F,stat}\) is the Fock-transfer amplitude in Eq.~\eqref{eq:fock-amplitude}.  This is the branch in which the residual oscillator is left in vacuum and the ancilla returns to \(|g\rangle\).  The remaining contributions come from branches that leave either photons in oscillator 1 or one excitation in the ancilla, but still have nonzero overlap with the target state after those degrees of freedom are traced out.  Define
\begin{align}
G_{N,p}^{\rm stat}(\tau) &= \frac{1}{N+1} \sum_{s=0}^{N-p} G_{p+s,s}^{\rm stat}(\tau), \quad
p=1,\ldots,N,
\label{eq:static-canonical-G}
\\
E_{N,p}^{\rm stat}(\tau) &= \frac{1}{N+1} \sum_{s=0}^{N-p-1} E_{p+s+1,s}^{\rm stat}(\tau),
\,\, p=0,\ldots,N-1 .
\label{eq:static-canonical-E}
\end{align}
The total residual contribution is then
\begin{equation}
P_{\Phi_N,{\rm stat}}^{\rm res}(\tau) = \sum_{p=1}^{N} \left| G_{N,p}^{\rm stat}(\tau) \right|^2
+ \sum_{p=0}^{N-1} \left| E_{N,p}^{\rm stat}(\tau) \right|^2 .
\label{eq:static-canonical-residual}
\end{equation}
Thus, the canonical finite-superposition fidelity is
\begin{equation}
F_{\Phi_N}^{\rm stat}(\tau) = \left| A_{\Phi_N}^{\rm stat}(\tau) \right|^2 + P_{\Phi_N,{\rm stat}}^{\rm res}(\tau).
\label{eq:static-canonical-fidelity}
\end{equation}
This form is exact.  It also makes the physical content clearer than the fully expanded double sum.  The first term is the coherent transfer of the desired finite superposition.  The second term collects all residual unobserved sectors that still contribute to the reduced target-mode fidelity.  The canonical state is therefore more stringent than a standard Fock state or a Fock-state qubit because it tests phase coherence across all photon-number sectors \(m=0,\ldots,N\), while also retaining sensitivity to how imperfect transfer distributes amplitude among residual oscillator and ancilla sectors.

\subsection{Static no-go, recurrence, and finite-time benchmarks}
\label{subsec:static-benchmarks}

The finite-sum expressions above show that perfect static transfer on a cutoff
requires all bright-mode return factors sampled by that cutoff to reproduce the parity pattern simultaneously. For \(0\leqslant n\leqslant N\), this requires
\begin{equation}
r_k^{\rm stat}(\tau) = \cos\!\left(\sqrt{2k}\,\tau\right) = (-1)^k, \quad k=1,\ldots,N .
\label{eq:static-condition-sectionIII}
\end{equation}
When this condition holds, each relevant cosine is \(\pm1\), and therefore
$\sin\!\left(\sqrt{2k}\,\tau\right)=0$.

Thus, Eq.~\eqref{eq:static-condition-sectionIII} enforces both the desired bright-mode parity phase and the return of the ancilla to \(|g\rangle\).

For \(N=1\), Eq.~\eqref{eq:static-condition-sectionIII} is satisfied at
\begin{equation}
\tau = \frac{(2q+1)\pi}{\sqrt{2}}, \qquad q\in\mathbb{Z}_{\geqslant 0}.
\label{eq:N1-perfect-time}
\end{equation}
For \(N\geqslant 2\), exact finite-time transfer is impossible.  The \(k=1\) and \(k=2\) conditions would require
\begin{equation}
\sqrt{2}\,\tau=(2m+1)\pi, \quad 2\tau=2q\pi ,
\label{eq:no-go-conditions}
\end{equation}
for integers \(m,q\).  Eliminating \(\tau\) gives
\[
q\sqrt{2}=2m+1,
\]
which has no integer solution.  Since the contradiction already appears in the \(N=2\) cutoff, it applies to every larger cutoff.

This no-go statement does not rule out arbitrarily accurate long-time recurrences on a fixed finite cutoff.  Because only finitely many bright-mode frequencies are involved, the corresponding phase vector can return arbitrarily close to the parity-compatible phase pattern, provided the target phases respect the rational relations among those frequencies.  Appendix~\ref{app:no-go-recurrence} verifies this compatibility explicitly.  Thus, the static resonant protocol has unit supremum on every finite cutoff, but for \(N\geqslant 2\) that supremum is not attained at a finite time.  The operational question is therefore not whether perfect recurrence exists asymptotically, but how well the protocol performs in a finite time window.

\begin{figure}[t]
\centering
\includegraphics[width=0.95\columnwidth]{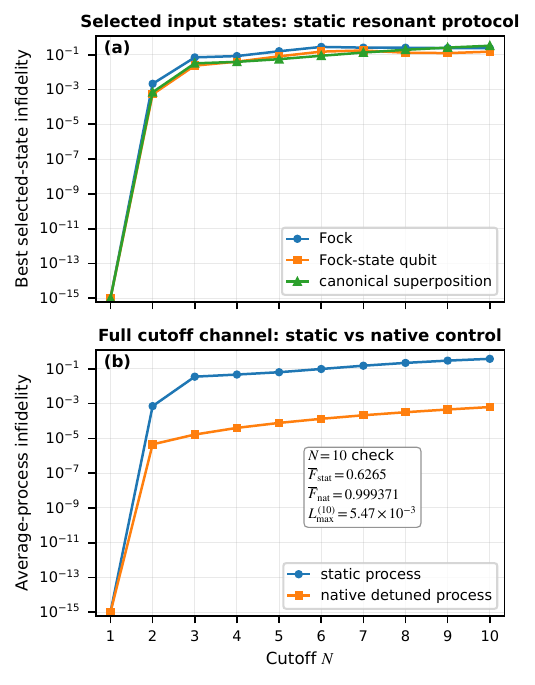}
\caption{
Finite-time transfer benchmarks over the window \(0\leqslant\tau\leqslant50\). For native detuned control, the optimization uses \(0\leqslant\delta\leqslant 100\). (a) Best selected-state infidelity for Fock states, Fock-state qubits, and canonical finite superposition inputs under the static resonant protocol. Each point is optimized independently for the corresponding selected input family, so this panel is not a process benchmark. (b) Average process infidelity on the full cutoff space.   The static resonant protocol is recurrence-limited because its bright-mode return factors oscillate as \(\cos(\sqrt{2k}\tau)\).
}
\label{fig:finite-time}
\end{figure}

The selected state fidelities derived above already give useful finite-window benchmarks.  Figure~\ref{fig:finite-time}(a) shows the best static resonant infidelity within \(0\leqslant\tau\leqslant50\) for Fock states, Fock-state qubits, and canonical finite-superposition states.  For each cutoff \(N\) and for each selected-state family, we plot the smallest infidelity obtained by choosing the best value of \(\tau\) in this window.  The three curves differ because the same imperfect bright-mode phases are sampled with different weights: population weights for Fock states, vacuum--\(n\)-photon coherence for Fock-state qubits, and many cross-sector coherences for canonical finite superpositions.

Selected state fidelities are diagnostic, but they do not certify the transfer operation on the full cutoff space.  A given parameter choice can transfer particular test states well while still failing on other superpositions in the same cutoff. For a channel-level benchmark, we therefore use the average process fidelity with respect to the ideal transfer channel.  On
 \begin{equation}
 {\cal H}_N=\mathrm{span}\{|0\rangle,\ldots,|N\rangle\}, \quad d=N+1 ,
 \label{eq:cutoff-channel-space}
 \end{equation}
the average process fidelity is related to the entanglement fidelity by the standard formula~\cite{Horodecki1999,Nielsen2002}
\begin{equation}
\overline F_N = \frac{dF_e+1}{d+1}.
\label{eq:average-fidelity-convention}
\end{equation}
Here, \(F_e\) is the entanglement fidelity
obtained by applying the actual channel, corrected by the inverse ideal transfer map, to one half of a maximally entangled state on \({\cal H}_N\).  Equivalently, if the actual transfer channel has Kraus operators \(K_j\) and the ideal transfer is \(U_{\rm ideal}\), then
\begin{equation}
F_e = \frac{1}{d^2} \sum_j \left| {\rm Tr} \left( U_{\rm ideal}^\dagger K_j  \right) \right|^2 .
\label{eq:entanglement-fidelity-kraus}
\end{equation}

For the static resonant protocol, only the no-residual ground-ancilla branch has nonzero trace overlap with the ideal transfer channel.  On the input basis state \(|m\rangle\), this branch contributes the diagonal transfer amplitude \(A_m^{\rm F,stat}(\tau)\).  Hence
\begin{equation}
F_{e,{\rm stat}}^{(N)}(\tau) = \frac{1}{(N+1)^2} \left| \sum_{m=0}^{N} A_m^{\rm F,stat}(\tau) \right|^2 ,
\label{eq:static-entanglement-fidelity}
\end{equation}
and therefore
\begin{equation}
\overline F_N^{\rm stat}(\tau) = \frac{(N+1)F_{e,{\rm stat}}^{(N)}(\tau)+1}{N+2}.
\label{eq:static-average-fidelity}
\end{equation}
This quantity is stricter than the selected-state fidelities in Eqs.~\eqref{eq:fock-fidelity}, \eqref{eq:fock-qubit-fidelity}, and \eqref{eq:static-canonical-fidelity}, because it tests coherent transfer of an arbitrary state in the full cutoff subspace.

Figure~\ref{fig:finite-time}(b) compares the corresponding channel-level benchmarks.  The native detuned curve, derived in Sec.~\ref{sec:native}, is included tHere, as a reference.  The contrast between the static and native curves follows from the analytical structure above.  In the static protocol, \(r_k^{\rm stat}(\tau)\) is real and oscillatory with frequency \(\sqrt{k}\), so high process fidelity requires recurrence of many square-root frequencies. In the detuned native protocol of Sec.~\ref{sec:native}, the return amplitude is complex, and its phase is approximately linear in \(k\), which targets the parity condition more directly.

\section{Native detuned synthesis of bright-mode parity}
\label{sec:native}

The static resonant protocol fails at finite time because it tries to synthesize bright-mode parity using real JC return factors whose frequencies scale as \(\sqrt{k}\).  The missing resource is not population exchange itself, but a controllable photon-number phase on the bright mode.  We now show that the same antisymmetric single-ancilla interface supplies such a phase when the ancilla is operated off resonance.

We consider the detuned bright-mode Hamiltonian
\begin{equation}
H_\Delta = \Delta |e\rangle\langle e| + i\sqrt{2}g
\left( a_-\sigma_+ - a_-^\dagger\sigma_- \right),
\label{eq:Hdetuned}
\end{equation}
where, \(\Delta\) is the detuning between the ancilla and the bright mode in the rotating frame of Sec.~\ref{sec:model}.  The resonant protocol is recovered at \(\Delta=0\).  We use the dimensionless detuning $\delta=\Delta/g$.
The purpose of Eq.~\eqref{eq:Hdetuned} is not to assume an ideal number-phase Hamiltonian.  Rather, it asks how accurately the native bright-mode JC interaction itself can approximate the parity target when only two control parameters, \(\delta\) and \(\tau\), are available. Here, “native” means that no additional number-selective or optimal-control drive is assumed; the approximation is produced by the detuned single-ancilla JC evolution.

The Hamiltonian remains block diagonal.  For each bright excitation number \(k\geqslant 1\), the relevant doublet is still
$\{|k,g\rangle_-,|k-1,e\rangle_-\}$.
In this basis, the dimensionless doublet Hamiltonian is
\begin{equation}
H_\delta^{(k)} =
\begin{pmatrix}
0 & -i\sqrt{2k}\\
i\sqrt{2k} & \delta
\end{pmatrix},
\label{eq:native-doublet-matrix}
\end{equation}
so that the doublet propagator is $U_k(\delta,\tau) = \exp[-iH_\delta^{(k)} \tau]$.
The exact ground-ancilla return amplitude is now of the form
\begin{equation}
r_k^{\rm nat}(\delta,\tau) = e^{-i\delta\tau/2}
\left[ \cos\!\left(\tfrac{\Lambda_k\tau}{2}\right) + i\tfrac{\delta}{\Lambda_k} \sin\!\left(\tfrac{\Lambda_k\tau}{2}\right) \right],
\label{eq:native-rk}
\end{equation}
where, $\Lambda_k=\sqrt{\delta^2+8k}$.
For \(k=0\), \(r_0^{\rm nat}=1\).  The corresponding excited-ancilla amplitude is
\begin{equation}
e_k^{\rm nat}(\delta,\tau) = \tfrac{2\sqrt{2k}}{\Lambda_k}
e^{-i\delta\tau/2} \sin\!\left(\tfrac{\Lambda_k\tau}{2}\right), \qquad k\geqslant 1,
\label{eq:native-ek}
\end{equation}
with \(e_0^{\rm nat}=0\).  This phase convention is chosen so that \(e_k^{\rm nat}\to e_k^{\rm stat}\) at \(\delta=0\).  The probability that the same bright component leaves the ancilla excited is
\begin{equation}
L_k(\delta,\tau) = \left| e_k^{\rm nat}(\delta,\tau) \right|^2 = \frac{8k}{\Lambda_k^2} \sin^2\!\left(\frac{\Lambda_k\tau}{2}\right).
\label{eq:native-Lk}
\end{equation}
We monitor leakage on the cutoff \(N\) using
\begin{equation}
L_{\max}^{(N)}(\delta,\tau) = \max_{1\leqslant k\leqslant N}L_k(\delta,\tau).
\label{eq:native-Lmax}
\end{equation}
This quantity is evaluated separately from the coherent process-fidelity
objective.

Equations~\eqref{eq:native-rk} and \eqref{eq:native-Lk} explain why detuning helps.  In the large-detuning regime, the ancilla is only weakly populated and the bright mode acquires an approximately photon-number-linear phase.  More explicitly, after removing a \(k\)-independent phase,
\begin{equation}
\arg r_k^{\rm nat}(\delta,\tau) \simeq \frac{2k}{\delta}\tau + O\!\left(\frac{k^2\tau}{\delta^3}\right),
\label{eq:native-dispersive-phase}
\end{equation}
while the leakage envelope obeys
\begin{equation}
L_k(\delta,\tau) \leqslant \frac{8k}{\delta^2+8k} \simeq \frac{8k}{\delta^2}.
\label{eq:native-leakage-envelope}
\end{equation}
Thus, detuning changes the character of the control problem.  The resonant protocol waits for a recurrence of several \(\sqrt{k}\)-dependent phases, whereas the detuned protocol directly generates the approximately linear phase structure required for bright-mode parity.

The selected-state fidelities of Sec.~\ref{sec:static} carry over with the same physical-to-normal-mode coefficients and the same bookkeeping of oscillator 1 and ancilla branches.  The only dynamical change is
\[
r_k^{\rm stat}(\tau),\,e_k^{\rm stat}(\tau)
\longrightarrow
r_k^{\rm nat}(\delta,\tau),\,e_k^{\rm nat}(\delta,\tau).
\]
Because the native amplitudes are generally complex, coherent terms such as the vacuum--Fock interference in Eq.~\eqref{eq:fock-qubit-fidelity} must be kept as modulus squares, and the phase convention of \(e_k^{\rm nat}\) must be used consistently.  Since the native benchmarks below are channel-level benchmarks, we now turn directly to the finite-cutoff process fidelity.

\subsection{Finite-cutoff process optimization}
\label{subsec:native-process}

For the channel-level benchmarks, we use the same cutoff-channel convention and average-fidelity normalization defined in Eqs.~\eqref{eq:cutoff-channel-space} and \eqref{eq:average-fidelity-convention}.  The entanglement-fidelity trace with respect to the ideal transfer channel is carried by the no-residual, ground-ancilla transfer branch.  Branches that leave photons in oscillator 1 or excitation in the ancilla have zero trace overlap with the ideal transfer operation and are diagnosed separately through Eq.~\eqref{eq:native-Lmax}.  Thus
\begin{equation}
F_{e,{\rm nat}}^{(N)}(\delta,\tau) = \frac{1}{(N+1)^2} \left| \sum_{m=0}^{N} A_m^{\rm F,nat}(\delta,\tau) \right|^2 ,
\label{eq:native-entanglement-fidelity}
\end{equation}
and the corresponding average process fidelity is
\begin{equation}
\overline F_N^{\rm nat}(\delta,\tau) = \frac{(N+1)F_{e,{\rm nat}}^{(N)}(\delta,\tau)+1}{N+2}.
\label{eq:native-average-fidelity}
\end{equation}
The optimization maximizes Eq.~\eqref{eq:native-average-fidelity} only. The coherent phase synthesis and ancilla leakage remain separate diagnostics.

The main-text benchmarks use the finite search window $0\leqslant \tau\leqslant50$, $0\leqslant\delta\leqslant 100$.
This window defines an operational finite-time comparison, not a fundamental limit of the model.  Appendix~\ref{app:window-dependence} checks alternative time and detuning windows, and Appendix~\ref{app:native-reproducibility} gives the numerical procedure, including grid search, local refinement, and matrix-exponential consistency checks.

The channel-level improvement over the static resonant protocol is shown in Fig.~\ref{fig:finite-time}(b).  The distinction follows directly from the bright-mode return factors: \(r_k^{\rm stat}\) is a real \(\sqrt{k}\)-dependent recurrence factor, whereas finite detuning makes \(r_k^{\rm nat}\) complex with an approximately linear phase over the optimized cutoff.

For the \(N=10\) cutoff, the best-found finite-window optimum from the stated search protocol is
\begin{equation}
\tau_{\rm opt}=49.968175624787,
\,\,
\delta_{\rm opt}=31.434701781850 .
\label{eq:native-N10-optimum}
\end{equation}
At this point,
\begin{align}
\overline F_{10}^{\rm nat} &= 0.999370508862, \nonumber\\
1-\overline F_{10}^{\rm nat} &= 6.29\times10^{-4}, \nonumber\\
L_{\max}^{(10)} &= 5.47\times10^{-3}.
\label{eq:native-N10-values}
\end{align}
These are ideal closed-system finite-window benchmarks.  They quantify the coherent phase-synthesis error of the restricted single-ancilla protocol without loss and decoherence.  They show that the detuned interface can approximate the bright-mode parity target with high finite-cutoff process fidelity.  They do not imply an absolute speed advantage over an ideal beam splitter Hamiltonian. 

The native detuned evolution is most directly diagnosed by the bright-mode return spectrum.  For a cutoff \(N\), define the best-fit global phase
\begin{equation}
\chi = \arg \left[ \sum_{k=0}^{N} (-1)^k r_k^{\rm nat}(\delta_{\rm opt},\tau_{\rm opt}) \right].
\label{eq:global-phase-fit}
\end{equation}
The aligned bright-mode phase is
\begin{equation}
\widetilde\phi_k = \arg \left[ e^{-i\chi} r_k^{\rm nat}(\delta_{\rm opt},\tau_{\rm opt}) \right],
\label{eq:aligned-phase}
\end{equation}
and the residual phase error relative to ideal parity is
\begin{equation}
\Delta\phi_k = \arg \left[ e^{-i\chi} r_k^{\rm nat}(\delta_{\rm opt},\tau_{\rm opt}) e^{-i\pi k} \right].
\label{eq:residual-phase}
\end{equation}
Exact bright-mode parity up to a global phase would give \(\Delta\phi_k=0\) and \(L_k=0\) for every \(k\leqslant N\).

\begin{figure}[t]
\centering
\includegraphics[width=\columnwidth]{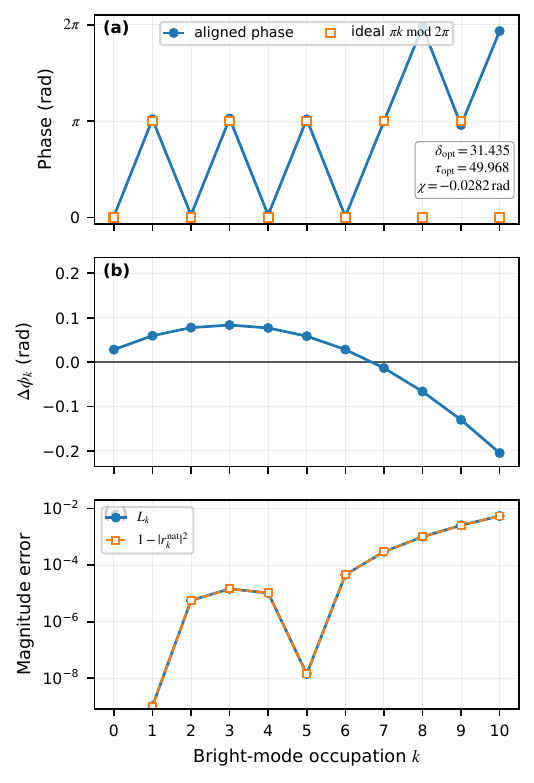}
\caption{
Phase residual spectrum for the best-found native detuned evolution at cutoff \(N=10\), using \(0\leqslant\tau\leqslant50\) and \(0\leqslant\delta\leqslant 100\). (a) Aligned phase of \(r_k^{\rm nat}\), after removing the best-fit global phase \(\chi\), compared with the ideal parity pattern \(\pi k\) modulo \(2\pi\).  The phase is displayed in \([0,2\pi)\).  (b) Residual phase error \(\Delta\phi_k\); exact parity would give zero for all \(k\leqslant N\). (c) Magnitude error, shown as both \(L_k\) and \(1-|r_k^{\rm nat}|^2\), which coincide by doublet unitarity.  The largest residual phase and magnitude errors occur near the cutoff.
}
\label{fig:phase-spectrum}
\end{figure}

Figure~\ref{fig:phase-spectrum} diagnoses the best-found \(N=10\) native detuned evolution.  The aligned phases follow the parity pattern, while the residual phase and magnitude errors remain smallest at low and intermediate \(k\) and grow near the cutoff.  This explains why encoded states with appreciable high-photon support are more sensitive to the native finite-cutoff approximation.

The phase-spectrum diagnostic provides the bridge to encoded state transfer. The process fidelity in Eq.~\eqref{eq:native-average-fidelity} averages residual phase and leakage errors over the full cutoff channel.  A particular bosonic code samples those errors according to its own photon-number support. Compact low-photon support is therefore less demanding, whereas broad or high-photon support is more sensitive to the residual \(\Delta\phi_k\) and \(L_k\) near the cutoff.  Section~\ref{sec:codes} makes this connection explicit for representative RSBC subspaces with different Fock-support patterns.

\section{Encoded-state benchmarks: Fock support and phase errors}
\label{sec:codes}

The phase-spectrum diagnostic of Sec.~\ref{sec:native} shows what the native detuned evolution synthesizes: an approximate bright-mode parity phase over a finite cutoff, with residual phase and leakage errors that grow near the upper end of the bright-mode occupation range.  Encoded oscillator states sample those residual errors according to their photon-number support.  A compact low-photon encoding can therefore be less demanding than an encoding with comparable weight on widely separated high-photon components, even when both lie inside the same cutoff.  This support dependence is especially important for non-Gaussian bosonic encodings, whose logical information is carried by structured Fock amplitudes and relative phases rather than by Gaussian first and second moments.  We use three finite-dimensional diagnostic subspaces to make this dependence explicit.

Our purpose is to compare three representative Fock-support patterns of RSBCs: a compact lowest-order binomial subspace, a factorially weighted truncated cat subspace, and an equal-weight fourfold rotation-symmetric subspace with appreciable high-photon support.  These are diagnostic subspaces, not complete fault-tolerant transfer protocols.  We do not include active recovery, repeated syndrome extraction, or code-specific optimization.  Instead, each subspace is evaluated under the same process-optimized transfer map studied in Secs.~\ref{sec:static} and \ref{sec:native}.  This keeps the comparison clean: the transfer map is optimized once for the full \(N=10\) cutoff channel and is then restricted to encoded subspaces with different photon-number structure.

The first example is the lowest-order binomial code~\cite{Michael2016,laha_binomial_2026},
\begin{equation}
|0^{\rm bin}_L\rangle = \frac{|0\rangle+|4\rangle}{\sqrt{2}}, \qquad |1^{\rm bin}_L\rangle = |2\rangle .
\label{eq:binomial-code}
\end{equation}
Although more general binomial codewords can be prepared using multiphoton spin--boson interactions~\cite{laha_binomial_2026}, we use this lowest-order code only as a compact diagnostic subspace.  Its relation to the state benchmarks of Sec.~\ref{sec:static} is transparent.  The logical zero is exactly the vacuum--four-photon Fock-state qubit \(|\psi^{\rm FQ}_{4}\rangle\), while the logical one is the Fock-state \(|2\rangle\).  Thus, this subspace tests both ingredients already isolated in the static theory: phase-sensitive vacuum--Fock coherence through \(|0^{\rm bin}_L\rangle\), and ordinary Fock-state transfer through \(|1^{\rm bin}_L\rangle\).  Although the support is compact, the balanced superposition makes the logical zero sensitive to residual bright-mode phase errors.

The second example is a truncated even/odd cat subspace.  For real coherent-state amplitude \(\alpha=\sqrt2\) and cutoff \(n\leqslant10\),  we define
\begin{align}
|C^+_\alpha\rangle_{\rm tr} &= {\cal N}_+ \sum_{m=0}^{5} \frac{\alpha^{2m}}{\sqrt{(2m)!}} |2m\rangle ,
\label{eq:even-cat-truncated} \\
|C^-_\alpha\rangle_{\rm tr} &= {\cal N}_- \sum_{m=0}^{4} \frac{\alpha^{2m+1}}{\sqrt{(2m+1)!}} |2m+1\rangle ,
\label{eq:odd-cat-truncated}
\end{align}
where, \({\cal N}_\pm\) are normalization constants.  This subspace samples more photon numbers than the lowest-order binomial code, but the high-\(n\) components are suppressed by the factorial weights.  It therefore probes an intermediate regime: the support is broader, but the transfer fidelity remains dominated by low- and intermediate-Fock components.

The third example is an equal-weight fourfold rotation-symmetric subspace,
\begin{equation}
|0^{\rm 4fold}_L\rangle = \frac{|0\rangle+|4\rangle+|8\rangle}{\sqrt{3}}, \quad |1^{\rm 4fold}_L\rangle = \frac{|2\rangle+|6\rangle+|10\rangle}{\sqrt{3}} .
\label{eq:rotation-code}
\end{equation}
This deliberately simple example isolates the feature most demanding for the present transfer primitive: appreciable weight on separated high-photon components.  In particular, \(|1^{\rm 4fold}_L\rangle\) contains the cutoff component \(|10\rangle\) with weight \(1/3\), and \(|0^{\rm 4fold}_L\rangle\) contains \(|8\rangle\) with the same weight. Residual phase or leakage errors that are largest near the cutoff therefore directly affect this subspace.

\begin{figure}[t]
\centering
\includegraphics[width=\columnwidth]{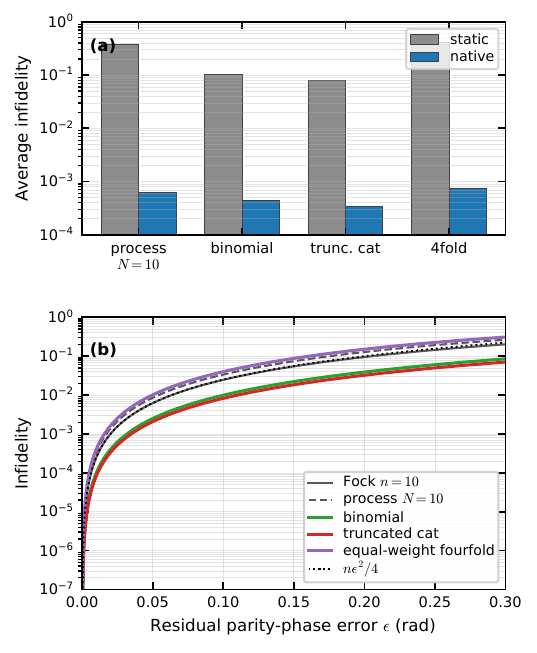}
\caption{
Encoded state transfer benchmarks and residual phase error sensitivity. (a) Average infidelity for the full \(N=10\) cutoff process and for the three logical subspaces in Eqs.~\eqref{eq:binomial-code}--\eqref{eq:rotation-code}. The static resonant evolution is optimized over \(0\leqslant\tau\leqslant50\) for the \(N=10\) process objective.  The native detuned evolution uses the best-found main-window process optimum \((\tau_{\rm opt},\delta_{\rm opt})=(49.968,31.435)\), obtained over \(0\leqslant\tau\leqslant50\) and \(0\leqslant\delta\leqslant 100\).  The same process-optimized transfer map is evaluated on each subspace; no code-specific reoptimization is used.  (b) Infidelity under the uniform residual bright-mode phase error \(U_\epsilon=\exp[i(\pi+\epsilon)a_-^\dagger a_-]\).  The dashed line shows the small-error Fock-state scaling \(1-P_n^{(\epsilon)}\simeq n\epsilon^2/4\) for \(n=10\).  The ordering of the curves follows from photon-number support and the small-\(\epsilon\) coefficients in Table~\ref{tab:logical-error-decomposition}.
}
\label{fig:phaseerror}
\end{figure}

For a two-dimensional logical subspace with basis
\(\{|0_L\rangle,|1_L\rangle\}\), we evaluate the reduced transfer channel after projection onto the target logical subspace.  Let \(K_j\) be the Kraus operators of the oscillator transfer map, where, \(j\) labels the unobserved output branch: the residual photon number in oscillator 1 and, when present, the final ancilla state.  After identifying the input and target logical bases, define the projected logical Kraus matrices
\begin{equation}
B_j = P_LK_jP_L ,
\label{eq:projected-kraus}
\end{equation}
with
\[
P_L = |0_L\rangle\langle0_L| + |1_L\rangle\langle1_L| .
\]
Using the standard Haar-averaged fidelity identity for a two-dimensional logical subspace, derived in Appendix~\ref{app:logical-average-fidelity}, the corresponding logical average fidelity is
\begin{equation}
\overline F_L = \frac{1}{6} \sum_j \left[ {\rm Tr}\!\left(B_j^\dagger B_j\right) + \left|{\rm Tr}\,B_j\right|^2 \right].
\label{eq:logical-average-fidelity}
\end{equation}
The projected map is not renormalized.  Thus, Eq.~\eqref{eq:logical-average-fidelity} penalizes both coherent errors inside the target code space and population that leaves that space. 

Figure~\ref{fig:phaseerror}(a) summarizes the code-level transfer benchmarks. The same process-optimized transfer maps are used for all entries.  The static resonant evolution is optimized over \(0\leqslant\tau\leqslant50\) for the full \(N=10\) process objective, giving \(\tau_{\rm stat}=6.617\).  The native detuned evolution uses the best-found main-window process optimum of Sec.~\ref{subsec:native-process}, \((\tau_{\rm opt},\delta_{\rm opt})=(49.968,31.435)\).  No code-specific reoptimization is performed.  The resulting average infidelities are listed in Table~\ref{tab:code-transfer-benchmarks}.

\begin{table}[t]
\centering
\caption{
Average infidelities for the full \(N=10\) cutoff process and for the three encoded subspaces in Fig.~\ref{fig:phaseerror}(a).  The static resonant evolution is optimized over \(0\leqslant\tau\leqslant50\) for the \(N=10\) process objective.  The native detuned evolution uses the best-found main-window process optimum \((\tau_{\rm opt},\delta_{\rm opt})=(49.968,31.435)\).
}
\label{tab:code-transfer-benchmarks}
\begin{ruledtabular}
\begin{tabular}{lcc}
Benchmark & Static & Native detuned \\ \hline
\(N=10\) process & \(3.73\times10^{-1}\) & \(6.29\times10^{-4}\) \\
Lowest-order binomial & \(1.01\times10^{-1}\) & \(4.46\times10^{-4}\) \\
Truncated cat & \(8.05\times10^{-2}\) & \(3.43\times10^{-4}\) \\
Equal-weight fourfold & \(4.18\times10^{-1}\) & \(7.62\times10^{-4}\)
\end{tabular}
\end{ruledtabular}
\end{table}

The code-level native infidelities can be smaller than the full \(N=10\) native process infidelity \(1-\overline F_{10}^{\rm nat}\), because the code states occupy structured subspaces of the cutoff rather than sampling all Fock components uniformly.  The equal-weight fourfold subspace is the hardest of the three because it places substantial weight near the top of the cutoff, where, Fig.~\ref{fig:phase-spectrum} shows that the residual phase error \(\Delta\phi_k\) and leakage \(L_k\) are largest.

The ordering of the code sensitivities can be understood analytically from a uniform residual phase error model.  Suppose the implemented bright-mode operation is not exactly \(\Pi_-\), but instead
\begin{equation}
U_\epsilon = \exp\!\left[i(\pi+\epsilon)a_-^\dagger a_-\right].
\label{eq:phase-error-unitary}
\end{equation}
This model is not meant to reproduce every detail of the optimized native detuned evolution, whose residual phase error is \(k\)-dependent as shown in Fig.~\ref{fig:phase-spectrum}.  Rather, it isolates a universal support-dependent mechanism: a small error in the bright-mode parity phase is amplified by photon-number support.

For a physical Fock input \(|n\rangle_1|0\rangle_2\), the no-residual transfer amplitude under Eq.~\eqref{eq:phase-error-unitary} is
\begin{equation}
A_n^{(\epsilon)} = \frac{1}{2^n} \sum_{k=0}^{n} \binom{n}{k} e^{ik\epsilon} = e^{in\epsilon/2} \cos^n\!\left(\frac{\epsilon}{2}\right).
\label{eq:phase error-amplitude}
\end{equation}
Therefore
\begin{equation}
P_n^{(\epsilon)} = \left|
A_n^{(\epsilon)} \right|^2 = \cos^{2n}\!\left(\frac{\epsilon}{2}\right).
\label{eq:phase error-fock}
\end{equation}
For small \(\epsilon\),
\begin{equation}
1-P_n^{(\epsilon)} = \frac{n\epsilon^2}{4} + O(\epsilon^4).
\label{eq:phase error-small}
\end{equation}
Thus, a fixed residual parity-phase error produces an infidelity that grows linearly with photon number at leading order.  This is the basic reason why high-\(n\) support is costly.


For the lowest-order binomial code, the same point can be seen almost by inspection.  Define
\[
c=\cos(\epsilon/2), \qquad A_2^{(\epsilon)}=e^{i\epsilon}c^2, \qquad A_4^{(\epsilon)}=e^{2i\epsilon}c^4 .
\]
The logical one is the Fock state \(|2\rangle\), so its projected survival is
\begin{equation}
S_1^{\rm bin}(\epsilon) = \left| A_2^{(\epsilon)} \right|^2 + O(\epsilon^4) = 1-\frac{\epsilon^2}{2} + O(\epsilon^4).
\label{eq:binomial-S1}
\end{equation}
The logical zero is the Fock-state qubit \((|0\rangle+|4\rangle)/\sqrt2\).  In the no-residual branch it is mapped to
\[
\frac{|0\rangle+A_4^{(\epsilon)}|4\rangle}{\sqrt2},
\]
which is no longer exactly parallel to \(|0_L^{\rm bin}\rangle\) when \(A_4^{(\epsilon)}\neq1\).  Projection back onto the logical zero gives
\begin{equation}
S_0^{\rm bin}(\epsilon) = \left| \tfrac{1+A_4^{(\epsilon)}}{2} \right|^2 + O(\epsilon^4) = 1-\tfrac{3}{2}\epsilon^2 + O(\epsilon^4).
\label{eq:binomial-S0}
\end{equation}
Equations~\eqref{eq:binomial-S1} and \eqref{eq:binomial-S0} explain why even this compact code has a visible phase error penalty: the \(|0\rangle\)--\(|4\rangle\) coherence of \(|0_L^{\rm bin}\rangle\) is rotated partly out of the logical subspace.  Averaging the two logical basis states gives the leading logical leakage coefficient \(C_{\rm leak}=1\), consistent with Table~\ref{tab:logical-error-decomposition}.

For the truncated cat and equal-weight fourfold examples, the same calculation is most compactly summarized by small-\(\epsilon\) coefficients.  We write
\begin{equation}
1-\overline F_L^{(\epsilon)} = \chi_L\epsilon^2 + O(\epsilon^4),
\label{eq:logical-infidelity-expansion}
\end{equation}
where, \(\overline F_L^{(\epsilon)}\) is the logical average fidelity under the residual phase operation in Eq.~\eqref{eq:phase-error-unitary}.  We also separate the error into projected logical leakage, conditional coherence loss, and coherent logical phase.

For logical basis state \(|b_L\rangle\), define
\begin{equation}
S_b(\epsilon) = \sum_j \left\| B_j^{(\epsilon)}|b_L\rangle
\right\|^2, \qquad b=0,1,
\label{eq:logical-survival}
\end{equation}
and the basis-averaged projected leakage
\begin{equation}
{\cal L}_L(\epsilon) = 1- \frac{S_0(\epsilon)+S_1(\epsilon)}{2}.
\label{eq:logical-leakage}
\end{equation}
The projected off-diagonal transfer coefficient is
\begin{equation}
\mu_L(\epsilon) = \langle 0_L| \left[ P_L{\cal E}_\epsilon \!\left(|0_L\rangle\langle1_L|\right) P_L \right] |1_L\rangle ,
\label{eq:logical-coherence}
\end{equation}
where, \({\cal E}_\epsilon\) is the reduced oscillator channel generated by Eq.~\eqref{eq:phase-error-unitary}.  We define the conditional coherence loss and coherent logical phase as
\begin{align}
\label{eq:logical-dephasing}
{\cal D}_L(\epsilon) &= 1- \frac{|\mu_L(\epsilon)|} {\sqrt{S_0(\epsilon)S_1(\epsilon)}} ,\\
\varphi_L(\epsilon) &= \arg\mu_L(\epsilon).
\label{eq:logical-phase}
\end{align}
Their leading small-error behavior is
\begin{align}
\label{eq:logical-error-expansion}
{\cal L}_L(\epsilon) &= C_{\rm leak}\, \epsilon^2 + O(\epsilon^4), \\
\label{eq:logical-error-expansion}
{\cal D}_L(\epsilon) &= C_{\rm coh}\, \epsilon^2 + O(\epsilon^4),\\
\varphi_L(\epsilon) &= -\eta_Z\, \epsilon + O(\epsilon^3).
\label{eq:logical-z-expansion}
\end{align}

The coefficients are listed in Table~\ref{tab:logical-error-decomposition}. They show that photon-number support matters through both range and weight. The truncated cat subspace samples more Fock levels than the lowest-order binomial code, but its high-\(n\) weights are factorially suppressed; correspondingly, its leading logical infidelity coefficient is slightly smaller.  The equal-weight fourfold subspace is different: the high-photon components have large weights, so the residual phase error produces strong leakage from the chosen finite code space and a coherent logical \(Z\)-phase. The coherent phase can in principle be tracked or corrected by a logical frame update, but the leakage term cannot be removed by such a frame change.

Figure~\ref{fig:phaseerror}(b) plots the same residual-error model beyond the infinitesimal regime.  The dashed reference line is the Fock state result
\[
1-P_{10}^{(\epsilon)} \simeq \frac{10\epsilon^2}{4}.
\]
The full \(N=10\) process curve lies above the Fock-state curve at small \(\epsilon\) because the process benchmark coherently averages phase errors across all photon numbers \(0,\ldots,10\), not only the single \(n=10\) input. The code curves follow the coefficients in Table~\ref{tab:logical-error-decomposition}.

The two panels of Fig.~\ref{fig:phaseerror} therefore support the same interpretation. The static resonant protocol is limited by imperfect recurrence of \(\sqrt{k}\)-dependent bright-mode phases, so its logical performance depends strongly on which photon numbers the code samples.  The native detuned evolution improves all three encoded benchmarks because it approximates bright-mode parity over the \(N=10\) cutoff.  Nevertheless, the remaining error is still governed by Fock support: compact or strongly low-photon-weighted subspaces are easier, while subspaces with appreciable high-photon support require a more accurate parity phase across the entire cutoff. Appendix~\ref{app:residual-phase-channel} gives the finite-sum channel formulas used to generate the residual error curves and the coefficients in Table~\ref{tab:logical-error-decomposition}.

\begin{table}[t]
\centering
\caption{
Small-\(\epsilon\) coefficients for the residual bright-mode phase error in Eq.~\eqref{eq:phase-error-unitary}.  The coefficient \(\chi_L\) controls \(1-\overline F_L^{(\epsilon)}\), while \(C_{\rm leak}\), \(C_{\rm coh}\), and \(\eta_Z\) characterize projected leakage, conditional coherence loss, and coherent logical phase.  The cat values use \(\alpha=\sqrt2\) and cutoff \(n\leqslant10\).
}
\label{tab:logical-error-decomposition}
\begin{ruledtabular}
\begin{tabular}{lcccc}
Example & \(\chi_L\) & \(C_{\rm leak}\) & \(C_{\rm coh}\) & \(\eta_Z\)
\\ \hline
Lowest-order binomial & \(1.000\) & \(1.000\) & \(0\) & \(0\) \\
Truncated cat & \(0.832\) & \(0.498\) & \(0.500\) & \(0.073\) \\
Equal-weight fourfold & \(4.083\) & \(3.917\) & \(0\) & \(1.000\)
\end{tabular}
\end{ruledtabular}
\end{table}

\section{Conclusion}
\label{sec:conclusion}

We have shown that bosonic state transfer through an antisymmetric single-ancilla interface is most naturally understood as a normal-mode phase-synthesis problem.  The symmetric mode is exactly dark, the antisymmetric mode is bright, and the physical transfer operation is equivalent to applying parity to the bright mode.  This equivalence is the organizing principle of the work: it identifies the actual controlled degree of freedom, fixes the target operation, and turns finite-dimensional transfer into the task of synthesizing the phase pattern \((-1)^k\) on bright-mode Fock components.

This viewpoint exposes a sharp limitation of resonant single-ancilla transfer.  A resonant JC evolution can implement the required operation exactly in the single-photon sector, but for any cutoff \(N\geqslant 2\) exact finite-time transfer would require incompatible alignments of the \(\sqrt{k}\)-dependent Rabi phases.  The static protocol is therefore not a finite-time swap on a bosonic cutoff; it is a recurrence problem.  The exact finite-sum formulas derived here make this statement operational by showing how Fock states, Fock-state qubits, and arbitrary finite superpositions weight the same underlying bright-mode return amplitudes.

Detuning supplies the missing resource.  In the detuned bright-mode JC model, the return amplitude acquires an approximately photon-number-linear phase while ancilla excitation remains perturbatively suppressed.  With only two native parameters, the detuning and evolution time, this gives a reproducible finite-cutoff approximation to bright-mode parity.  In the main benchmark window \(0\leqslant\tau\leqslant50\), \(0\leqslant\delta\leqslant 100\), the optimized \(N=10\) evolution reaches $\overline F_{10}=0.9994$ and $L_{\max}^{(10)} = 5.47\times10^{-3}$.
These numbers are closed-system finite-window benchmarks, but they demonstrate that the restricted single-ancilla primitive can approach the correct normal-mode target without assuming a direct mode--mode exchange Hamiltonian.

The encoded-state examples show why the phase-synthesis formulation matters beyond population transfer.  Logical transfer fidelity is governed by the Fock support of the encoded state and by the residual bright-mode phase error of the implemented evolution.  Compact low-photon support is comparatively forgiving, while subspaces with appreciable high-photon weight require a more accurate parity phase across the cutoff.




The framework is complementary to an ideal beam splitter swap, which remains the natural benchmark when a high-quality direct exchange Hamiltonian is available.  A comparison with that benchmark and with possible implementation routes is given in Appendix~\ref{app:beam splitter-implementation}.

The same framework also gives clear targets for future implementations.  The minimal open-system estimate in Appendix~\ref{app:open-system} shows that, for the best-found \(N=10\) native evolution, oscillator loss is the dominant dissipative limitation over the rates considered, while ancilla relaxation and dephasing are substantially less damaging because the detuned protocol keeps the ancilla weakly populated.  More realistic device models should also include Kerr and cross-Kerr phases, evolution distortions, non-Markovian noise, and slow calibration drift.  More flexible controls, including dispersive phase gates, photon-number-selective corrections, or optimal-control waveforms, can be benchmarked directly against the bright-mode parity condition and the residual phase error diagnostics derived here.  The broader lesson is that coherent bosonic transfer is not merely excitation transport; it is collective-mode phase engineering.

\begin{acknowledgments}
  We thank Radim Filip and Darren W. Moore for discussions. We acknowledge funding from the BMFTR in Germany for support via PhotonQ, QR.N, QuKuK, QuaPhySI, and also from the EU project CLUSTEC (Grant Agreement No. 101080173).
\end{acknowledgments}




\appendix

\section{Coupling sign and the transfer target}
\label{app:phase-conventions}

This appendix clarifies the sign convention used in the main text.  Two separate choices should not be confused.  The first is the overall phase of the JC coupling, which can be absorbed into a phase rotation of the two-level ancilla.  The second is the relative sign with which the two oscillators couple to the ancilla.  Only the latter choice determines which collective oscillator mode is bright, and therefore whether bright-mode parity gives the bare swap or the swap followed by a known output parity.

Let
\begin{equation}
b_\eta=\left(a_1+\eta a_2\right)/\sqrt{2},
\qquad d_\eta=\left(a_1-\eta a_2\right)/\sqrt{2},
\label{eq:app-bright-dark-eta}
\end{equation}
with $\eta=\pm1$.
Here, \(b_\eta\) is the mode coupled to the ancilla and \(d_\eta\) is the corresponding dark mode.  The physical modes are
\begin{equation}
a_1=\left(b_\eta+d_\eta\right)/\sqrt{2},
\qquad
a_2=\eta\,\left(b_\eta-d_\eta\right)/\sqrt{2} .
\label{eq:app-inverse-eta}
\end{equation}
For \(\eta=-1\), the bright mode is the antisymmetric mode, $b_{-}=a_-=\left(a_1-a_2\right)/\sqrt{2}$,
and the dark mode is \(d_{-}=a_+\).  For \(\eta=+1\), the bright mode is instead the symmetric mode \(b_+=a_+\), and the dark mode is \(d_+=a_-\).

Now synthesize parity on the bright mode,
\begin{equation}
\Pi_{b_\eta} = \exp\!\left(i\pi b_\eta^\dagger b_\eta\right).
\label{eq:app-bright-parity-eta}
\end{equation}
This operation flips the bright mode and leaves the dark mode unchanged
\begin{equation}
\Pi_{b_\eta}^\dagger b_\eta \Pi_{b_\eta}=-b_\eta,
\qquad
\Pi_{b_\eta}^\dagger d_\eta \Pi_{b_\eta}=d_\eta .
\label{eq:app-bright-dark-action}
\end{equation}
Using Eq.~\eqref{eq:app-inverse-eta}, the corresponding action on the physical modes is
\begin{equation}
\Pi_{b_\eta}^\dagger a_1\Pi_{b_\eta} = -\eta a_2,
\qquad \Pi_{b_\eta}^\dagger a_2\Pi_{b_\eta} = -\eta a_1 .
\label{eq:app-physical-action-eta}
\end{equation}
Thus, the antisymmetric convention used in the main text, \(\eta=-1\), gives
\begin{equation}
a_1\longrightarrow a_2, \qquad
a_2\longrightarrow a_1 .
\label{eq:app-antisymmetric-bare-swap}
\end{equation}
Bright-mode parity is then exactly the bare physical-mode swap.  By contrast, the symmetric convention, \(\eta=+1\), gives
\begin{equation}
a_1\longrightarrow -a_2,
\qquad
a_2\longrightarrow -a_1 .
\label{eq:app-symmetric-signed-swap}
\end{equation}
This is still a swap, but with an additional \(\pi\)-phase rotation on the output mode.

The difference is invisible in fixed-Fock-state transfer probabilities but not in superpositions across photon number.  For an arbitrary finite input state in oscillator 1, $|\psi\rangle_1 = \sum_n c_n |n\rangle_1$ and $|0\rangle_2 \ {\rm in\ oscillator\ 2}$
bright-mode parity gives
\begin{equation}
\Pi_{b_\eta} \left( \sum_n c_n |n\rangle_1 |0\rangle_2 \right) = |0\rangle_1 \sum_n (-\eta)^n c_n |n\rangle_2 .
\label{eq:app-state-action-eta}
\end{equation}
For \(\eta=-1\), the factor \((-\eta)^n\) is unity for every \(n\), so the state is transferred without an additional phase.  For \(\eta=+1\), the output is
\begin{equation}
|0\rangle_1 \sum_n (-1)^n c_n |n\rangle_2 = |0\rangle_1\,\Pi_2|\psi\rangle_2 ,
\label{eq:app-symmetric-output-parity}
\end{equation}
where, $\Pi_2=\exp\!\left(i\pi a_2^\dagger a_2\right)$
is the local parity operator of the target oscillator.  Equivalently, a vacuum--\(n\)-photon superposition is transferred as
\begin{equation}
\frac{|0\rangle+|n\rangle}{\sqrt{2}}
\longrightarrow
\frac{|0\rangle+(-1)^n|n\rangle}{\sqrt{2}}
\label{eq:app-vacuum-fock-phase}
\end{equation}
under the symmetric convention, relative to the bare transfer target.

Finally, changing the overall phase of the JC interaction does not change this conclusion.  For example, the interaction Hamiltonians $\sqrt{2}g \left( b_\eta\sigma_+ + b_\eta^\dagger\sigma_- \right)$ and $i\sqrt{2}g \left( b_\eta\sigma_+ - b_\eta^\dagger\sigma_- \right)$
are related by a phase redefinition of the ancilla states.  They have the same bright mode, the same JC doublet frequencies, and the same leakage probabilities.  The physically relevant choice for the transfer target is therefore not this overall phase, but the relative oscillator sign \(\eta\).

The antisymmetric coupling used in the main text is the cleanest convention because the coupled mode is \(a_-\), and parity on that mode is exactly the desired physical transfer.  A symmetric coupling is analytically equivalent only after redefining the target by a known local parity phase, or after applying the deterministic correction \(\Pi_2\) to the output oscillator.

\section{Normal-mode coefficients and static transition amplitudes}
\label{app:transition-amplitudes}

This appendix records the fixed-photon-number basis transformation used in Sec.~\ref{sec:static} and derives the ground- and excited-ancilla transition amplitudes for the static resonant protocol.  The transformation is purely kinematic; the dynamics enter only through the bright-mode doublet factors \(r_k^{\rm stat}\) and \(e_k^{\rm stat}\).

With
\[
a_\pm=\left(a_1\pm a_2\right)/\sqrt2,
\qquad
a_{1,2}=\left(a_+ \pm a_-\right)/\sqrt2,
\]
a physical two-mode Fock-state with total photon number \(m\) can be written as
\begin{align}
|m-s\rangle_1|s\rangle_2 &= \frac{ (a_1^\dagger)^{m-s} (a_2^\dagger)^s}{ \sqrt{(m-s)!\,s!}} |0,0\rangle \nonumber\\
&= \frac{ (a_+^\dagger+a_-^\dagger)^{m-s} (a_+^\dagger-a_-^\dagger)^s }{2^{m/2}\sqrt{(m-s)!\,s!}} |0,0\rangle .
\label{eq:app-physical-creation-form}
\end{align}
Expanding both binomials gives
\begin{align}
|m-s\rangle_1|s\rangle_2 &= \frac{1}{2^{m/2}\sqrt{(m-s)!\,s!}} \sum_{r=0}^{m-s} \sum_{q=0}^{s} (-1)^q \binom{m-s}{r} \nonumber\\
&\quad \times \binom{s}{q}(a_+^\dagger)^{m-r-q} (a_-^\dagger)^{r+q} |0,0\rangle .
\label{eq:app-double-binomial}
\end{align}
To collect the coefficient of a fixed bright-mode occupation \(k\), set \(k=r+q\), or equivalently \(q=k-r\).  The allowed values of \(r\) are those for which \(0\leqslant r\leqslant m-s\) and \(0\leqslant k-r\leqslant s\), namely
\begin{equation}
\max(0,k-s) \leqslant r\leqslant \min(k,m-s).
\label{eq:app-r-range}
\end{equation}
For these terms,
\[
(a_+^\dagger)^{m-k} (a_-^\dagger)^k |0,0\rangle = \sqrt{(m-k)!\,k!}\, |m-k\rangle_+|k\rangle_- .
\]
Therefore
\begin{equation}
|m-s\rangle_1|s\rangle_2 = \sum_{k=0}^{m} U_{sk}^{(m)} |m-k\rangle_+|k\rangle_- ,
\label{eq:app-physical-state}
\end{equation}
with
\begin{equation}
U_{sk}^{(m)}
= \frac{1}{2^{m/2}} \sqrt{ \frac{(m-k)!\,k!}{(m-s)!\,s!} } \sum_{r} (-1)^{k-r} \binom{m-s}{r} \binom{s}{k-r}.
\label{eq:app-U-coeff}
\end{equation}
These are the real fixed-\(m\) matrix elements of the balanced physical-to-normal-mode transformation, equivalently the photon-number representation of a \(50{:}50\) lossless beam splitter~\cite{CamposSalehTeich1989}. Since the transformation from \((a_1,a_2)\) to \((a_+,a_-)\) is unitary,
\(U^{(m)}\) is real orthogonal:
\begin{equation}
\sum_{k=0}^{m}
U_{sk}^{(m)} U_{s'k}^{(m)} = \delta_{ss'}, \qquad \sum_{s=0}^{m} U_{sk}^{(m)} U_{sk'}^{(m)} = \delta_{kk'} .
\label{eq:app-U-orthogonality}
\end{equation}

The two physical states that define transfer correspond to the first and last rows of this matrix.  For the input state localized in oscillator 1, \(s=0\), Eq.~\eqref{eq:app-U-coeff} gives
\begin{equation}
U_{0k}^{(m)} = \frac{1}{2^{m/2}} \sqrt{\binom{m}{k}}.
\label{eq:app-U0}
\end{equation}
For the fully transferred state, \(s=m\), the same formula gives
\begin{equation}
U_{mk}^{(m)} = \frac{(-1)^k}{2^{m/2}} \sqrt{\binom{m}{k}} .
\label{eq:app-Um}
\end{equation}
Thus, the input and transferred states have identical normal-mode weights, but the transferred state carries an additional factor \((-1)^k\) on the \(k\)-photon bright-mode component.  This alternating sign is the normal-mode signature of transfer and is precisely the bright-mode parity phase.

For the static resonant protocol, the bright-mode dynamics enter only through the doublet factors \(r_k^{\rm stat}\) and \(e_k^{\rm stat}\) defined in Eqs.~\eqref{eq:static-return-factor} and \eqref{eq:static-excited-factor}.  Starting from \(|m\rangle_1|0\rangle_2|g\rangle\), Eq.~\eqref{eq:app-U0} gives
\begin{equation}
|m\rangle_1|0\rangle_2|g\rangle = \sum_{k=0}^{m} U_{0k}^{(m)} |m-k\rangle_+|k,g\rangle_- .
\label{eq:app-initial-expansion}
\end{equation}
After the static evolution, the ground-ancilla branch is transformed back to the physical basis using the same orthogonal matrix.  The amplitude to end with \(s\) photons in oscillator 2, \(m-s\) photons in oscillator 1, and the ancilla in \(|g\rangle\) is
\begin{equation}
G_{m,s}^{\rm stat}(\tau) = \sum_{k=0}^{m} U_{sk}^{(m)} U_{0k}^{(m)} r_k^{\rm stat}(\tau), \,\,\,\,\,\, 0\leqslant s\leqslant m .
\label{eq:app-static-Gms}
\end{equation}
The excited-ancilla branch contains one fewer oscillator photon and is expanded in the \((m-1)\)-photon manifold.  This gives
\begin{equation}
E_{m,s}^{\rm stat}(\tau) = \sum_{k=1}^{m} U_{s,k-1}^{(m-1)} U_{0k}^{(m)} e_k^{\rm stat}(\tau), \,\,\,\, 0\leqslant s\leqslant m-1 .
\label{eq:app-static-Ems}
\end{equation}
Equations~\eqref{eq:app-static-Gms} and
\eqref{eq:app-static-Ems} are the branch amplitudes used in
Sec.~\ref{sec:static}.

The orthogonality of \(U^{(m)}\), together with
\[
|r_0^{\rm stat}|^2=1, \qquad |r_k^{\rm stat}|^2+|e_k^{\rm stat}|^2=1 \quad (k\geqslant 1),
\]
gives the fixed-\(m\) normalization check
\begin{equation}
\sum_{s=0}^{m} \left| G_{m,s}^{\rm stat}(\tau) \right|^2 + \sum_{s=0}^{m-1} \left| E_{m,s}^{\rm stat}(\tau) \right|^2 = 1 .
\label{eq:app-normalization-check}
\end{equation}
This identity confirms that the ground- and excited-ancilla branches account for the full closed-system probability in each fixed-\(m\) sector.

Setting \(s=m\) in Eq.~\eqref{eq:app-static-Gms} and using
Eqs.~\eqref{eq:app-U0} and \eqref{eq:app-Um} gives the Fock-transfer amplitude
\(A_m^{\rm F,stat}=G_{m,m}^{\rm stat}\) used in
Eq.~\eqref{eq:fock-amplitude}.

\section{Static no-go theorem and finite-cutoff recurrence}
\label{app:no-go-recurrence}

This appendix proves the two statements about static resonant transfer used in Sec.~\ref{sec:static}. First, an exactly resonant single-ancilla evolution cannot implement perfect transfer at any finite time once the cutoff includes the two-photon sector. Second, this finite-time obstruction does not prevent arbitrarily accurate long-time recurrences on any fixed finite cutoff. The distinction is important: the static protocol has unit supremum on a finite cutoff, but that supremum is not attained at finite time for \(N\geqslant 2\).

For exact transfer on the cutoff ${\cal H}_N=\mathrm{span}\{|0\rangle,\ldots,|N\rangle\}$,
the resonant bright-mode return factors must reproduce the parity pattern
\begin{equation}
\cos(\sqrt{2k}\,\tau)=(-1)^k, \qquad k=1,\ldots,N .
\label{eq:appC-parity-condition}
\end{equation}
Equivalently, the bright-mode phases must satisfy
\begin{equation}
\sqrt{2k}\,\tau = \pi k \pmod{2\pi}, \qquad k=1,\ldots,N .
\label{eq:appC-phase-condition}
\end{equation}
The recurrence statement is an application of Kronecker's simultaneous approximation theorem: a linear flow on a finite torus is dense in the subtorus allowed by the rational relations among its frequencies~\cite{HardyWright2008,Cassels1957}. Thus, to prove recurrence on the finite cutoff, it is enough to check that the target phases in Eq.~\eqref{eq:appC-phase-condition} are compatible with all rational dependencies among the frequencies \(\{\sqrt{2k}\}_{k=1}^N\).

For \(N=1\), Eq.~\eqref{eq:appC-phase-condition} is satisfied by
\begin{equation}
\tau = (2q+1) \tfrac{\pi}{\sqrt{2}}, \qquad q\in\mathbb{Z}_{\geqslant 0}.
\label{eq:appC-single-photon-time}
\end{equation}
Thus, the one-photon sector can be transferred exactly.  The obstruction appears as soon as \(N\geqslant 2\).  The \(k=1\) and \(k=2\) conditions require
\begin{equation}
\sqrt{2}\,\tau=(2m+1)\pi, \qquad 2\tau=2q\pi ,
\label{eq:appC-two-conditions}
\end{equation}
with \(m,q\in\mathbb{Z}\).  Eliminating \(\tau\) gives
\begin{equation}
q\sqrt{2}=2m+1 .
\label{eq:appC-irrational-contradiction}
\end{equation}
The left-hand side is irrational for nonzero integer \(q\), while the right-hand side is an integer.  Hence tHere, is no finite resonant time that satisfies both conditions.  Since this contradiction already occurs in the two-photon sector, exact finite-time transfer is impossible on every cutoff \(N\geqslant 2\).

This no-go result is a statement about exact finite-time alignment, not about the closure of the dynamics.  To describe the latter, define the finite set of resonant frequencies
\begin{equation}
\omega_k=\sqrt{2k},
\qquad
k=1,\ldots,N,
\label{eq:appC-frequencies}
\end{equation}
and the desired phase vector
\begin{equation}
\theta_k=\pi k
\pmod{2\pi}.
\label{eq:appC-target-phases}
\end{equation}
We want to know whether the one-parameter orbit
\[
\{\omega_k\tau\}_{k=1}^{N}
\]
can approach the target phases \(\{\theta_k\}_{k=1}^{N}\) modulo \(2\pi\).

The only possible obstruction to such an approximation comes from rational dependencies among the frequencies.  We therefore write each frequency in squarefree form,
\begin{equation}
\omega_k=q_k\sqrt{d_k},
\qquad
2k=q_k^2 d_k,
\label{eq:appC-squarefree}
\end{equation}
where, \(d_k\) is squarefree and \(q_k\) is a positive integer.  Frequencies with different squarefree parts belong to different rational-dependence classes, while frequencies with the same squarefree part are integer multiples of the same radical.  Let \(D_N\) be the finite set of squarefree parts that appear for \(1\leqslant k\leqslant N\).

For a fixed \(d\in D_N\), the frequencies in that class have the form
\begin{equation}
\omega_{d,q}=q\sqrt{d},
\qquad
2k=q^2 d .
\label{eq:appC-class-frequency}
\end{equation}
The corresponding parity target phase is
\begin{equation}
\theta_{d,q}
=
\pi k
=
\frac{\pi q^2 d}{2}
\pmod{2\pi}.
\label{eq:appC-class-target}
\end{equation}
This target is generated by a single phase for each squarefree class.  Define
\begin{equation}
\alpha_d=
\begin{cases}
0, & d\ {\rm odd},\\
\pi, & d\ {\rm even}.
\end{cases}
\label{eq:appC-class-phase}
\end{equation}
Then
\begin{equation}
\theta_{d,q}=q\alpha_d
\pmod{2\pi}.
\label{eq:appC-target-generated}
\end{equation}
To see this, first suppose \(d\) is odd.  Since \(q^2d=2k\) is even, \(q\) must be even.  Equation~\eqref{eq:appC-class-target} is then an integer multiple of \(2\pi\), so \(\theta_{d,q}=0=q\alpha_d\) modulo \(2\pi\).  If \(d\) is even and squarefree, write \(d=2d'\) with \(d'\) odd.  Then
\begin{equation}
\theta_{d,q}=\pi q^2 d'
\equiv
\pi q
=
q\alpha_d
\pmod{2\pi},
\label{eq:appC-even-class}
\end{equation}
because \(q^2d'\) and \(q\) have the same parity.

Equation~\eqref{eq:appC-target-generated} is the compatibility condition needed for recurrence.  Indeed, any integer relation among the frequencies can be written as
\begin{equation}
\sum_{d\in D_N}\sum_q n_{d,q}\,q\sqrt d=0,
\qquad
n_{d,q}\in\mathbb{Z}.
\label{eq:appC-integer-relation}
\end{equation}
The distinct squarefree radicals are rationally independent, so Eq.~\eqref{eq:appC-integer-relation} implies
\begin{equation}
\sum_q n_{d,q}q=0
\qquad
{\rm for\ each}\ d .
\label{eq:appC-class-relation}
\end{equation}
Using Eq.~\eqref{eq:appC-target-generated}, the same relation gives
\begin{equation}
\sum_{d,q} n_{d,q}\theta_{d,q}
=
\sum_{d\in D_N}
\alpha_d
\sum_q n_{d,q}q
=
0
\pmod{2\pi}.
\label{eq:appC-target-compatible}
\end{equation}
Thus, the desired parity phases obey every rational constraint obeyed by the resonant frequencies.

Kronecker's simultaneous approximation theorem then implies that the target phases lie in the closure of the resonant orbit.  Equivalently, for every tolerance \(\eta>0\), tHere, exists a time \(\tau\) such that
\begin{equation}
{\rm dist}_{2\pi}
\!\left(
\sqrt d\,\tau,\alpha_d
\right)
<\eta,
\qquad
d\in D_N ,
\label{eq:appC-kronecker}
\end{equation}
where
\begin{equation}
{\rm dist}_{2\pi}(x,y)
=
\min_{n\in\mathbb{Z}}|x-y-2\pi n|
\label{eq:appC-distance}
\end{equation}
is the distance on the circle.  Let
\begin{equation}
q_{\max}=\max_{1\leqslant k\leqslant N}q_k .
\label{eq:appC-qmax}
\end{equation}
Choosing \(\eta=\epsilon/q_{\max}\) gives, for every \(k\leqslant N\),
\begin{align}
{\rm dist}_{2\pi}
\!\left(
\omega_k\tau,\theta_k
\right)
&=
{\rm dist}_{2\pi}
\!\left(
q_k\sqrt{d_k}\,\tau,q_k\alpha_{d_k}
\right)
\nonumber\\
&\le
q_k\,
{\rm dist}_{2\pi}
\!\left(
\sqrt{d_k}\,\tau,\alpha_{d_k}
\right)
<\epsilon .
\label{eq:appC-all-phases-close}
\end{align}
Therefore, for any fixed cutoff \(N\) and any \(\epsilon>0\), tHere, exists a resonant time \(\tau\) such that
\begin{equation}
{\rm dist}_{2\pi}
\!\left(
\sqrt{2k}\,\tau,\pi k
\right)
<\epsilon,
\qquad
k=1,\ldots,N .
\label{eq:appC-approx-parity}
\end{equation}
At such a time,
\begin{equation}
\cos(\sqrt{2k}\,\tau)\approx(-1)^k,
\quad
\sin(\sqrt{2k}\,\tau)\approx0 .
\label{eq:appC-cos-sin-close}
\end{equation}
The ground-ancilla return amplitudes therefore approach the ideal bright-mode parity phases, while the excited-ancilla branches vanish.

Because all finite-cutoff transfer amplitudes are finite sums of these return and leakage amplitudes, Eq.~\eqref{eq:appC-cos-sin-close} implies that the static resonant channel can approach the ideal transfer channel arbitrarily closely on any fixed finite cutoff.  Thus
\begin{equation}
\sup_{\tau\geqslant 0}\overline F_N^{\rm static}(\tau)=1
\qquad
{\rm for\ every\ finite}\ N .
\label{eq:appC-unit-supremum}
\end{equation}
For \(N\geqslant 2\), however, the supremum is not attained at finite \(\tau\), as shown by the exact no-go argument above.  This is the sense in which static resonant transfer is recurrence-limited.  Long-time quasiperiodic returns exist in principle, but finite-window process fidelity is the operational benchmark relevant to the main text.

\begin{figure*}[t]
\centering
\includegraphics[width=\textwidth]{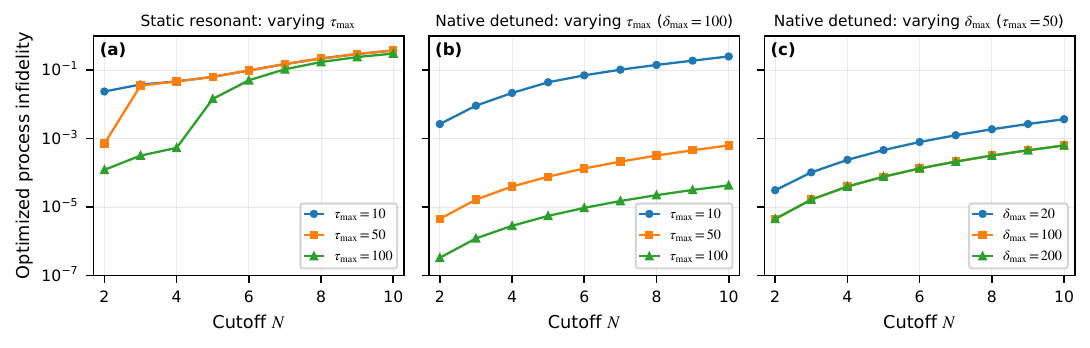}
\caption{
Finite-window dependence of the optimized average-process infidelity. All three panels use the same vertical scale, so the relative improvement from enlarging the time or detuning window can be compared directly. (a) Static resonant transfer for three time windows, \(\tau_{\max}=10,50,100\).  Increasing the window improves the optimum only through better recurrence of the \(\sqrt{k}\)-dependent resonant phases. (b) Native detuned transfer for the same time windows at fixed \(\delta_{\max}=100\).  The main-text window \(\tau_{\max}=50\) already gives high-fidelity finite-cutoff parity synthesis, while \(\tau_{\max}=100\) provides the expected additional improvement. (c) Native detuned transfer for three detuning ranges, \(\delta_{\max}=20,100,200\), at fixed \(\tau_{\max}=50\).  The range \(\delta_{\max}=100\) contains the relevant best-found optima for the cutoffs shown; increasing to \(\delta_{\max}=200\) does not change the plotted best-found solutions.
}
\label{fig:window-dependence}
\end{figure*}

\section{Finite-window dependence of the transfer benchmarks}
\label{app:window-dependence}

The main-text benchmarks use the finite search window \(0\leqslant \tau\leqslant 50\) and \(0\leqslant \delta\leqslant 100\), where, \(\tau=gt\) and \(\delta=\Delta/g\).  These bounds define an operational comparison window, not a fundamental limit of the model.  This appendix checks that the qualitative separation between recurrence-limited static transfer and detuned native parity synthesis is not tied to a single choice of window.

For the static resonant protocol, we define the optimized finite-window process fidelity
\begin{equation}
\overline F_{N,{\rm opt}}^{\rm stat}(\tau_{\max}) = \max_{0\leqslant \tau\leqslant \tau_{\max}} \overline F_N^{\rm stat}(\tau).
\label{eq:appD-static-window-objective}
\end{equation}
For the native detuned protocol, we define
\begin{equation}
\overline F_{N,{\rm opt}}^{\rm nat}(\tau_{\max},\delta_{\max}) = \max_{\substack{0\leqslant \tau\leqslant \tau_{\max}\\ 0\leqslant \delta\leqslant \delta_{\max}}}
\overline F_N^{\rm nat}(\delta,\tau).
\label{eq:appD-native-window-objective}
\end{equation}
The plotted quantity in Fig.~\ref{fig:window-dependence} is the corresponding optimized average-process infidelity.  For the native protocol, \(L_{\max}^{(N)}\) is evaluated at the same best-found point but is not included as a penalty in Eq.~\eqref{eq:appD-native-window-objective}.

Figure~\ref{fig:window-dependence}(a) varies the static time window, \(\tau_{\max}=10,50,100\).  Enlarging the window can improve the optimum, but only by allowing better recurrence of the \(\sqrt{k}\)-dependent resonant phases.  For \(N=10\), increasing the window from \(\tau_{\max}=50\) to \(\tau_{\max}=100\) reduces the optimized process infidelity from \(3.73\times10^{-1}\) to \(3.04\times10^{-1}\).

Figure~\ref{fig:window-dependence}(b) varies the native time window at fixed \(\delta_{\max}=100\).  The short window \(\tau_{\max}=10\) is too restrictive at high cutoff, while the main-text window already gives high-fidelity finite-cutoff parity synthesis.  For \(N=10\),
\begin{align}
\overline F_{10,{\rm opt}}^{\rm nat}(10,100) &= 0.7499, \nonumber\\
\overline F_{10,{\rm opt}}^{\rm nat}(50,100) &= 0.999370508862, \nonumber\\
\overline F_{10,{\rm opt}}^{\rm nat}(100,100) &= 0.999956843440 .
\label{eq:appD-native-time-N10-values}
\end{align}
Thus, the main-text value should be read as a finite-window benchmark, not as a global optimum over all evolution times.

Figure~\ref{fig:window-dependence}(c) varies the detuning range at fixed \(\tau_{\max}=50\).  For  \(N=10\),
\begin{align}
\overline F_{10,{\rm opt}}^{\rm nat}(50,20) &= 0.996334913300, \nonumber\\
\overline F_{10,{\rm opt}}^{\rm nat}(50,100) &= 0.999370508862, \nonumber\\
\overline F_{10,{\rm opt}}^{\rm nat}(50,200) &= 0.999370508862 .
\label{eq:appD-native-detuning-N10-values}
\end{align}
Thus, \(\delta_{\max}=20\) is too narrow to reach the best-found main-window native solution, whereas \(\delta_{\max}=100\) already contains the relevant best-found optimum.  Extending to \(\delta_{\max}=200\) does not change the best-found \(N=10\) solution at \(\tau_{\max}=50\), whose coordinates are
\begin{equation*}
\tau_{\rm opt}=49.968175624787,
\quad
\delta_{\rm opt}=31.434701781850 .
\label{eq:appD-N10-main-optimum}
\end{equation*}

These checks support the use of the finite search window \(0\leqslant \tau\leqslant 50\), \(0\leqslant \delta\leqslant 100\) in the main text.  Enlarging the windows changes the numerical optima in the expected direction, but it does not change the physical interpretation: static resonant transfer improves only through quasiperiodic recurrence, whereas detuned native control directly targets the finite-cutoff bright-mode parity phase.

\section{Numerical reproducibility of the process benchmarks}
\label{app:native-reproducibility}

\begin{figure}[t]
\centering
\includegraphics[width=\columnwidth]{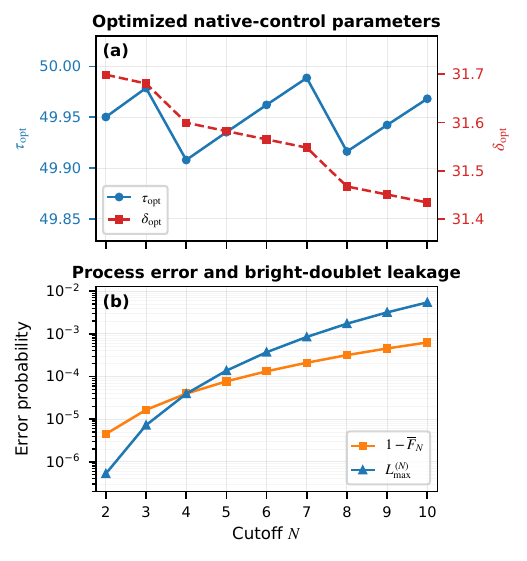}
\caption{
Best-found native detuned optima over the main-text finite search window \(0\leqslant\tau\leqslant50\), \(0\leqslant\delta\leqslant 100\).  The trivial \(N=1\) case is omitted because exact transfer is possible and the optimum is non-unique. (a) Best-found evolution time \(\tau_{\rm opt}\) and detuning \(\delta_{\rm opt}\) for the nontrivial cutoffs \(N=2,\ldots,10\).  The plotted points form a high-fidelity branch within the stated search protocol and should not be interpreted as unique global optima.  (b) Average process infidelity \(1-\overline F_N^{\rm nat}\) and maximum bright-doublet leakage \(L_{\max}^{(N)}\), evaluated at the same best-found points.  The leakage is reported separately from the coherent process-fidelity objective and is not included as an optimization penalty.
}
\label{fig:native-optima}
\end{figure}

\begin{figure}[t]
\centering
\includegraphics[width=\columnwidth]{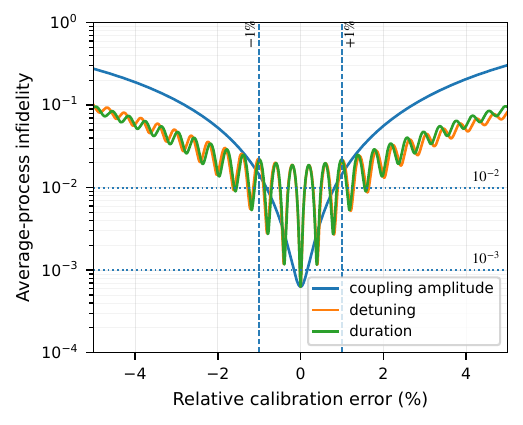}
\caption{
Calibration sensitivity of the best-found native detuned \(N=10\) evolution.  The curves show the average-process infidelity on the full cutoff space after applying a fractional calibration error \(\epsilon\) to one control parameter at a time.  The evolution is centered at the best-found main-window optimum \((\tau_{\rm opt},\delta_{\rm opt})=(49.968,31.435)\), for which \(1-\overline F_{10}^{\rm nat}=6.29\times10^{-4}\).  A duration error is implemented as \(\tau\rightarrow(1+\epsilon)\tau_{\rm opt}\) at fixed \(\delta\); a detuning error as \(\delta\rightarrow(1+\epsilon)\delta_{\rm opt}\) at fixed \(\tau\); and a coupling-amplitude error by keeping the laboratory detuning and evolution time fixed, so that \(\tau\rightarrow(1+\epsilon)\tau_{\rm opt}\) and \(\delta\rightarrow\delta_{\rm opt}/(1+\epsilon)\).  Vertical dashed lines mark \(\pm1\%\) calibration error, and horizontal dotted lines mark \(10^{-2}\) and \(10^{-3}\) process infidelity.  The oscillatory structure is a coherent finite-doublet phase-wrapping effect rather than a stochastic noise model.
}
\label{fig:robust}
\end{figure}

This appendix gives the numerical procedure used to generate the finite-window process benchmarks in Figs.~\ref{fig:finite-time}, \ref{fig:phase-spectrum}, \ref{fig:native-optima}, and \ref{fig:robust}.  The analytical ingredients are defined in the main text: the static amplitudes \(A_m^{\rm F,stat}\) in Sec.~\ref{sec:static}, the native amplitudes \(A_m^{\rm F,nat}\), \(r_k^{\rm nat}\), and \(L_k=|e_k^{\rm nat}|^2\) in Sec.~\ref{sec:native}, and the average-process fidelity convention in Eqs.~\eqref{eq:cutoff-channel-space}--\eqref{eq:average-fidelity-convention}. All numerical values below are obtained from these closed-form finite sums.

For the static resonant protocol, the objective is \(\overline F_N^{\rm stat}(\tau)\), maximized by a one-dimensional search over the stated time window.  In the main window \(0\leqslant\tau\leqslant50\), the \(N=10\) best-found point is
\begin{equation}
\tau_{\rm stat}=6.616988607594,
\quad
\overline F_{10}^{\rm stat}=0.626505334395 .
\label{eq:appE-static-N10}
\end{equation}

For the native detuned protocol, the objective is \(\overline F_N^{\rm nat}(\delta,\tau)\) in Eq.~\eqref{eq:native-average-fidelity}, maximized over \(0\leqslant\tau\leqslant50\) and \(0\leqslant\delta\leqslant 100\).  The leakage \(L_{\max}^{(N)}\) is evaluated at the resulting point but is not included as a penalty in the objective.  We evaluate the objective on a deterministic grid, refine the best candidates by local continuous optimization, and recompute all reported quantities from the closed-form finite sums at the final best-found point.  The initial grid uses
\begin{equation}
\Delta\delta=\Delta\tau=0.25 .
\label{eq:appE-coarse-grid}
\end{equation}
The leading grid candidates are refined by Nelder--Mead optimization with tolerances
\begin{equation}
x_{\rm tol}=10^{-12},
\qquad
f_{\rm tol}=10^{-13},
\label{eq:appE-NM-tolerances}
\end{equation}
and checked against additional seeds from a refined local grid,
\begin{equation}
\Delta\delta=\Delta\tau=0.125 .
\label{eq:appE-refined-grid}
\end{equation}
The values quoted in the text and figures are obtained only after a final evaluation of \(A_m^{\rm F,nat}\), \(\overline F_N^{\rm nat}\), and \(L_{\max}^{(N)}\) at the refined best-found point.

Figure~\ref{fig:native-optima} shows the best-found native-control branch within the finite search window mentioned earlier.  The trivial \(N=1\) case is omitted because exact transfer is possible and the optimum is non-unique.  For \(N=2,\ldots,10\), the best-found finite-window optima lie near the upper end of the allowed evolution time window and vary smoothly with cutoff.  Since the finite-window objective is multimodal, the plotted parameters should be interpreted as best-found process optima within the stated deterministic search and local-refinement procedure, not as unique global optima.  The corresponding process infidelities and leakage diagnostics are listed in Table~\ref{tab:appE-native-optima}.

\begin{table}[t]
\centering
\caption{
Best-found native detuned finite-window optima over
\(0\leqslant\tau\leqslant50\) and \(0\leqslant\delta\leqslant 100\).  The objective is \(\overline F_N^{\rm nat}\).  The leakage \(L_{\max}^{(N)}\) is evaluated at the same point but is not included as an optimization penalty. 
}
\label{tab:appE-native-optima}
\setlength{\tabcolsep}{3.2pt}
\begin{ruledtabular}
\begin{tabular}{ccccc}
\(N\) & \(\tau_{\rm opt}\) & \(\delta_{\rm opt}\) & \(1-\overline F_N^{\rm nat}\) & \( L_{\max}^{(N)}\) \\ \hline
2  & 49.950 & 31.699 & \(4.48{\times}10^{-6}\) & \(5.27{\times}10^{-7}\) \\
3  & 49.979 & 31.681 & \(1.65{\times}10^{-5}\) & \(7.22{\times}10^{-6}\) \\
4  & 49.908 & 31.600 & \(3.97{\times}10^{-5}\) & \(3.93{\times}10^{-5}\) \\
5  & 49.935 & 31.582 & \(7.71{\times}10^{-5}\) & \(1.37{\times}10^{-4}\) \\
6  & 49.962 & 31.565 & \(1.33{\times}10^{-4}\) & \(3.71{\times}10^{-4}\) \\
7  & 49.989 & 31.548 & \(2.11{\times}10^{-4}\) & \(8.47{\times}10^{-4}\) \\
8  & 49.916 & 31.468 & \(3.18{\times}10^{-4}\) & \(1.73{\times}10^{-3}\) \\
9  & 49.942 & 31.451 & \(4.55{\times}10^{-4}\) & \(3.19{\times}10^{-3}\) \\
10 & 49.968 & 31.435 & \(6.30{\times}10^{-4}\) & \(5.47{\times}10^{-3}\)
\end{tabular}
\end{ruledtabular}
\end{table}

As an independent check, the analytical native doublet amplitudes are compared with direct matrix exponentiation of the \(2\times2\) doublet Hamiltonians in Eq.~\eqref{eq:native-doublet-matrix}.  At the \(N=10\) best-found point, the maximum discrepancy between the closed-form return amplitude \(r_k^{\rm nat}\) and the corresponding matrix-exponential element, over \(1\leqslant k\leqslant10\), is \(1.30\times10^{-13}\).  The maximum discrepancy between the closed-form leakage \(L_k=|e_k^{\rm nat}|^2\) and the matrix-exponential excited-state probability is \(1.86\times10^{-15}\). Across all native optima in Table~\ref{tab:appE-native-optima}, the corresponding maximum discrepancies remain below \(1.76\times10^{-13}\) and \(3.60\times10^{-15}\), respectively.

Figure~\ref{fig:robust} tests the calibration sensitivity of the same \(N=10\) operating point.  A duration error changes the scaled time while holding the detuning fixed,
\begin{equation}
\tau\rightarrow(1+\epsilon)\tau_{\rm opt},
\qquad
\delta\rightarrow\delta_{\rm opt}.
\label{eq:duration-error}
\end{equation}
A detuning error changes the scaled detuning while holding the evolution time fixed,
\begin{equation}
\delta\rightarrow(1+\epsilon)\delta_{\rm opt},
\qquad
\tau\rightarrow\tau_{\rm opt}.
\label{eq:detuning-error}
\end{equation}
A coupling-amplitude error is different because the laboratory detuning \(\Delta\) and physical evolution time \(t\) are fixed while \(g\rightarrow(1+\epsilon)g\).  In the scaled variables, this gives
\begin{equation}
\tau\rightarrow(1+\epsilon)\tau_{\rm opt},
\qquad
\delta\rightarrow
\frac{\delta_{\rm opt}}{1+\epsilon}.
\label{eq:coupling-error}
\end{equation}
The robustness curves are coherent miscalibration tests, not stochastic-noise simulations.  The oscillations arise from phase wrapping of the finite set of detuned doublet amplitudes.

The calibration analysis above treats coherent control errors.  Dissipative errors are considered separately in Appendix~\ref{app:open-system}, where, we add oscillator loss, ancilla relaxation, and ancilla dephasing to the same best-found \(N=10\) evolution time.  That calculation provides a dimensionless scale for when the closed-system finite-window benchmark remains predictive.

\section{Finite-sum formulas for residual phase-error diagnostics}
\label{app:residual-phase-channel}

This appendix records the finite-sum channel formulas used for the residual phase-error curves in Fig.~\ref{fig:phaseerror}(b) and for the coefficients in Table~\ref{tab:logical-error-decomposition}.  The residual model is the uniform bright-mode phase error
\begin{equation}
U_\epsilon =
\exp\!\left[i(\pi+\epsilon)a_-^\dagger a_-\right],
\label{eq:appF-Ueps}
\end{equation}
with ideal transfer at \(\epsilon=0\).

Let \(U_{sk}^{(m)}\) be the physical-to-normal-mode coefficients derived in Appendix~\ref{app:transition-amplitudes}.  Under Eq.~\eqref{eq:appF-Ueps}, an input \(|m\rangle_1|0\rangle_2\) has amplitude
\begin{equation}
G_{m,s}^{(\epsilon)} = \sum_{k=0}^{m} U_{sk}^{(m)} U_{0k}^{(m)} e^{i(\pi+\epsilon)k}
\label{eq:appF-Geps}
\end{equation}
to end with \(s\) photons in oscillator 2.  The remaining \(p=m-s\) photons are left in oscillator 1.  On a cutoff \(0\leqslant m\leqslant M\), the resulting reduced channel on oscillator 2 can therefore be written as
\begin{equation}
{\cal E}_\epsilon(\rho) = \sum_{p=0}^{M} K_p^{(\epsilon)} \rho K_p^{(\epsilon)\dagger},
\label{eq:appF-channel}
\end{equation}
where
\begin{equation}
K_p^{(\epsilon)} = \sum_{s=0}^{M-p} G_{p+s,s}^{(\epsilon)} |s\rangle\langle p+s|, \qquad p=0,\ldots,M .
\label{eq:appF-Kp}
\end{equation}
At \(\epsilon=0\), \(G_{m,s}^{(0)}=\delta_{s,m}\), so only \(p=0\) remains and \({\cal E}_0\) is the ideal transfer channel.

For reference, the no-residual Fock-state transfer amplitude is the special case
\begin{equation}
A_n^{(\epsilon)} = G_{n,n}^{(\epsilon)} = \frac{1}{2^n} \sum_{k=0}^{n} \binom{n}{k} e^{ik\epsilon} = e^{in\epsilon/2} \cos^n\!\left(\frac{\epsilon}{2}\right).
\label{eq:appF-An}
\end{equation}
Thus
\begin{equation}
P_n^{(\epsilon)} = \left| A_n^{(\epsilon)} \right|^2 = \cos^{2n}\!\left(\frac{\epsilon}{2}\right),
\label{eq:appF-Pn}
\end{equation}
and the full cutoff-process reference curve is
\begin{align}
F_e^{(N,\epsilon)} &= \frac{1}{(N+1)^2} \left| \sum_{n=0}^{N} A_n^{(\epsilon)} \right|^2,
\nonumber\\
\overline F_N^{(\epsilon)} &= \frac{(N+1)F_e^{(N,\epsilon)}+1}{N+2}.
\label{eq:appF-process-curve}
\end{align}

For a logical code
\begin{equation}
|b_L\rangle = \sum_{n=0}^{M} c_n^{(b)}|n\rangle, \qquad b=0,1,
\label{eq:appF-code-coefficients}
\end{equation}
with projector $P_L = |0_L\rangle\langle0_L| + |1_L\rangle\langle1_L|$
the projected logical Kraus matrices are
\begin{equation}
B_p^{(\epsilon)} = P_LK_p^{(\epsilon)}P_L .
\label{eq:appF-Bp}
\end{equation}
Their logical-basis matrix elements are
\begin{equation}
\left[B_p^{(\epsilon)}\right]_{ab} = \sum_{s=0}^{M-p} \left(c_s^{(a)}\right)^* c_{p+s}^{(b)} G_{p+s,s}^{(\epsilon)},
\qquad a,b=0,1 .
\label{eq:appF-Bp-elements}
\end{equation}
The logical average fidelity plotted in Fig.~\ref{fig:phaseerror}(b) is
\begin{equation}
\overline F_L^{(\epsilon)} = \frac{1}{6} \sum_{p=0}^{M} \left[ {\rm Tr}\!\left( B_p^{(\epsilon)\dagger}B_p^{(\epsilon)} \right) + \left| {\rm Tr}\,B_p^{(\epsilon)} \right|^2 \right].
\label{eq:appF-logical-Favg}
\end{equation}
A derivation of this two-dimensional average-fidelity identity is given in Appendix~\ref{app:logical-average-fidelity}.

The small-\(\epsilon\) coefficients in Table~\ref{tab:logical-error-decomposition} are obtained by Taylor expanding Eqs.~\eqref{eq:appF-Bp-elements} and \eqref{eq:appF-logical-Favg}.  In particular,
\begin{equation}
1-\overline F_L^{(\epsilon)} = \chi_L\epsilon^2 + O(\epsilon^4),
\label{eq:appF-chi-expansion}
\end{equation}
while \(C_{\rm leak}\), \(C_{\rm coh}\), and \(\eta_Z\) are extracted from the small-\(\epsilon\) expansions of \({\cal L}_L(\epsilon)\), \({\cal D}_L(\epsilon)\), and \(\varphi_L(\epsilon)\) defined in Sec.~\ref{sec:codes}.

For the lowest-order binomial code, the finite sums give
\begin{equation}
\chi_L=1, \quad
C_{\rm leak}=1, \quad
C_{\rm coh}=0, \quad
\eta_Z=0 .
\label{eq:appF-binomial-coefficients}
\end{equation}
For the equal-weight fourfold subspace,
\begin{equation}
\chi_L=\frac{49}{12}, \quad
C_{\rm leak}=\frac{47}{12}, \quad
C_{\rm coh}=0, \quad
\eta_Z=1 .
\label{eq:appF-fourfold-coefficients}
\end{equation}
For the truncated cat subspace, the coefficients are obtained by inserting the normalized truncated even/odd coefficient vectors at \(\alpha=\sqrt2\) and cutoff \(n\leqslant10\) into Eq.~\eqref{eq:appF-Bp-elements}.  The finite-\(\epsilon\) curves in Fig.~\ref{fig:phaseerror}(b) are generated from Eq.~\eqref{eq:appF-logical-Favg}, not from the small-\(\epsilon\) expansions.

\section{Logical average fidelity for projected code subspaces}
\label{app:logical-average-fidelity}

This appendix records the average-fidelity identity used for the projected logical channels in Sec.~\ref{sec:codes}.  Let \({\cal E}_L\) denote the reduced transfer map after projection onto a two-dimensional logical target subspace,
\begin{equation}
{\cal E}_L(\rho) = \sum_j B_j \rho B_j^\dagger ,
\label{eq:app-logical-channel}
\end{equation}
where, the projected Kraus matrices \(B_j=P_LK_jP_L\) act on the logical basis \(\{|0_L\rangle,|1_L\rangle\}\).  The map in Eq.~\eqref{eq:app-logical-channel} need not be trace preserving, because population can leave the target logical subspace.

For a pure logical input \(|\psi\rangle\), after identifying the input and target logical bases, the fidelity with the ideal logical transfer is
\begin{equation}
F_L(\psi) = \langle\psi| {\cal E}_L(|\psi\rangle\langle\psi|) |\psi\rangle = \sum_j \left| \langle\psi|B_j|\psi\rangle \right|^2 .
\label{eq:app-logical-state-fidelity}
\end{equation}
The logical average fidelity is the uniform average over all pure states in the two-dimensional logical space,
\begin{equation}
\overline F_L = \int d\psi\, F_L(\psi).
\label{eq:app-logical-average-definition}
\end{equation}
Using the standard Haar identity
\begin{equation}
\int d\psi\, \left(|\psi\rangle\langle\psi|\right)^{\otimes2} = \frac{I+S}{d(d+1)},
\label{eq:app-haar-identity}
\end{equation}
where, \(S\) swaps the two copies and \(d\) is the Hilbert-space dimension, one finds for each Kraus matrix~\cite{Horodecki1999,Nielsen2002}
\begin{equation}
\int d\psi\, \left| \langle\psi|B_j|\psi\rangle \right|^2 =
\frac{{\rm Tr}(B_j^\dagger B_j) + \left|{\rm Tr}\,B_j\right|^2 }{d(d+1)} .
\label{eq:app-average-kraus-identity}
\end{equation}
For a logical qubit, \(d=2\).  Therefore
\begin{equation}
\overline F_L = \frac{1}{6} \sum_j \left[ {\rm Tr}\!\left(B_j^\dagger B_j\right) + \left| {\rm Tr}\,B_j \right|^2 \right].
\label{eq:app-logical-average-fidelity}
\end{equation}
This is Eq.~\eqref{eq:logical-average-fidelity} of the main text.  The first term accounts for the average probability retained inside the projected logical subspace, while the second term measures coherent agreement with the ideal logical transfer.  For an ideal logical transfer, \(B=I_2\), giving \(\overline F_L=(2+4)/6=1\), as required.

\section{Minimal open-system feasibility estimate}
\label{app:open-system}

The benchmarks in the main text are closed-system coherent-control results.  To estimate the scale on which those benchmarks remain predictive, we add a minimal Markovian noise model to the best-found native \(N=10\) evolution in Eq.~\eqref{eq:native-N10-optimum}.
We consider equal normal-mode loss, ancilla relaxation, and ancilla pure dephasing,
\begin{align}
\dot\rho &= -i[H_\Delta,\rho] + \kappa_+{\cal D}[a_+]\rho + \kappa_-{\cal D}[a_-]\rho \nonumber\\
&\quad + \gamma_1{\cal D}[\sigma_-]\rho + \gamma_\phi{\cal D}[\sigma_z]\rho ,
\label{eq:app-open-master}
\end{align}
with \(\kappa_+=\kappa_-=\kappa\) in the numerical estimates below.  Here
\begin{equation}
{\cal D}[O]\rho = O\rho O^\dagger - \tfrac{1}{2}\{O^\dagger O,\rho\}.
\label{eq:app-open-dissipator}
\end{equation}
The dephasing convention is therefore \(\gamma_\phi{\cal D}[\sigma_z]\).  For equal independent loss rates of the two physical oscillators, \(\kappa{\cal D}[a_1]+\kappa{\cal D}[a_2]\) is equivalent to \(\kappa{\cal D}[a_+]+\kappa{\cal D}[a_-]\).  Unequal physical loss rates would introduce device-specific normal-mode mixing terms and are not considered here.

The figure of merit is the same \(N=10\) average process fidelity used in the main text.  For the ideal-transfer overlap entering the entanglement fidelity, trajectories with photon-loss or ancilla-relaxation jumps end in sectors orthogonal to the ideal no-loss, ground-ancilla branch.  Their contribution to this overlap is therefore captured by the corresponding non-Hermitian no-jump evolution.  Ancilla dephasing does not remove excitation and is included directly in the bright-doublet operator evolution.  In the absence of noise, this block method reproduces the closed-form native result,
\begin{align}
\overline F_{10,{\rm block}}^{\rm nat} &= 0.999370508861683, \nonumber\\
\overline F_{10,{\rm closed}}^{\rm nat} &= 0.999370508861687,
\label{eq:app-open-sanity}
\end{align}
with a difference of \(4.1\times10^{-15}\).

\begin{figure}[t]
\centering
\includegraphics[width=\columnwidth]{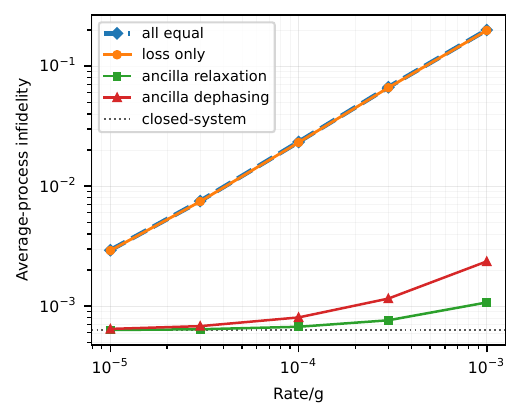}
\caption{
Minimal open-system estimate for the best-found native \(N=10\) evolution.  The curves show the average-process infidelity as a function of the dimensionless noise rate.  The horizontal dotted line is the coherent closed-system native infidelity \(6.29\times10^{-4}\).  The ``all equal'' curve is shown as a thick dashed line and nearly overlaps the loss-only curve, indicating that oscillator loss dominates the open-system error budget over the rates shown.
}
\label{fig:open-system-feasibility}
\end{figure}

Table~\ref{tab:open-system-feasibility} and Fig.~\ref{fig:open-system-feasibility} show the resulting open-system estimates.  The coherent closed-system native infidelity is
\[
1-\overline F_{10}^{\rm nat} = 6.29\times10^{-4}.
\]

\begin{table}[t]
\centering
\caption{
Minimal open-system feasibility estimate for the best-found native \(N=10\) evolution.  Entries are average-process infidelities.  The columns turn on equal normal-mode loss, ancilla relaxation, ancilla dephasing, and all three rates set equal. Here, ``Loss'' denotes \(\kappa_+=\kappa_-=\kappa\), ``Relaxation'' denotes ancilla relaxation at rate \(\gamma_1\), ``Dephasing'' denotes ancilla dephasing \(\gamma_\phi{\cal D}[\sigma_z]\), and ``All'' sets the three rates equal.
}
\label{tab:open-system-feasibility}
\setlength{\tabcolsep}{3.0pt}
\begin{ruledtabular}
\begin{tabular}{ccccc}
Rate/\(g\) & Loss & Relaxation & Dephasing & All \\ \hline
\(1.0{\times}10^{-5}\) & \(2.91{\times}10^{-3}\) & \(6.34{\times}10^{-4}\) & \(6.47{\times}10^{-4}\) & \(2.93{\times}10^{-3}\) \\
\(3.0{\times}10^{-5}\) & \(7.45{\times}10^{-3}\) & \(6.43{\times}10^{-4}\) & \(6.83{\times}10^{-4}\) & \(7.52{\times}10^{-3}\) \\
\(1.0{\times}10^{-4}\) & \(2.31{\times}10^{-2}\) & \(6.74{\times}10^{-4}\) & \(8.07{\times}10^{-4}\) & \(2.34{\times}10^{-2}\) \\
\(3.0{\times}10^{-4}\) & \(6.62{\times}10^{-2}\) & \(7.64{\times}10^{-4}\) & \(1.16{\times}10^{-3}\) & \(6.68{\times}10^{-2}\) \\
\(1.0{\times}10^{-3}\) & \(1.98{\times}10^{-1}\) & \(1.08{\times}10^{-3}\) & \(2.36{\times}10^{-3}\) & \(2.00{\times}10^{-1}\)
\end{tabular}
\end{ruledtabular}
\end{table}


The near overlap between the loss-only and all-equal curves shows that oscillator loss dominates the dissipative error budget over the rates shown. This is expected because the best-found operating point occurs at
\(\tau_{\rm opt}\simeq50\) and the process benchmark samples photon numbers up to \(N=10\).  The ancilla channels are much less harmful because the detuned protocol keeps the real ancilla population small.  In particular, at rate/\(g=10^{-3}\), the loss-only infidelity is \(1.98\times10^{-1}\), whereas ancilla relaxation and dephasing alone give \(1.08\times10^{-3}\) and \(2.36\times10^{-3}\), respectively.

This estimate is not a device-level noise model.  It omits Kerr and cross-Kerr phases, control imperfections, non-Markovian noise, and slow calibration drift. Coherent nonlinearities such as
\[
K_+ n_+^2 + K_- n_-^2 + \chi_{+-}n_+n_-
\]
must be calibrated or echoed when their accumulated phases over the evolution time \(t_{\rm evo}=\tau_{\rm opt}/g\) become comparable to the target error budget.  The purpose of the present calculation is narrower: it gives a dimensionless feasibility scale for the closed-system native benchmark.  In particular, keeping oscillator-loss error at the few-\(10^{-3}\) level for this \(N=10\), \(\tau\simeq50\) evolution requires roughly \(\kappa/g\lesssim10^{-5}\), whereas the detuned protocol is substantially less sensitive to ancilla relaxation and dephasing over the same range.

\begin{table*}[t]
\centering
\caption{
Comparison of transfer and bright-mode parity-synthesis routes.  The beam splitter is the direct-exchange benchmark; the remaining routes address the restricted single-ancilla setting considered in this work.
}
\label{tab:route-comparison}
\small
\begin{ruledtabular}
\begin{tabular}{p{0.18\textwidth}p{0.25\textwidth}p{0.28\textwidth}p{0.21\textwidth}}
Route
&
Resource
&
Role
&
Main limitation
\\
\hline
Ideal beam splitter
&
Direct exchange \(a_1^\dagger a_2+a_2^\dagger a_1\)
&
Implements the relative normal-mode phase directly; gives a swap up to known
frame phases
&
Requires high-quality mode--mode exchange
\\
Static resonant bright-mode JC
&
Resonant ancilla--bright-mode doublets
&
Exactly solvable single-ancilla baseline
&
\(\sqrt{k}\)-dependent Rabi phases give recurrence-limited transfer
\\
Native detuned bright-mode JC
&
Off-resonant ancilla--bright-mode coupling
&
Two-parameter finite-cutoff approximation to bright-mode parity
&
Residual nonlinear phase, leakage, and calibration sensitivity
\\
Dispersive bright-mode phase
&
Effective \(\chi_- a_-^\dagger a_- \sigma_z\)
&
Direct accumulation of a bright-mode number phase
&
Unwanted dark-mode phases, Kerr, cross-Kerr, and echo/calibration requirements
\\
Number-selective phase control
&
Programmed phases \(\theta_k=\pi k\)
&
Exact parity on a chosen finite cutoff
&
Spectral selectivity, evolution time, and \(O(N)\) phase calibration
\end{tabular}
\end{ruledtabular}
\end{table*}

\section{Beam splitter benchmark and implementation routes}
\label{app:beam splitter-implementation}
The bright-mode parity target is closely related to an ideal beam splitter swap. This relation is the same permutation–parity exchange structure identified for a balanced beam splitter~\cite{Campos_PRA_2005}, and is equivalent to writing SWAP as parity in the antisymmetric normal mode~\cite{Karaev2026}.
A direct beam splitter Hamiltonian,
\begin{equation}
H_{\rm BS} = J\left(a_1^\dagger a_2+a_2^\dagger a_1\right),
\label{eq:appG-HBS}
\end{equation}
is diagonal in the normal-mode basis $a_\pm = (a_1\pm a_2)/\sqrt{2}$,
where
\begin{equation}
H_{\rm BS}
=
J\left(a_+^\dagger a_+-a_-^\dagger a_-\right).
\label{eq:appG-HBS-normal}
\end{equation}
Thus, a beam splitter directly generates a relative phase between the symmetric
and antisymmetric modes:
\begin{equation}
a_+\rightarrow e^{-iJt}a_+,
\qquad
a_-\rightarrow e^{+iJt}a_- .
\label{eq:appG-BS-Heisenberg}
\end{equation}
At \(Jt=\pi/2\), this gives
\begin{equation}
a_+\rightarrow -i a_+,
\qquad
a_-\rightarrow i a_- = (-i)(-a_-).
\label{eq:appG-BS-swap-time}
\end{equation}
The common phase is a known oscillator-frame rotation, while the relative
phase is precisely the bright-mode parity phase.  Therefore, when a high-quality
direct exchange Hamiltonian is available, the beam splitter is the natural
closed-system swap benchmark.

The restricted single-ancilla setting studied in this work is different.  In
that case the relative normal-mode phase is not directly available as an
exchange Hamiltonian; it must be synthesized through the bright-mode dynamics
generated by the ancilla.  Table~\ref{tab:route-comparison} summarizes the
main routes to the same parity target.

For the native detuned route, the return amplitude \(r_k(\delta,\tau)\) is given in Eq.~\eqref{eq:native-rk}.  In the large-detuning regime, its phase is approximately linear in \(k\), while the ancilla population is suppressed. This is the minimal mechanism exploited in Sec.~\ref{sec:native}.  More elaborate routes can target the same operation more directly.  For example, a dispersive interaction can generate parity when
\begin{equation}
\chi_- t=\pi \pmod{2\pi},
\label{eq:appG-dispersive-parity-condition}
\end{equation}
provided unwanted dark-mode, Kerr, and cross-Kerr phases are corrected or echoed.  Such dispersive number-dependent phase accumulation is a standard resource in cavity and circuit QED~\cite{Blais2004,Schuster2007}.  Alternatively, number-selective phase control can program
\begin{equation}
U_{\rm SNAP}^{(N)} = \sum_{k=0}^{N} e^{i\theta_k}|k\rangle_-\langle k|, \qquad \theta_k=\pi k,
\label{eq:appG-SNAP}
\end{equation}
which is exactly bright-mode parity on the cutoff \(0\leqslant k\leqslant N\).  This is the finite-cutoff SNAP route to the same target~\cite{Heeres2015,Krastanov2015,Eickbusch2022}.

A hardware implementation should be judged by how accurately it realizes $|k\rangle_- \rightarrow (-1)^k |k\rangle_-$
on the relevant bright-mode Fock components while leaving the dark mode with only known correctable phases and returning the ancilla unentangled with the oscillators.  The relevant error scales are platform dependent: photon loss \(\kappa_\pm\), ancilla relaxation and dephasing \(\gamma_1,\gamma_\phi\), self-Kerr and cross-Kerr terms \(K_\pm,\chi_{+-}\), leakage \(L_{\max}^{(N)}\), and calibration errors in \(g\), \(\Delta\), and the evolution duration.  The present work treats the closed-system coherent-control limit.  Appendix~\ref{app:open-system} gives a minimal Markovian feasibility estimate for the best-found native \(N=10\) process.  A device-specific proposal must add platform-dependent Kerr, cross-Kerr, control-imperfection, non-Markovian, and drift effects to this common parity target.

\bibliography{references}

\begin{thebibliography}{68}%
\makeatletter
\providecommand \@ifxundefined [1]{%
 \@ifx{#1\undefined}
}%
\providecommand \@ifnum [1]{%
 \ifnum #1\expandafter \@firstoftwo
 \else \expandafter \@secondoftwo
 \fi
}%
\providecommand \@ifx [1]{%
 \ifx #1\expandafter \@firstoftwo
 \else \expandafter \@secondoftwo
 \fi
}%
\providecommand \natexlab [1]{#1}%
\providecommand \enquote  [1]{``#1''}%
\providecommand \bibnamefont  [1]{#1}%
\providecommand \bibfnamefont [1]{#1}%
\providecommand \citenamefont [1]{#1}%
\providecommand \href@noop [0]{\@secondoftwo}%
\providecommand \href [0]{\begingroup \@sanitize@url \@href}%
\providecommand \@href[1]{\@@startlink{#1}\@@href}%
\providecommand \@@href[1]{\endgroup#1\@@endlink}%
\providecommand \@sanitize@url [0]{\catcode `\\12\catcode `\$12\catcode
  `\&12\catcode `\#12\catcode `\^12\catcode `\_12\catcode `\%12\relax}%
\providecommand \@@startlink[1]{}%
\providecommand \@@endlink[0]{}%
\providecommand \url  [0]{\begingroup\@sanitize@url \@url }%
\providecommand \@url [1]{\endgroup\@href {#1}{\urlprefix }}%
\providecommand \urlprefix  [0]{URL }%
\providecommand \Eprint [0]{\href }%
\providecommand \doibase [0]{https://doi.org/}%
\providecommand \selectlanguage [0]{\@gobble}%
\providecommand \bibinfo  [0]{\@secondoftwo}%
\providecommand \bibfield  [0]{\@secondoftwo}%
\providecommand \translation [1]{[#1]}%
\providecommand \BibitemOpen [0]{}%
\providecommand \bibitemStop [0]{}%
\providecommand \bibitemNoStop [0]{.\EOS\space}%
\providecommand \EOS [0]{\spacefactor3000\relax}%
\providecommand \BibitemShut  [1]{\csname bibitem#1\endcsname}%
\let\auto@bib@innerbib\@empty
\bibitem [{\citenamefont {Braunstein}\ and\ \citenamefont {van
  Loock}(2005)}]{Braunstein2005}%
  \BibitemOpen
  \bibfield  {author} {\bibinfo {author} {\bibfnamefont {S.~L.}\ \bibnamefont
  {Braunstein}}\ and\ \bibinfo {author} {\bibfnamefont {P.}~\bibnamefont {van
  Loock}},\ }\bibfield  {title} {\bibinfo {title} {Quantum information with
  continuous variables},\ }\href {https://doi.org/10.1103/RevModPhys.77.513}
  {\bibfield  {journal} {\bibinfo  {journal} {Rev. Mod. Phys.}\ }\textbf
  {\bibinfo {volume} {77}},\ \bibinfo {pages} {513} (\bibinfo {year}
  {2005})}\BibitemShut {NoStop}%
\bibitem [{\citenamefont {Weedbrook}\ \emph {et~al.}(2012)\citenamefont
  {Weedbrook}, \citenamefont {Pirandola}, \citenamefont {Garc\'{\i}a-Patr\'on},
  \citenamefont {Cerf}, \citenamefont {Ralph}, \citenamefont {Shapiro},\ and\
  \citenamefont {Lloyd}}]{Weedbrook_RMP_2012}%
  \BibitemOpen
  \bibfield  {author} {\bibinfo {author} {\bibfnamefont {C.}~\bibnamefont
  {Weedbrook}}, \bibinfo {author} {\bibfnamefont {S.}~\bibnamefont
  {Pirandola}}, \bibinfo {author} {\bibfnamefont {R.}~\bibnamefont
  {Garc\'{\i}a-Patr\'on}}, \bibinfo {author} {\bibfnamefont {N.~J.}\
  \bibnamefont {Cerf}}, \bibinfo {author} {\bibfnamefont {T.~C.}\ \bibnamefont
  {Ralph}}, \bibinfo {author} {\bibfnamefont {J.~H.}\ \bibnamefont {Shapiro}},\
  and\ \bibinfo {author} {\bibfnamefont {S.}~\bibnamefont {Lloyd}},\ }\bibfield
   {title} {\bibinfo {title} {Gaussian quantum information},\ }\href
  {https://doi.org/10.1103/RevModPhys.84.621} {\bibfield  {journal} {\bibinfo
  {journal} {Rev. Mod. Phys.}\ }\textbf {\bibinfo {volume} {84}},\ \bibinfo
  {pages} {621} (\bibinfo {year} {2012})}\BibitemShut {NoStop}%
\bibitem [{\citenamefont {Serafini}(2017)}]{Serafini_2017}%
  \BibitemOpen
  \bibfield  {author} {\bibinfo {author} {\bibfnamefont {A.}~\bibnamefont
  {Serafini}},\ }\href {https://doi.org/https://doi.org/10.1201/9781315118727}
  {\emph {\bibinfo {title} {Quantum Continuous Variables}}}\ (\bibinfo
  {publisher} {CRC Press},\ \bibinfo {year} {2017})\BibitemShut {NoStop}%
\bibitem [{\citenamefont {Albert}\ \emph {et~al.}(2018)\citenamefont {Albert}
  \emph {et~al.}}]{Albert2018}%
  \BibitemOpen
  \bibfield  {author} {\bibinfo {author} {\bibfnamefont {V.~V.}\ \bibnamefont
  {Albert}} \emph {et~al.},\ }\bibfield  {title} {\bibinfo {title} {Performance
  and structure of single-mode bosonic codes},\ }\href
  {https://doi.org/10.1103/PhysRevA.97.032346} {\bibfield  {journal} {\bibinfo
  {journal} {Phys. Rev. A}\ }\textbf {\bibinfo {volume} {97}},\ \bibinfo
  {pages} {032346} (\bibinfo {year} {2018})}\BibitemShut {NoStop}%
\bibitem [{\citenamefont {Haroche}\ and\ \citenamefont
  {Raimond}(2006)}]{Haroche2006}%
  \BibitemOpen
  \bibfield  {author} {\bibinfo {author} {\bibfnamefont {S.}~\bibnamefont
  {Haroche}}\ and\ \bibinfo {author} {\bibfnamefont {J.-M.}\ \bibnamefont
  {Raimond}},\ }\href
  {https://doi.org/10.1093/acprof:oso/9780198509141.001.0001} {\emph {\bibinfo
  {title} {Exploring the Quantum: Atoms, Cavities, and Photons}}}\ (\bibinfo
  {publisher} {Oxford University Press},\ \bibinfo {address} {Oxford},\
  \bibinfo {year} {2006})\BibitemShut {NoStop}%
\bibitem [{\citenamefont {Li}\ \emph {et~al.}(2024)\citenamefont {Li},
  \citenamefont {Ni}, \citenamefont {Zhang}, \citenamefont {Cai}, \citenamefont
  {Mai}, \citenamefont {Wen}, \citenamefont {Zheng}, \citenamefont {Deng},
  \citenamefont {Liu}, \citenamefont {Xu},\ and\ \citenamefont
  {Yu}}]{Li_PRL_2024}%
  \BibitemOpen
  \bibfield  {author} {\bibinfo {author} {\bibfnamefont {S.}~\bibnamefont
  {Li}}, \bibinfo {author} {\bibfnamefont {Z.}~\bibnamefont {Ni}}, \bibinfo
  {author} {\bibfnamefont {L.}~\bibnamefont {Zhang}}, \bibinfo {author}
  {\bibfnamefont {Y.}~\bibnamefont {Cai}}, \bibinfo {author} {\bibfnamefont
  {J.}~\bibnamefont {Mai}}, \bibinfo {author} {\bibfnamefont {S.}~\bibnamefont
  {Wen}}, \bibinfo {author} {\bibfnamefont {P.}~\bibnamefont {Zheng}}, \bibinfo
  {author} {\bibfnamefont {X.}~\bibnamefont {Deng}}, \bibinfo {author}
  {\bibfnamefont {S.}~\bibnamefont {Liu}}, \bibinfo {author} {\bibfnamefont
  {Y.}~\bibnamefont {Xu}},\ and\ \bibinfo {author} {\bibfnamefont
  {D.}~\bibnamefont {Yu}},\ }\bibfield  {title} {\bibinfo {title} {Autonomous
  stabilization of fock states in an oscillator against multiphoton losses},\
  }\href {https://doi.org/10.1103/PhysRevLett.132.203602} {\bibfield  {journal}
  {\bibinfo  {journal} {Phys. Rev. Lett.}\ }\textbf {\bibinfo {volume} {132}},\
  \bibinfo {pages} {203602} (\bibinfo {year} {2024})}\BibitemShut {NoStop}%
\bibitem [{\citenamefont {Cochrane}\ \emph {et~al.}(1999)\citenamefont
  {Cochrane}, \citenamefont {Milburn},\ and\ \citenamefont
  {Munro}}]{Cochrane_PRA_1999}%
  \BibitemOpen
  \bibfield  {author} {\bibinfo {author} {\bibfnamefont {P.~T.}\ \bibnamefont
  {Cochrane}}, \bibinfo {author} {\bibfnamefont {G.~J.}\ \bibnamefont
  {Milburn}},\ and\ \bibinfo {author} {\bibfnamefont {W.~J.}\ \bibnamefont
  {Munro}},\ }\bibfield  {title} {\bibinfo {title} {Macroscopically distinct
  quantum-superposition states as a bosonic code for amplitude damping},\
  }\href {https://doi.org/10.1103/PhysRevA.59.2631} {\bibfield  {journal}
  {\bibinfo  {journal} {Phys. Rev. A}\ }\textbf {\bibinfo {volume} {59}},\
  \bibinfo {pages} {2631} (\bibinfo {year} {1999})}\BibitemShut {NoStop}%
\bibitem [{\citenamefont {Leghtas}\ \emph {et~al.}(2013)\citenamefont
  {Leghtas}, \citenamefont {Kirchmair}, \citenamefont {Vlastakis},
  \citenamefont {Schoelkopf}, \citenamefont {Devoret},\ and\ \citenamefont
  {Mirrahimi}}]{Leghtas2013}%
  \BibitemOpen
  \bibfield  {author} {\bibinfo {author} {\bibfnamefont {Z.}~\bibnamefont
  {Leghtas}}, \bibinfo {author} {\bibfnamefont {G.}~\bibnamefont {Kirchmair}},
  \bibinfo {author} {\bibfnamefont {B.}~\bibnamefont {Vlastakis}}, \bibinfo
  {author} {\bibfnamefont {R.~J.}\ \bibnamefont {Schoelkopf}}, \bibinfo
  {author} {\bibfnamefont {M.~H.}\ \bibnamefont {Devoret}},\ and\ \bibinfo
  {author} {\bibfnamefont {M.}~\bibnamefont {Mirrahimi}},\ }\bibfield  {title}
  {\bibinfo {title} {Hardware-efficient autonomous quantum memory protection},\
  }\href {https://doi.org/10.1103/PhysRevLett.111.120501} {\bibfield  {journal}
  {\bibinfo  {journal} {Phys. Rev. Lett.}\ }\textbf {\bibinfo {volume} {111}},\
  \bibinfo {pages} {120501} (\bibinfo {year} {2013})}\BibitemShut {NoStop}%
\bibitem [{\citenamefont {Leghtas}\ \emph {et~al.}(2015)\citenamefont {Leghtas}
  \emph {et~al.}}]{Leghtas2015}%
  \BibitemOpen
  \bibfield  {author} {\bibinfo {author} {\bibfnamefont {Z.}~\bibnamefont
  {Leghtas}} \emph {et~al.},\ }\bibfield  {title} {\bibinfo {title} {Confining
  the state of light to a quantum manifold by engineered two-photon loss},\
  }\href {https://doi.org/10.1126/science.aaa2085} {\bibfield  {journal}
  {\bibinfo  {journal} {Science}\ }\textbf {\bibinfo {volume} {347}},\ \bibinfo
  {pages} {853} (\bibinfo {year} {2015})}\BibitemShut {NoStop}%
\bibitem [{\citenamefont {Mirrahimi}\ \emph {et~al.}(2014)\citenamefont
  {Mirrahimi}, \citenamefont {Leghtas}, \citenamefont {Albert}, \citenamefont
  {Touzard}, \citenamefont {Schoelkopf}, \citenamefont {Jiang},\ and\
  \citenamefont {Devoret}}]{Mirrahimi2014}%
  \BibitemOpen
  \bibfield  {author} {\bibinfo {author} {\bibfnamefont {M.}~\bibnamefont
  {Mirrahimi}}, \bibinfo {author} {\bibfnamefont {Z.}~\bibnamefont {Leghtas}},
  \bibinfo {author} {\bibfnamefont {V.~V.}\ \bibnamefont {Albert}}, \bibinfo
  {author} {\bibfnamefont {S.}~\bibnamefont {Touzard}}, \bibinfo {author}
  {\bibfnamefont {R.~J.}\ \bibnamefont {Schoelkopf}}, \bibinfo {author}
  {\bibfnamefont {L.}~\bibnamefont {Jiang}},\ and\ \bibinfo {author}
  {\bibfnamefont {M.~H.}\ \bibnamefont {Devoret}},\ }\bibfield  {title}
  {\bibinfo {title} {Dynamically protected cat-qubits: A new paradigm for
  universal quantum computation},\ }\href
  {https://doi.org/10.1088/1367-2630/16/4/045014} {\bibfield  {journal}
  {\bibinfo  {journal} {New J. Phys.}\ }\textbf {\bibinfo {volume} {16}},\
  \bibinfo {pages} {045014} (\bibinfo {year} {2014})}\BibitemShut {NoStop}%
\bibitem [{\citenamefont {Michael}\ \emph {et~al.}(2016)\citenamefont
  {Michael}, \citenamefont {Silveri}, \citenamefont {Brierley}, \citenamefont
  {Albert}, \citenamefont {Salmilehto}, \citenamefont {Jiang},\ and\
  \citenamefont {Girvin}}]{Michael2016}%
  \BibitemOpen
  \bibfield  {author} {\bibinfo {author} {\bibfnamefont {M.~H.}\ \bibnamefont
  {Michael}}, \bibinfo {author} {\bibfnamefont {M.}~\bibnamefont {Silveri}},
  \bibinfo {author} {\bibfnamefont {R.~T.}\ \bibnamefont {Brierley}}, \bibinfo
  {author} {\bibfnamefont {V.~V.}\ \bibnamefont {Albert}}, \bibinfo {author}
  {\bibfnamefont {J.}~\bibnamefont {Salmilehto}}, \bibinfo {author}
  {\bibfnamefont {L.}~\bibnamefont {Jiang}},\ and\ \bibinfo {author}
  {\bibfnamefont {S.~M.}\ \bibnamefont {Girvin}},\ }\bibfield  {title}
  {\bibinfo {title} {New class of quantum error-correcting codes for a bosonic
  mode},\ }\href {https://doi.org/10.1103/PhysRevX.6.031006} {\bibfield
  {journal} {\bibinfo  {journal} {Phys. Rev. X}\ }\textbf {\bibinfo {volume}
  {6}},\ \bibinfo {pages} {031006} (\bibinfo {year} {2016})}\BibitemShut
  {NoStop}%
\bibitem [{\citenamefont {Chen}\ \emph {et~al.}(2021)\citenamefont {Chen},
  \citenamefont {Qin}, \citenamefont {Stassi}, \citenamefont {Wang},\ and\
  \citenamefont {Nori}}]{Chen_PRR_2021}%
  \BibitemOpen
  \bibfield  {author} {\bibinfo {author} {\bibfnamefont {Y.-H.}\ \bibnamefont
  {Chen}}, \bibinfo {author} {\bibfnamefont {W.}~\bibnamefont {Qin}}, \bibinfo
  {author} {\bibfnamefont {R.}~\bibnamefont {Stassi}}, \bibinfo {author}
  {\bibfnamefont {X.}~\bibnamefont {Wang}},\ and\ \bibinfo {author}
  {\bibfnamefont {F.}~\bibnamefont {Nori}},\ }\bibfield  {title} {\bibinfo
  {title} {Fast binomial-code holonomic quantum computation with ultrastrong
  light-matter coupling},\ }\href
  {https://doi.org/10.1103/PhysRevResearch.3.033275} {\bibfield  {journal}
  {\bibinfo  {journal} {Phys. Rev. Res.}\ }\textbf {\bibinfo {volume} {3}},\
  \bibinfo {pages} {033275} (\bibinfo {year} {2021})}\BibitemShut {NoStop}%
\bibitem [{\citenamefont {Chang}(2025)}]{chang_2025}%
  \BibitemOpen
  \bibfield  {author} {\bibinfo {author} {\bibfnamefont {E.-J.}\ \bibnamefont
  {Chang}},\ }\bibfield  {title} {\bibinfo {title} {High-rate extended binomial
  codes for multiqubit encoding},\ }\href {https://doi.org/10.1103/hwfz-c6vy}
  {\bibfield  {journal} {\bibinfo  {journal} {Phys. Rev. A}\ }\textbf {\bibinfo
  {volume} {112}},\ \bibinfo {pages} {032419} (\bibinfo {year}
  {2025})}\BibitemShut {NoStop}%
\bibitem [{\citenamefont {Gottesman}\ \emph {et~al.}(2001)\citenamefont
  {Gottesman}, \citenamefont {Kitaev},\ and\ \citenamefont
  {Preskill}}]{GKP2001}%
  \BibitemOpen
  \bibfield  {author} {\bibinfo {author} {\bibfnamefont {D.}~\bibnamefont
  {Gottesman}}, \bibinfo {author} {\bibfnamefont {A.}~\bibnamefont {Kitaev}},\
  and\ \bibinfo {author} {\bibfnamefont {J.}~\bibnamefont {Preskill}},\
  }\bibfield  {title} {\bibinfo {title} {Encoding a qubit in an oscillator},\
  }\href {https://doi.org/10.1103/PhysRevA.64.012310} {\bibfield  {journal}
  {\bibinfo  {journal} {Phys. Rev. A}\ }\textbf {\bibinfo {volume} {64}},\
  \bibinfo {pages} {012310} (\bibinfo {year} {2001})}\BibitemShut {NoStop}%
\bibitem [{\citenamefont {Baragiola}\ \emph {et~al.}(2019)\citenamefont
  {Baragiola}, \citenamefont {Pantaleoni}, \citenamefont {Alexander},
  \citenamefont {Karanjai},\ and\ \citenamefont
  {Menicucci}}]{Baragiola_PRL_2019}%
  \BibitemOpen
  \bibfield  {author} {\bibinfo {author} {\bibfnamefont {B.~Q.}\ \bibnamefont
  {Baragiola}}, \bibinfo {author} {\bibfnamefont {G.}~\bibnamefont
  {Pantaleoni}}, \bibinfo {author} {\bibfnamefont {R.~N.}\ \bibnamefont
  {Alexander}}, \bibinfo {author} {\bibfnamefont {A.}~\bibnamefont
  {Karanjai}},\ and\ \bibinfo {author} {\bibfnamefont {N.~C.}\ \bibnamefont
  {Menicucci}},\ }\bibfield  {title} {\bibinfo {title} {All-gaussian
  universality and fault tolerance with the gottesman-kitaev-preskill code},\
  }\href {https://doi.org/10.1103/PhysRevLett.123.200502} {\bibfield  {journal}
  {\bibinfo  {journal} {Phys. Rev. Lett.}\ }\textbf {\bibinfo {volume} {123}},\
  \bibinfo {pages} {200502} (\bibinfo {year} {2019})}\BibitemShut {NoStop}%
\bibitem [{\citenamefont {Noh}\ \emph {et~al.}(2019)\citenamefont {Noh},
  \citenamefont {Albert},\ and\ \citenamefont {Jiang}}]{Noh_IEEE_2019}%
  \BibitemOpen
  \bibfield  {author} {\bibinfo {author} {\bibfnamefont {K.}~\bibnamefont
  {Noh}}, \bibinfo {author} {\bibfnamefont {V.~V.}\ \bibnamefont {Albert}},\
  and\ \bibinfo {author} {\bibfnamefont {L.}~\bibnamefont {Jiang}},\ }\bibfield
   {title} {\bibinfo {title} {Quantum capacity bounds of gaussian thermal loss
  channels and achievable rates with {Gottesman-Kitaev-Preskill} codes},\
  }\href {https://doi.org/10.1109/TIT.2018.2873764} {\bibfield  {journal}
  {\bibinfo  {journal} {IEEE Trans. Inf. Theory}\ }\textbf {\bibinfo {volume}
  {65}},\ \bibinfo {pages} {2563} (\bibinfo {year} {2019})}\BibitemShut
  {NoStop}%
\bibitem [{\citenamefont {Campagne-Ibarcq}\ \emph {et~al.}(2020)\citenamefont
  {Campagne-Ibarcq} \emph {et~al.}}]{CampagneIbarcq2020}%
  \BibitemOpen
  \bibfield  {author} {\bibinfo {author} {\bibfnamefont {P.}~\bibnamefont
  {Campagne-Ibarcq}} \emph {et~al.},\ }\bibfield  {title} {\bibinfo {title}
  {Quantum error correction of a qubit encoded in grid states of an
  oscillator},\ }\href {https://doi.org/10.1038/s41586-020-2603-3} {\bibfield
  {journal} {\bibinfo  {journal} {Nature}\ }\textbf {\bibinfo {volume} {584}},\
  \bibinfo {pages} {368} (\bibinfo {year} {2020})}\BibitemShut {NoStop}%
\bibitem [{\citenamefont {Noh}\ and\ \citenamefont
  {Chamberland}(2020)}]{Noh_PRA_2020}%
  \BibitemOpen
  \bibfield  {author} {\bibinfo {author} {\bibfnamefont {K.}~\bibnamefont
  {Noh}}\ and\ \bibinfo {author} {\bibfnamefont {C.}~\bibnamefont
  {Chamberland}},\ }\bibfield  {title} {\bibinfo {title} {Fault-tolerant
  bosonic quantum error correction with the
  surface--{Gottesman-Kitaev-Preskill} code},\ }\href
  {https://doi.org/10.1103/PhysRevA.101.012316} {\bibfield  {journal} {\bibinfo
   {journal} {Phys. Rev. A}\ }\textbf {\bibinfo {volume} {101}},\ \bibinfo
  {pages} {012316} (\bibinfo {year} {2020})}\BibitemShut {NoStop}%
\bibitem [{\citenamefont {Noh}\ \emph {et~al.}(2022)\citenamefont {Noh},
  \citenamefont {Chamberland},\ and\ \citenamefont
  {Brand\~ao}}]{Noh_PRXQ_2022}%
  \BibitemOpen
  \bibfield  {author} {\bibinfo {author} {\bibfnamefont {K.}~\bibnamefont
  {Noh}}, \bibinfo {author} {\bibfnamefont {C.}~\bibnamefont {Chamberland}},\
  and\ \bibinfo {author} {\bibfnamefont {F.~G.}\ \bibnamefont {Brand\~ao}},\
  }\bibfield  {title} {\bibinfo {title} {Low-overhead fault-tolerant quantum
  error correction with the surface-{GKP} code},\ }\href
  {https://doi.org/10.1103/PRXQuantum.3.010315} {\bibfield  {journal} {\bibinfo
   {journal} {PRX Quantum}\ }\textbf {\bibinfo {volume} {3}},\ \bibinfo {pages}
  {010315} (\bibinfo {year} {2022})}\BibitemShut {NoStop}%
\bibitem [{\citenamefont {Soule}\ \emph {et~al.}(2024)\citenamefont {Soule},
  \citenamefont {Doherty},\ and\ \citenamefont {Grimsmo}}]{Juliette_2024}%
  \BibitemOpen
  \bibfield  {author} {\bibinfo {author} {\bibfnamefont {J.}~\bibnamefont
  {Soule}}, \bibinfo {author} {\bibfnamefont {A.~C.}\ \bibnamefont {Doherty}},\
  and\ \bibinfo {author} {\bibfnamefont {A.~L.}\ \bibnamefont {Grimsmo}},\
  }\href {https://arxiv.org/abs/2312.14390} {\bibinfo {title} {Concatenating
  binomial codes with the planar code}} (\bibinfo {year} {2024}),\ \Eprint
  {https://arxiv.org/abs/2312.14390} {arXiv:2312.14390 [quant-ph]} \BibitemShut
  {NoStop}%
\bibitem [{\citenamefont {Conrad}(2024)}]{Conrad_2024}%
  \BibitemOpen
  \bibfield  {author} {\bibinfo {author} {\bibfnamefont {J.}~\bibnamefont
  {Conrad}},\ }\emph {\bibinfo {title} {The fabulous world of {GKP} codes}},\
  \href {http://dx.doi.org/10.17169/refubium-45505} {\bibinfo {type}
  {Dissertation}} (\bibinfo {year} {2024})\BibitemShut {NoStop}%
\bibitem [{\citenamefont {Eickbusch}\ \emph {et~al.}(2022)\citenamefont
  {Eickbusch}, \citenamefont {Sivak}, \citenamefont {Ding}, \citenamefont
  {Elder}, \citenamefont {Jha}, \citenamefont {Venkatraman}, \citenamefont
  {Royer}, \citenamefont {Girvin}, \citenamefont {Schoelkopf},\ and\
  \citenamefont {Devoret}}]{Eickbusch2022}%
  \BibitemOpen
  \bibfield  {author} {\bibinfo {author} {\bibfnamefont {A.}~\bibnamefont
  {Eickbusch}}, \bibinfo {author} {\bibfnamefont {V.}~\bibnamefont {Sivak}},
  \bibinfo {author} {\bibfnamefont {A.~Z.}\ \bibnamefont {Ding}}, \bibinfo
  {author} {\bibfnamefont {S.~S.}\ \bibnamefont {Elder}}, \bibinfo {author}
  {\bibfnamefont {S.~R.}\ \bibnamefont {Jha}}, \bibinfo {author} {\bibfnamefont
  {J.}~\bibnamefont {Venkatraman}}, \bibinfo {author} {\bibfnamefont
  {B.}~\bibnamefont {Royer}}, \bibinfo {author} {\bibfnamefont {S.~M.}\
  \bibnamefont {Girvin}}, \bibinfo {author} {\bibfnamefont {R.~J.}\
  \bibnamefont {Schoelkopf}},\ and\ \bibinfo {author} {\bibfnamefont {M.~H.}\
  \bibnamefont {Devoret}},\ }\bibfield  {title} {\bibinfo {title} {Fast
  universal control of an oscillator with weak dispersive coupling to a
  qubit},\ }\href {https://doi.org/10.1038/s41567-022-01776-9} {\bibfield
  {journal} {\bibinfo  {journal} {Nat. Phys.}\ }\textbf {\bibinfo {volume}
  {18}},\ \bibinfo {pages} {1464} (\bibinfo {year} {2022})}\BibitemShut
  {NoStop}%
\bibitem [{\citenamefont {Grimsmo}\ \emph {et~al.}(2020)\citenamefont
  {Grimsmo}, \citenamefont {Combes},\ and\ \citenamefont
  {Baragiola}}]{Grimsmo2020}%
  \BibitemOpen
  \bibfield  {author} {\bibinfo {author} {\bibfnamefont {A.~L.}\ \bibnamefont
  {Grimsmo}}, \bibinfo {author} {\bibfnamefont {J.}~\bibnamefont {Combes}},\
  and\ \bibinfo {author} {\bibfnamefont {B.~Q.}\ \bibnamefont {Baragiola}},\
  }\bibfield  {title} {\bibinfo {title} {Quantum computing with
  rotation-symmetric bosonic codes},\ }\href
  {https://doi.org/10.1103/PhysRevX.10.011058} {\bibfield  {journal} {\bibinfo
  {journal} {Phys. Rev. X}\ }\textbf {\bibinfo {volume} {10}},\ \bibinfo
  {pages} {011058} (\bibinfo {year} {2020})}\BibitemShut {NoStop}%
\bibitem [{\citenamefont {Xu}\ \emph {et~al.}(2025)\citenamefont {Xu},
  \citenamefont {Wang}, \citenamefont {Vuillot},\ and\ \citenamefont
  {Albert}}]{xu_2024}%
  \BibitemOpen
  \bibfield  {author} {\bibinfo {author} {\bibfnamefont {Y.}~\bibnamefont
  {Xu}}, \bibinfo {author} {\bibfnamefont {Y.}~\bibnamefont {Wang}}, \bibinfo
  {author} {\bibfnamefont {C.}~\bibnamefont {Vuillot}},\ and\ \bibinfo {author}
  {\bibfnamefont {V.~V.}\ \bibnamefont {Albert}},\ }\bibfield  {title}
  {\bibinfo {title} {Letting the tiger out of its cage: Bosonic coding without
  concatenation},\ }\href {https://doi.org/10.1103/ls5r-vj7r} {\bibfield
  {journal} {\bibinfo  {journal} {Phys. Rev. X}\ }\textbf {\bibinfo {volume}
  {15}},\ \bibinfo {pages} {041025} (\bibinfo {year} {2025})}\BibitemShut
  {NoStop}%
\bibitem [{\citenamefont {Cai}\ \emph {et~al.}(2021)\citenamefont {Cai},
  \citenamefont {Ma}, \citenamefont {Wang}, \citenamefont {Zou},\ and\
  \citenamefont {Sun}}]{Cai2021}%
  \BibitemOpen
  \bibfield  {author} {\bibinfo {author} {\bibfnamefont {W.}~\bibnamefont
  {Cai}}, \bibinfo {author} {\bibfnamefont {Y.}~\bibnamefont {Ma}}, \bibinfo
  {author} {\bibfnamefont {W.}~\bibnamefont {Wang}}, \bibinfo {author}
  {\bibfnamefont {C.-L.}\ \bibnamefont {Zou}},\ and\ \bibinfo {author}
  {\bibfnamefont {L.}~\bibnamefont {Sun}},\ }\bibfield  {title} {\bibinfo
  {title} {Bosonic quantum error correction codes in superconducting quantum
  circuits},\ }\href {https://doi.org/10.1016/j.fmre.2020.12.006} {\bibfield
  {journal} {\bibinfo  {journal} {Fundam. Res.}\ }\textbf {\bibinfo {volume}
  {1}},\ \bibinfo {pages} {50} (\bibinfo {year} {2021})}\BibitemShut {NoStop}%
\bibitem [{\citenamefont {Ma}\ \emph {et~al.}(2021)\citenamefont {Ma},
  \citenamefont {Puri}, \citenamefont {Schoelkopf}, \citenamefont {Devoret},
  \citenamefont {Girvin},\ and\ \citenamefont {Jiang}}]{Ma2021}%
  \BibitemOpen
  \bibfield  {author} {\bibinfo {author} {\bibfnamefont {W.-L.}\ \bibnamefont
  {Ma}}, \bibinfo {author} {\bibfnamefont {S.}~\bibnamefont {Puri}}, \bibinfo
  {author} {\bibfnamefont {R.~J.}\ \bibnamefont {Schoelkopf}}, \bibinfo
  {author} {\bibfnamefont {M.~H.}\ \bibnamefont {Devoret}}, \bibinfo {author}
  {\bibfnamefont {S.~M.}\ \bibnamefont {Girvin}},\ and\ \bibinfo {author}
  {\bibfnamefont {L.}~\bibnamefont {Jiang}},\ }\bibfield  {title} {\bibinfo
  {title} {Quantum control of bosonic modes with superconducting circuits},\
  }\href {https://doi.org/10.1016/j.scib.2021.05.024} {\bibfield  {journal}
  {\bibinfo  {journal} {Sci. Bull.}\ }\textbf {\bibinfo {volume} {66}},\
  \bibinfo {pages} {1789} (\bibinfo {year} {2021})}\BibitemShut {NoStop}%
\bibitem [{\citenamefont {Albert}(2025)}]{Albert2025}%
  \BibitemOpen
  \bibfield  {author} {\bibinfo {author} {\bibfnamefont {V.~V.}\ \bibnamefont
  {Albert}},\ }\bibfield  {title} {\bibinfo {title} {Bosonic coding:
  Introduction and use cases},\ }in\ \href {https://doi.org/10.3254/ENFI250007}
  {\emph {\bibinfo {booktitle} {Quantum Error Correction}}},\ \bibinfo {series}
  {Proceedings of the International School of Physics ``Enrico Fermi''}, Vol.\
  \bibinfo {volume} {209}\ (\bibinfo  {publisher} {IOS Press},\ \bibinfo {year}
  {2025})\ pp.\ \bibinfo {pages} {1--46}\BibitemShut {NoStop}%
\bibitem [{\citenamefont {Putterman}\ \emph {et~al.}(2025)\citenamefont
  {Putterman} \emph {et~al.}}]{Putterman2025}%
  \BibitemOpen
  \bibfield  {author} {\bibinfo {author} {\bibfnamefont {H.}~\bibnamefont
  {Putterman}} \emph {et~al.},\ }\bibfield  {title} {\bibinfo {title}
  {Hardware-efficient quantum error correction via concatenated bosonic
  qubits},\ }\href {https://doi.org/10.1038/s41586-025-08642-7} {\bibfield
  {journal} {\bibinfo  {journal} {Nature}\ }\textbf {\bibinfo {volume} {638}},\
  \bibinfo {pages} {927} (\bibinfo {year} {2025})}\BibitemShut {NoStop}%
\bibitem [{\citenamefont {Hann}\ \emph {et~al.}(2025)\citenamefont {Hann},
  \citenamefont {Noh}, \citenamefont {Putterman}, \citenamefont {Matheny},
  \citenamefont {Iverson}, \citenamefont {Fang}, \citenamefont {Chamberland},
  \citenamefont {Painter},\ and\ \citenamefont {Brand{\~a}o}}]{Hann2025}%
  \BibitemOpen
  \bibfield  {author} {\bibinfo {author} {\bibfnamefont {C.~T.}\ \bibnamefont
  {Hann}}, \bibinfo {author} {\bibfnamefont {K.}~\bibnamefont {Noh}}, \bibinfo
  {author} {\bibfnamefont {H.}~\bibnamefont {Putterman}}, \bibinfo {author}
  {\bibfnamefont {M.~H.}\ \bibnamefont {Matheny}}, \bibinfo {author}
  {\bibfnamefont {J.~K.}\ \bibnamefont {Iverson}}, \bibinfo {author}
  {\bibfnamefont {M.~T.}\ \bibnamefont {Fang}}, \bibinfo {author}
  {\bibfnamefont {C.}~\bibnamefont {Chamberland}}, \bibinfo {author}
  {\bibfnamefont {O.}~\bibnamefont {Painter}},\ and\ \bibinfo {author}
  {\bibfnamefont {F.~G. S.~L.}\ \bibnamefont {Brand{\~a}o}},\ }\bibfield
  {title} {\bibinfo {title} {Hybrid cat-transmon architecture for scalable,
  hardware-efficient quantum error correction},\ }\href
  {https://doi.org/10.1103/75x7-5ysv} {\bibfield  {journal} {\bibinfo
  {journal} {PRX Quantum}\ }\textbf {\bibinfo {volume} {6}},\ \bibinfo {pages}
  {030305} (\bibinfo {year} {2025})}\BibitemShut {NoStop}%
\bibitem [{\citenamefont {Christandl}\ \emph {et~al.}(2004)\citenamefont
  {Christandl}, \citenamefont {Datta}, \citenamefont {Ekert},\ and\
  \citenamefont {Landahl}}]{Christandl_PRL_2004}%
  \BibitemOpen
  \bibfield  {author} {\bibinfo {author} {\bibfnamefont {M.}~\bibnamefont
  {Christandl}}, \bibinfo {author} {\bibfnamefont {N.}~\bibnamefont {Datta}},
  \bibinfo {author} {\bibfnamefont {A.}~\bibnamefont {Ekert}},\ and\ \bibinfo
  {author} {\bibfnamefont {A.~J.}\ \bibnamefont {Landahl}},\ }\bibfield
  {title} {\bibinfo {title} {Perfect state transfer in quantum spin networks},\
  }\href {https://doi.org/10.1103/PhysRevLett.92.187902} {\bibfield  {journal}
  {\bibinfo  {journal} {Phys. Rev. Lett.}\ }\textbf {\bibinfo {volume} {92}},\
  \bibinfo {pages} {187902} (\bibinfo {year} {2004})}\BibitemShut {NoStop}%
\bibitem [{\citenamefont {Axline}\ \emph {et~al.}(2018)\citenamefont {Axline}
  \emph {et~al.}}]{Axline2018}%
  \BibitemOpen
  \bibfield  {author} {\bibinfo {author} {\bibfnamefont {C.~J.}\ \bibnamefont
  {Axline}} \emph {et~al.},\ }\bibfield  {title} {\bibinfo {title} {On-demand
  quantum state transfer and entanglement between remote microwave cavity
  memories},\ }\href {https://doi.org/10.1038/s41567-018-0115-y} {\bibfield
  {journal} {\bibinfo  {journal} {Nat. Phys.}\ }\textbf {\bibinfo {volume}
  {14}},\ \bibinfo {pages} {705} (\bibinfo {year} {2018})}\BibitemShut
  {NoStop}%
\bibitem [{\citenamefont {Kurpiers}\ \emph {et~al.}(2018)\citenamefont
  {Kurpiers} \emph {et~al.}}]{Kurpiers2018}%
  \BibitemOpen
  \bibfield  {author} {\bibinfo {author} {\bibfnamefont {P.}~\bibnamefont
  {Kurpiers}} \emph {et~al.},\ }\bibfield  {title} {\bibinfo {title}
  {Deterministic quantum state transfer and generation of remote entanglement
  using microwave photons},\ }\href {https://doi.org/10.1038/s41586-018-0195-y}
  {\bibfield  {journal} {\bibinfo  {journal} {Nature}\ }\textbf {\bibinfo
  {volume} {558}},\ \bibinfo {pages} {264} (\bibinfo {year}
  {2018})}\BibitemShut {NoStop}%
\bibitem [{\citenamefont {Campagne-Ibarcq}\ \emph {et~al.}(2018)\citenamefont
  {Campagne-Ibarcq} \emph {et~al.}}]{CampagneIbarcq2018}%
  \BibitemOpen
  \bibfield  {author} {\bibinfo {author} {\bibfnamefont {P.}~\bibnamefont
  {Campagne-Ibarcq}} \emph {et~al.},\ }\bibfield  {title} {\bibinfo {title}
  {Deterministic remote entanglement of superconducting circuits through
  microwave two-photon transitions},\ }\href
  {https://doi.org/10.1103/PhysRevLett.120.200501} {\bibfield  {journal}
  {\bibinfo  {journal} {Phys. Rev. Lett.}\ }\textbf {\bibinfo {volume} {120}},\
  \bibinfo {pages} {200501} (\bibinfo {year} {2018})}\BibitemShut {NoStop}%
\bibitem [{\citenamefont {Xu}\ \emph {et~al.}(2023)\citenamefont {Xu},
  \citenamefont {Zhu}, \citenamefont {Sun}, \citenamefont {He},\ and\
  \citenamefont {Zhang}}]{XuEtAl2023NJP}%
  \BibitemOpen
  \bibfield  {author} {\bibinfo {author} {\bibfnamefont {Y.}~\bibnamefont
  {Xu}}, \bibinfo {author} {\bibfnamefont {D.}~\bibnamefont {Zhu}}, \bibinfo
  {author} {\bibfnamefont {F.-X.}\ \bibnamefont {Sun}}, \bibinfo {author}
  {\bibfnamefont {Q.}~\bibnamefont {He}},\ and\ \bibinfo {author}
  {\bibfnamefont {W.}~\bibnamefont {Zhang}},\ }\bibfield  {title} {\bibinfo
  {title} {Fast quantum state transfer and entanglement preparation in strongly
  coupled bosonic systems},\ }\href {https://doi.org/10.1088/1367-2630/ad08f2}
  {\bibfield  {journal} {\bibinfo  {journal} {New J. Phys.}\ }\textbf {\bibinfo
  {volume} {25}},\ \bibinfo {pages} {113015} (\bibinfo {year}
  {2023})}\BibitemShut {NoStop}%
\bibitem [{\citenamefont {He}\ and\ \citenamefont
  {Zhang}(2025)}]{HeZhang2025PRL}%
  \BibitemOpen
  \bibfield  {author} {\bibinfo {author} {\bibfnamefont {Y.}~\bibnamefont
  {He}}\ and\ \bibinfo {author} {\bibfnamefont {Y.-X.}\ \bibnamefont {Zhang}},\
  }\bibfield  {title} {\bibinfo {title} {Quantum state transfer via a multimode
  resonator},\ }\href {https://doi.org/10.1103/PhysRevLett.134.023602}
  {\bibfield  {journal} {\bibinfo  {journal} {Phys. Rev. Lett.}\ }\textbf
  {\bibinfo {volume} {134}},\ \bibinfo {pages} {023602} (\bibinfo {year}
  {2025})}\BibitemShut {NoStop}%
\bibitem [{\citenamefont {Xiang}\ \emph {et~al.}(2024)\citenamefont {Xiang},
  \citenamefont {Chen}, \citenamefont {Zhu}, \citenamefont {Song},
  \citenamefont {Bao}, \citenamefont {Zhu}, \citenamefont {Jin}, \citenamefont
  {Wang}, \citenamefont {Xu}, \citenamefont {Zou}, \citenamefont {Li},
  \citenamefont {Wang}, \citenamefont {Song}, \citenamefont {Yue},
  \citenamefont {Partridge}, \citenamefont {Guo}, \citenamefont {Mondaini},
  \citenamefont {Wang},\ and\ \citenamefont
  {Scalettar}}]{XiangEtAl2024NatCommun}%
  \BibitemOpen
  \bibfield  {author} {\bibinfo {author} {\bibfnamefont {L.}~\bibnamefont
  {Xiang}}, \bibinfo {author} {\bibfnamefont {J.}~\bibnamefont {Chen}},
  \bibinfo {author} {\bibfnamefont {Z.}~\bibnamefont {Zhu}}, \bibinfo {author}
  {\bibfnamefont {Z.}~\bibnamefont {Song}}, \bibinfo {author} {\bibfnamefont
  {Z.}~\bibnamefont {Bao}}, \bibinfo {author} {\bibfnamefont {X.}~\bibnamefont
  {Zhu}}, \bibinfo {author} {\bibfnamefont {F.}~\bibnamefont {Jin}}, \bibinfo
  {author} {\bibfnamefont {K.}~\bibnamefont {Wang}}, \bibinfo {author}
  {\bibfnamefont {S.}~\bibnamefont {Xu}}, \bibinfo {author} {\bibfnamefont
  {Y.}~\bibnamefont {Zou}}, \bibinfo {author} {\bibfnamefont {H.}~\bibnamefont
  {Li}}, \bibinfo {author} {\bibfnamefont {Z.}~\bibnamefont {Wang}}, \bibinfo
  {author} {\bibfnamefont {C.}~\bibnamefont {Song}}, \bibinfo {author}
  {\bibfnamefont {A.}~\bibnamefont {Yue}}, \bibinfo {author} {\bibfnamefont
  {J.}~\bibnamefont {Partridge}}, \bibinfo {author} {\bibfnamefont
  {Q.}~\bibnamefont {Guo}}, \bibinfo {author} {\bibfnamefont {R.}~\bibnamefont
  {Mondaini}}, \bibinfo {author} {\bibfnamefont {H.}~\bibnamefont {Wang}},\
  and\ \bibinfo {author} {\bibfnamefont {R.~T.}\ \bibnamefont {Scalettar}},\
  }\bibfield  {title} {\bibinfo {title} {Enhanced quantum state transfer by
  circumventing quantum chaotic behavior},\ }\href
  {https://doi.org/10.1038/s41467-024-48791-3} {\bibfield  {journal} {\bibinfo
  {journal} {Nat. Commun.}\ }\textbf {\bibinfo {volume} {15}},\ \bibinfo
  {pages} {4918} (\bibinfo {year} {2024})}\BibitemShut {NoStop}%
\bibitem [{\citenamefont {Tian}\ \emph {et~al.}(2008)\citenamefont {Tian},
  \citenamefont {Allman},\ and\ \citenamefont {Simmonds}}]{Tian2008}%
  \BibitemOpen
  \bibfield  {author} {\bibinfo {author} {\bibfnamefont {L.}~\bibnamefont
  {Tian}}, \bibinfo {author} {\bibfnamefont {M.~S.}\ \bibnamefont {Allman}},\
  and\ \bibinfo {author} {\bibfnamefont {R.~W.}\ \bibnamefont {Simmonds}},\
  }\bibfield  {title} {\bibinfo {title} {Parametric coupling between
  macroscopic quantum resonators},\ }\href
  {https://doi.org/10.1088/1367-2630/10/11/115001} {\bibfield  {journal}
  {\bibinfo  {journal} {New J. Phys.}\ }\textbf {\bibinfo {volume} {10}},\
  \bibinfo {pages} {115001} (\bibinfo {year} {2008})}\BibitemShut {NoStop}%
\bibitem [{\citenamefont {Basilewitsch}\ \emph {et~al.}(2022)\citenamefont
  {Basilewitsch}, \citenamefont {Zhang}, \citenamefont {Girvin},\ and\
  \citenamefont {Koch}}]{Basilewitsch2022}%
  \BibitemOpen
  \bibfield  {author} {\bibinfo {author} {\bibfnamefont {D.}~\bibnamefont
  {Basilewitsch}}, \bibinfo {author} {\bibfnamefont {Y.}~\bibnamefont {Zhang}},
  \bibinfo {author} {\bibfnamefont {S.~M.}\ \bibnamefont {Girvin}},\ and\
  \bibinfo {author} {\bibfnamefont {C.~P.}\ \bibnamefont {Koch}},\ }\bibfield
  {title} {\bibinfo {title} {Engineering strong beamsplitter interaction
  between bosonic modes via quantum optimal control theory},\ }\href
  {https://doi.org/10.1103/PhysRevResearch.4.023054} {\bibfield  {journal}
  {\bibinfo  {journal} {Phys. Rev. Res.}\ }\textbf {\bibinfo {volume} {4}},\
  \bibinfo {pages} {023054} (\bibinfo {year} {2022})}\BibitemShut {NoStop}%
\bibitem [{\citenamefont {Chapman}\ \emph {et~al.}(2023)\citenamefont {Chapman}
  \emph {et~al.}}]{Chapman2023}%
  \BibitemOpen
  \bibfield  {author} {\bibinfo {author} {\bibfnamefont {B.~J.}\ \bibnamefont
  {Chapman}} \emph {et~al.},\ }\bibfield  {title} {\bibinfo {title}
  {High-on-off-ratio beam-splitter interaction for gates on bosonically encoded
  qubits},\ }\href {https://doi.org/10.1103/PRXQuantum.4.020355} {\bibfield
  {journal} {\bibinfo  {journal} {PRX Quantum}\ }\textbf {\bibinfo {volume}
  {4}},\ \bibinfo {pages} {020355} (\bibinfo {year} {2023})}\BibitemShut
  {NoStop}%
\bibitem [{\citenamefont {Lu}\ \emph {et~al.}(2023)\citenamefont {Lu},
  \citenamefont {Maiti}, \citenamefont {Garmon}, \citenamefont {Ganjam},
  \citenamefont {Zhang}, \citenamefont {Claes}, \citenamefont {Frunzio},
  \citenamefont {Girvin},\ and\ \citenamefont {Schoelkopf}}]{Lu2023}%
  \BibitemOpen
  \bibfield  {author} {\bibinfo {author} {\bibfnamefont {Y.}~\bibnamefont
  {Lu}}, \bibinfo {author} {\bibfnamefont {A.}~\bibnamefont {Maiti}}, \bibinfo
  {author} {\bibfnamefont {J.~W.~O.}\ \bibnamefont {Garmon}}, \bibinfo {author}
  {\bibfnamefont {S.}~\bibnamefont {Ganjam}}, \bibinfo {author} {\bibfnamefont
  {Y.}~\bibnamefont {Zhang}}, \bibinfo {author} {\bibfnamefont
  {J.}~\bibnamefont {Claes}}, \bibinfo {author} {\bibfnamefont
  {L.}~\bibnamefont {Frunzio}}, \bibinfo {author} {\bibfnamefont {S.~M.}\
  \bibnamefont {Girvin}},\ and\ \bibinfo {author} {\bibfnamefont {R.~J.}\
  \bibnamefont {Schoelkopf}},\ }\bibfield  {title} {\bibinfo {title}
  {High-fidelity parametric beamsplitting with a parity-protected converter},\
  }\href {https://doi.org/10.1038/s41467-023-41104-0} {\bibfield  {journal}
  {\bibinfo  {journal} {Nat. Commun.}\ }\textbf {\bibinfo {volume} {14}},\
  \bibinfo {pages} {5767} (\bibinfo {year} {2023})}\BibitemShut {NoStop}%
\bibitem [{\citenamefont {Wang}\ and\ \citenamefont
  {Clerk}(2012)}]{WangClerk2012}%
  \BibitemOpen
  \bibfield  {author} {\bibinfo {author} {\bibfnamefont {Y.-D.}\ \bibnamefont
  {Wang}}\ and\ \bibinfo {author} {\bibfnamefont {A.~A.}\ \bibnamefont
  {Clerk}},\ }\bibfield  {title} {\bibinfo {title} {Using interference for high
  fidelity quantum state transfer in optomechanics},\ }\href
  {https://doi.org/10.1103/PhysRevLett.108.153603} {\bibfield  {journal}
  {\bibinfo  {journal} {Phys. Rev. Lett.}\ }\textbf {\bibinfo {volume} {108}},\
  \bibinfo {pages} {153603} (\bibinfo {year} {2012})}\BibitemShut {NoStop}%
\bibitem [{\citenamefont {Lau}\ and\ \citenamefont
  {Clerk}(2019)}]{LauClerk2019}%
  \BibitemOpen
  \bibfield  {author} {\bibinfo {author} {\bibfnamefont {H.-K.}\ \bibnamefont
  {Lau}}\ and\ \bibinfo {author} {\bibfnamefont {A.~A.}\ \bibnamefont
  {Clerk}},\ }\bibfield  {title} {\bibinfo {title} {High-fidelity bosonic
  quantum state transfer using imperfect transducers and interference},\ }\href
  {https://doi.org/10.1038/s41534-019-0143-1} {\bibfield  {journal} {\bibinfo
  {journal} {npj Quantum Inf.}\ }\textbf {\bibinfo {volume} {5}},\ \bibinfo
  {pages} {31} (\bibinfo {year} {2019})}\BibitemShut {NoStop}%
\bibitem [{\citenamefont {Burkhart}\ \emph {et~al.}(2021)\citenamefont
  {Burkhart} \emph {et~al.}}]{Burkhart2021}%
  \BibitemOpen
  \bibfield  {author} {\bibinfo {author} {\bibfnamefont {L.~D.}\ \bibnamefont
  {Burkhart}} \emph {et~al.},\ }\bibfield  {title} {\bibinfo {title}
  {Error-detected state transfer and entanglement in a superconducting quantum
  network},\ }\href {https://doi.org/10.1103/PRXQuantum.2.030321} {\bibfield
  {journal} {\bibinfo  {journal} {PRX Quantum}\ }\textbf {\bibinfo {volume}
  {2}},\ \bibinfo {pages} {030321} (\bibinfo {year} {2021})}\BibitemShut
  {NoStop}%
\bibitem [{\citenamefont {Zhou}\ \emph {et~al.}(2024)\citenamefont {Zhou} \emph
  {et~al.}}]{Zhou2024}%
  \BibitemOpen
  \bibfield  {author} {\bibinfo {author} {\bibfnamefont {J.}~\bibnamefont
  {Zhou}} \emph {et~al.},\ }\bibfield  {title} {\bibinfo {title} {Quantum state
  transfer between superconducting cavities via exchange-free interactions},\
  }\href {https://doi.org/10.1103/PhysRevLett.133.220801} {\bibfield  {journal}
  {\bibinfo  {journal} {Phys. Rev. Lett.}\ }\textbf {\bibinfo {volume} {133}},\
  \bibinfo {pages} {220801} (\bibinfo {year} {2024})}\BibitemShut {NoStop}%
\bibitem [{\citenamefont {Yue}\ \emph {et~al.}(2024)\citenamefont {Yue},
  \citenamefont {Mondaini}, \citenamefont {Guo},\ and\ \citenamefont
  {Scalettar}}]{YueEtAl2024PRB}%
  \BibitemOpen
  \bibfield  {author} {\bibinfo {author} {\bibfnamefont {A.}~\bibnamefont
  {Yue}}, \bibinfo {author} {\bibfnamefont {R.}~\bibnamefont {Mondaini}},
  \bibinfo {author} {\bibfnamefont {Q.}~\bibnamefont {Guo}},\ and\ \bibinfo
  {author} {\bibfnamefont {R.~T.}\ \bibnamefont {Scalettar}},\ }\bibfield
  {title} {\bibinfo {title} {Quantum state transfer in interacting
  multiple-excitation systems},\ }\href
  {https://doi.org/10.1103/PhysRevB.110.195410} {\bibfield  {journal} {\bibinfo
   {journal} {Phys. Rev. B}\ }\textbf {\bibinfo {volume} {110}},\ \bibinfo
  {pages} {195410} (\bibinfo {year} {2024})}\BibitemShut {NoStop}%
\bibitem [{\citenamefont {Bergmann}\ \emph {et~al.}(1998)\citenamefont
  {Bergmann}, \citenamefont {Theuer},\ and\ \citenamefont
  {Shore}}]{Bergmann1998}%
  \BibitemOpen
  \bibfield  {author} {\bibinfo {author} {\bibfnamefont {K.}~\bibnamefont
  {Bergmann}}, \bibinfo {author} {\bibfnamefont {H.}~\bibnamefont {Theuer}},\
  and\ \bibinfo {author} {\bibfnamefont {B.~W.}\ \bibnamefont {Shore}},\
  }\bibfield  {title} {\bibinfo {title} {Coherent population transfer among
  quantum states of atoms and molecules},\ }\href
  {https://doi.org/10.1103/RevModPhys.70.1003} {\bibfield  {journal} {\bibinfo
  {journal} {Rev. Mod. Phys.}\ }\textbf {\bibinfo {volume} {70}},\ \bibinfo
  {pages} {1003} (\bibinfo {year} {1998})}\BibitemShut {NoStop}%
\bibitem [{\citenamefont {Larson}\ and\ \citenamefont
  {Andersson}(2005)}]{Larson2005}%
  \BibitemOpen
  \bibfield  {author} {\bibinfo {author} {\bibfnamefont {J.}~\bibnamefont
  {Larson}}\ and\ \bibinfo {author} {\bibfnamefont {E.}~\bibnamefont
  {Andersson}},\ }\bibfield  {title} {\bibinfo {title} {Cavity-state
  preparation using adiabatic transfer},\ }\href
  {https://doi.org/10.1103/PhysRevA.71.053814} {\bibfield  {journal} {\bibinfo
  {journal} {Phys. Rev. A}\ }\textbf {\bibinfo {volume} {71}},\ \bibinfo
  {pages} {053814} (\bibinfo {year} {2005})}\BibitemShut {NoStop}%
\bibitem [{\citenamefont {Kumar}\ \emph {et~al.}(2016)\citenamefont {Kumar},
  \citenamefont {Veps{\"a}l{\"a}inen}, \citenamefont {Danilin},\ and\
  \citenamefont {Paraoanu}}]{Kumar2016}%
  \BibitemOpen
  \bibfield  {author} {\bibinfo {author} {\bibfnamefont {K.~S.}\ \bibnamefont
  {Kumar}}, \bibinfo {author} {\bibfnamefont {A.}~\bibnamefont
  {Veps{\"a}l{\"a}inen}}, \bibinfo {author} {\bibfnamefont {S.}~\bibnamefont
  {Danilin}},\ and\ \bibinfo {author} {\bibfnamefont {G.~S.}\ \bibnamefont
  {Paraoanu}},\ }\bibfield  {title} {\bibinfo {title} {Stimulated raman
  adiabatic passage in a three-level superconducting circuit},\ }\href
  {https://doi.org/10.1038/ncomms10628} {\bibfield  {journal} {\bibinfo
  {journal} {Nat. Commun.}\ }\textbf {\bibinfo {volume} {7}},\ \bibinfo {pages}
  {10628} (\bibinfo {year} {2016})}\BibitemShut {NoStop}%
\bibitem [{\citenamefont {Premaratne}\ \emph {et~al.}(2017)\citenamefont
  {Premaratne}, \citenamefont {Wellstood},\ and\ \citenamefont
  {Palmer}}]{Premaratne2017}%
  \BibitemOpen
  \bibfield  {author} {\bibinfo {author} {\bibfnamefont {S.~P.}\ \bibnamefont
  {Premaratne}}, \bibinfo {author} {\bibfnamefont {F.~C.}\ \bibnamefont
  {Wellstood}},\ and\ \bibinfo {author} {\bibfnamefont {B.~S.}\ \bibnamefont
  {Palmer}},\ }\bibfield  {title} {\bibinfo {title} {Microwave photon fock
  state generation by stimulated raman adiabatic passage},\ }\href
  {https://doi.org/10.1038/ncomms14148} {\bibfield  {journal} {\bibinfo
  {journal} {Nat. Commun.}\ }\textbf {\bibinfo {volume} {8}},\ \bibinfo {pages}
  {14148} (\bibinfo {year} {2017})}\BibitemShut {NoStop}%
\bibitem [{\citenamefont {Heeres}\ \emph {et~al.}(2015)\citenamefont {Heeres},
  \citenamefont {Vlastakis}, \citenamefont {Holland}, \citenamefont
  {Krastanov}, \citenamefont {Albert}, \citenamefont {Frunzio}, \citenamefont
  {Jiang},\ and\ \citenamefont {Schoelkopf}}]{Heeres2015}%
  \BibitemOpen
  \bibfield  {author} {\bibinfo {author} {\bibfnamefont {R.~W.}\ \bibnamefont
  {Heeres}}, \bibinfo {author} {\bibfnamefont {B.}~\bibnamefont {Vlastakis}},
  \bibinfo {author} {\bibfnamefont {E.}~\bibnamefont {Holland}}, \bibinfo
  {author} {\bibfnamefont {S.}~\bibnamefont {Krastanov}}, \bibinfo {author}
  {\bibfnamefont {V.~V.}\ \bibnamefont {Albert}}, \bibinfo {author}
  {\bibfnamefont {L.}~\bibnamefont {Frunzio}}, \bibinfo {author} {\bibfnamefont
  {L.}~\bibnamefont {Jiang}},\ and\ \bibinfo {author} {\bibfnamefont {R.~J.}\
  \bibnamefont {Schoelkopf}},\ }\bibfield  {title} {\bibinfo {title} {Cavity
  state manipulation using photon-number selective phase gates},\ }\href
  {https://doi.org/10.1103/PhysRevLett.115.137002} {\bibfield  {journal}
  {\bibinfo  {journal} {Phys. Rev. Lett.}\ }\textbf {\bibinfo {volume} {115}},\
  \bibinfo {pages} {137002} (\bibinfo {year} {2015})}\BibitemShut {NoStop}%
\bibitem [{\citenamefont {Krastanov}\ \emph {et~al.}(2015)\citenamefont
  {Krastanov}, \citenamefont {Albert}, \citenamefont {Shen}, \citenamefont
  {Zou}, \citenamefont {Heeres}, \citenamefont {Vlastakis}, \citenamefont
  {Schoelkopf},\ and\ \citenamefont {Jiang}}]{Krastanov2015}%
  \BibitemOpen
  \bibfield  {author} {\bibinfo {author} {\bibfnamefont {S.}~\bibnamefont
  {Krastanov}}, \bibinfo {author} {\bibfnamefont {V.~V.}\ \bibnamefont
  {Albert}}, \bibinfo {author} {\bibfnamefont {C.}~\bibnamefont {Shen}},
  \bibinfo {author} {\bibfnamefont {C.-L.}\ \bibnamefont {Zou}}, \bibinfo
  {author} {\bibfnamefont {R.~W.}\ \bibnamefont {Heeres}}, \bibinfo {author}
  {\bibfnamefont {B.}~\bibnamefont {Vlastakis}}, \bibinfo {author}
  {\bibfnamefont {R.~J.}\ \bibnamefont {Schoelkopf}},\ and\ \bibinfo {author}
  {\bibfnamefont {L.}~\bibnamefont {Jiang}},\ }\bibfield  {title} {\bibinfo
  {title} {Universal control of an oscillator with dispersive coupling to a
  qubit},\ }\href {https://doi.org/10.1103/PhysRevA.92.040303} {\bibfield
  {journal} {\bibinfo  {journal} {Phys. Rev. A}\ }\textbf {\bibinfo {volume}
  {92}},\ \bibinfo {pages} {040303(R)} (\bibinfo {year} {2015})}\BibitemShut
  {NoStop}%
\bibitem [{\citenamefont {Landgraf}\ \emph {et~al.}(2024)\citenamefont
  {Landgraf}, \citenamefont {Fl{\"u}hmann}, \citenamefont {F{\"o}sel},
  \citenamefont {Marquardt},\ and\ \citenamefont {Schoelkopf}}]{Landgraf2024}%
  \BibitemOpen
  \bibfield  {author} {\bibinfo {author} {\bibfnamefont {J.}~\bibnamefont
  {Landgraf}}, \bibinfo {author} {\bibfnamefont {C.}~\bibnamefont
  {Fl{\"u}hmann}}, \bibinfo {author} {\bibfnamefont {T.}~\bibnamefont
  {F{\"o}sel}}, \bibinfo {author} {\bibfnamefont {F.}~\bibnamefont
  {Marquardt}},\ and\ \bibinfo {author} {\bibfnamefont {R.~J.}\ \bibnamefont
  {Schoelkopf}},\ }\bibfield  {title} {\bibinfo {title} {Fast quantum control
  of cavities using an improved protocol without coherent errors},\ }\href
  {https://doi.org/10.1103/PhysRevLett.133.260802} {\bibfield  {journal}
  {\bibinfo  {journal} {Phys. Rev. Lett.}\ }\textbf {\bibinfo {volume} {133}},\
  \bibinfo {pages} {260802} (\bibinfo {year} {2024})}\BibitemShut {NoStop}%
\bibitem [{\citenamefont {Palao}\ and\ \citenamefont
  {Kosloff}(2003)}]{PalaoKosloff2003}%
  \BibitemOpen
  \bibfield  {author} {\bibinfo {author} {\bibfnamefont {J.~P.}\ \bibnamefont
  {Palao}}\ and\ \bibinfo {author} {\bibfnamefont {R.}~\bibnamefont
  {Kosloff}},\ }\bibfield  {title} {\bibinfo {title} {Optimal control theory
  for unitary transformations},\ }\href
  {https://doi.org/10.1103/PhysRevA.68.062308} {\bibfield  {journal} {\bibinfo
  {journal} {Phys. Rev. A}\ }\textbf {\bibinfo {volume} {68}},\ \bibinfo
  {pages} {062308} (\bibinfo {year} {2003})}\BibitemShut {NoStop}%
\bibitem [{\citenamefont {Khaneja}\ \emph {et~al.}(2005)\citenamefont
  {Khaneja}, \citenamefont {Reiss}, \citenamefont {Kehlet}, \citenamefont
  {Schulte-Herbr{\"u}ggen},\ and\ \citenamefont {Glaser}}]{Khaneja2005}%
  \BibitemOpen
  \bibfield  {author} {\bibinfo {author} {\bibfnamefont {N.}~\bibnamefont
  {Khaneja}}, \bibinfo {author} {\bibfnamefont {T.}~\bibnamefont {Reiss}},
  \bibinfo {author} {\bibfnamefont {C.}~\bibnamefont {Kehlet}}, \bibinfo
  {author} {\bibfnamefont {T.}~\bibnamefont {Schulte-Herbr{\"u}ggen}},\ and\
  \bibinfo {author} {\bibfnamefont {S.~J.}\ \bibnamefont {Glaser}},\ }\bibfield
   {title} {\bibinfo {title} {Optimal control of coupled spin dynamics: Design
  of {NMR} pulse sequences by gradient ascent algorithms},\ }\href
  {https://doi.org/10.1016/j.jmr.2004.11.004} {\bibfield  {journal} {\bibinfo
  {journal} {J. Magn. Reson.}\ }\textbf {\bibinfo {volume} {172}},\ \bibinfo
  {pages} {296} (\bibinfo {year} {2005})}\BibitemShut {NoStop}%
\bibitem [{\citenamefont {Jin}\ and\ \citenamefont
  {Jing}(2026)}]{JinJing2026PRA}%
  \BibitemOpen
  \bibfield  {author} {\bibinfo {author} {\bibfnamefont {Z.-Y.}\ \bibnamefont
  {Jin}}\ and\ \bibinfo {author} {\bibfnamefont {J.}~\bibnamefont {Jing}},\
  }\bibfield  {title} {\bibinfo {title} {Universal quantum control over bosonic
  networks},\ }\href {https://doi.org/10.1103/ghqr-ydxs} {\bibfield  {journal}
  {\bibinfo  {journal} {Phys. Rev. A}\ }\textbf {\bibinfo {volume} {113}},\
  \bibinfo {pages} {012426} (\bibinfo {year} {2026})}\BibitemShut {NoStop}%
\bibitem [{\citenamefont {Laha}\ \emph {et~al.}(2024)\citenamefont {Laha},
  \citenamefont {Yasir},\ and\ \citenamefont {van Loock}}]{laha_PRR_2024}%
  \BibitemOpen
  \bibfield  {author} {\bibinfo {author} {\bibfnamefont {P.}~\bibnamefont
  {Laha}}, \bibinfo {author} {\bibfnamefont {P.~A.~A.}\ \bibnamefont {Yasir}},\
  and\ \bibinfo {author} {\bibfnamefont {P.}~\bibnamefont {van Loock}},\
  }\bibfield  {title} {\bibinfo {title} {Genuine non-gaussian entanglement of
  light and quantum coherence for an atom from noisy multiphoton spin-boson
  interactions},\ }\href {https://doi.org/10.1103/PhysRevResearch.6.033302}
  {\bibfield  {journal} {\bibinfo  {journal} {Phys. Rev. Res.}\ }\textbf
  {\bibinfo {volume} {6}},\ \bibinfo {pages} {033302} (\bibinfo {year}
  {2024})}\BibitemShut {NoStop}%
\bibitem [{\citenamefont {Brif}\ \emph {et~al.}(2010)\citenamefont {Brif},
  \citenamefont {Chakrabarti},\ and\ \citenamefont {Rabitz}}]{Brif_NJP_2010}%
  \BibitemOpen
  \bibfield  {author} {\bibinfo {author} {\bibfnamefont {C.}~\bibnamefont
  {Brif}}, \bibinfo {author} {\bibfnamefont {R.}~\bibnamefont {Chakrabarti}},\
  and\ \bibinfo {author} {\bibfnamefont {H.}~\bibnamefont {Rabitz}},\
  }\bibfield  {title} {\bibinfo {title} {Control of quantum phenomena: past,
  present and future},\ }\href {https://doi.org/10.1088/1367-2630/12/7/075008}
  {\bibfield  {journal} {\bibinfo  {journal} {New J. Phys.}\ }\textbf {\bibinfo
  {volume} {12}},\ \bibinfo {pages} {075008} (\bibinfo {year}
  {2010})}\BibitemShut {NoStop}%
\bibitem [{\citenamefont {Koch}\ \emph {et~al.}(2022)\citenamefont {Koch},
  \citenamefont {Boscain}, \citenamefont {Calarco}, \citenamefont {Dirr},
  \citenamefont {Filipp}, \citenamefont {Glaser}, \citenamefont {Kosloff},
  \citenamefont {Montangero}, \citenamefont {Schulte-Herbr{\"u}ggen},
  \citenamefont {Sugny},\ and\ \citenamefont {Wilhelm}}]{Koch_EPJ_2022}%
  \BibitemOpen
  \bibfield  {author} {\bibinfo {author} {\bibfnamefont {C.~P.}\ \bibnamefont
  {Koch}}, \bibinfo {author} {\bibfnamefont {U.}~\bibnamefont {Boscain}},
  \bibinfo {author} {\bibfnamefont {T.}~\bibnamefont {Calarco}}, \bibinfo
  {author} {\bibfnamefont {G.}~\bibnamefont {Dirr}}, \bibinfo {author}
  {\bibfnamefont {S.}~\bibnamefont {Filipp}}, \bibinfo {author} {\bibfnamefont
  {S.~J.}\ \bibnamefont {Glaser}}, \bibinfo {author} {\bibfnamefont
  {R.}~\bibnamefont {Kosloff}}, \bibinfo {author} {\bibfnamefont
  {S.}~\bibnamefont {Montangero}}, \bibinfo {author} {\bibfnamefont
  {T.}~\bibnamefont {Schulte-Herbr{\"u}ggen}}, \bibinfo {author} {\bibfnamefont
  {D.}~\bibnamefont {Sugny}},\ and\ \bibinfo {author} {\bibfnamefont {F.~K.}\
  \bibnamefont {Wilhelm}},\ }\bibfield  {title} {\bibinfo {title} {Quantum
  optimal control in quantum technologies. strategic report on current status,
  visions and goals for research in europe},\ }\href
  {https://doi.org/10.1140/epjqt/s40507-022-00138-x} {\bibfield  {journal}
  {\bibinfo  {journal} {EPJ Quantum Technol.}\ }\textbf {\bibinfo {volume}
  {9}},\ \bibinfo {pages} {19} (\bibinfo {year} {2022})}\BibitemShut {NoStop}%
\bibitem [{\citenamefont {Campos}\ and\ \citenamefont
  {Gerry}(2005)}]{Campos_PRA_2005}%
  \BibitemOpen
  \bibfield  {author} {\bibinfo {author} {\bibfnamefont {R.~A.}\ \bibnamefont
  {Campos}}\ and\ \bibinfo {author} {\bibfnamefont {C.~C.}\ \bibnamefont
  {Gerry}},\ }\bibfield  {title} {\bibinfo {title} {Permutation-parity exchange
  at a beam splitter: Application to heisenberg-limited interferometry},\
  }\href {https://doi.org/10.1103/PhysRevA.72.065803} {\bibfield  {journal}
  {\bibinfo  {journal} {Phys. Rev. A}\ }\textbf {\bibinfo {volume} {72}},\
  \bibinfo {pages} {065803} (\bibinfo {year} {2005})}\BibitemShut {NoStop}%
\bibitem [{\citenamefont {Karaev}\ \emph {et~al.}(2026)\citenamefont {Karaev},
  \citenamefont {Blumenthal},\ and\ \citenamefont
  {Hacohen-Gourgy}}]{Karaev2026}%
  \BibitemOpen
  \bibfield  {author} {\bibinfo {author} {\bibfnamefont {N.}~\bibnamefont
  {Karaev}}, \bibinfo {author} {\bibfnamefont {E.}~\bibnamefont {Blumenthal}},\
  and\ \bibinfo {author} {\bibfnamefont {S.}~\bibnamefont {Hacohen-Gourgy}},\
  }\href {https://arxiv.org/abs/2604.07235} {\bibinfo {title} {Analytical
  fock-state generation and swap using a rabi-driven transmon}} (\bibinfo
  {year} {2026}),\ \Eprint {https://arxiv.org/abs/2604.07235} {arXiv:2604.07235
  [quant-ph]} \BibitemShut {NoStop}%
\bibitem [{\citenamefont {Horodecki}\ \emph {et~al.}(1999)\citenamefont
  {Horodecki}, \citenamefont {Horodecki},\ and\ \citenamefont
  {Horodecki}}]{Horodecki1999}%
  \BibitemOpen
  \bibfield  {author} {\bibinfo {author} {\bibfnamefont {M.}~\bibnamefont
  {Horodecki}}, \bibinfo {author} {\bibfnamefont {P.}~\bibnamefont
  {Horodecki}},\ and\ \bibinfo {author} {\bibfnamefont {R.}~\bibnamefont
  {Horodecki}},\ }\bibfield  {title} {\bibinfo {title} {General teleportation
  channel, singlet fraction, and quasidistillation},\ }\href
  {https://doi.org/10.1103/PhysRevA.60.1888} {\bibfield  {journal} {\bibinfo
  {journal} {Phys. Rev. A}\ }\textbf {\bibinfo {volume} {60}},\ \bibinfo
  {pages} {1888} (\bibinfo {year} {1999})}\BibitemShut {NoStop}%
\bibitem [{\citenamefont {Nielsen}(2002)}]{Nielsen2002}%
  \BibitemOpen
  \bibfield  {author} {\bibinfo {author} {\bibfnamefont {M.~A.}\ \bibnamefont
  {Nielsen}},\ }\bibfield  {title} {\bibinfo {title} {A simple formula for the
  average gate fidelity of a quantum dynamical operation},\ }\href
  {https://doi.org/10.1016/S0375-9601(02)01272-0} {\bibfield  {journal}
  {\bibinfo  {journal} {Phys. Lett. A}\ }\textbf {\bibinfo {volume} {303}},\
  \bibinfo {pages} {249} (\bibinfo {year} {2002})}\BibitemShut {NoStop}%
\bibitem [{\citenamefont {Laha}\ and\ \citenamefont {van
  Loock}(2026)}]{laha_binomial_2026}%
  \BibitemOpen
  \bibfield  {author} {\bibinfo {author} {\bibfnamefont {P.}~\bibnamefont
  {Laha}}\ and\ \bibinfo {author} {\bibfnamefont {P.}~\bibnamefont {van
  Loock}},\ }\bibfield  {title} {\bibinfo {title} {Arbitrary high-fidelity
  binomial codes from multiphoton spin-boson interactions},\ }\href
  {https://doi.org/10.1103/58jr-l1x6} {\bibfield  {journal} {\bibinfo
  {journal} {Phys. Rev. Res.}\ }\textbf {\bibinfo {volume} {8}},\ \bibinfo
  {pages} {013237} (\bibinfo {year} {2026})}\BibitemShut {NoStop}%
\bibitem [{\citenamefont {Campos}\ \emph {et~al.}(1989)\citenamefont {Campos},
  \citenamefont {Saleh},\ and\ \citenamefont {Teich}}]{CamposSalehTeich1989}%
  \BibitemOpen
  \bibfield  {author} {\bibinfo {author} {\bibfnamefont {R.~A.}\ \bibnamefont
  {Campos}}, \bibinfo {author} {\bibfnamefont {B.~E.~A.}\ \bibnamefont
  {Saleh}},\ and\ \bibinfo {author} {\bibfnamefont {M.~C.}\ \bibnamefont
  {Teich}},\ }\bibfield  {title} {\bibinfo {title} {Quantum-mechanical lossless
  beam splitter: {SU(2)} symmetry and photon statistics},\ }\href
  {https://doi.org/10.1103/PhysRevA.40.1371} {\bibfield  {journal} {\bibinfo
  {journal} {Phys. Rev. A}\ }\textbf {\bibinfo {volume} {40}},\ \bibinfo
  {pages} {1371} (\bibinfo {year} {1989})}\BibitemShut {NoStop}%
\bibitem [{\citenamefont {Hardy}\ and\ \citenamefont
  {Wright}(2008)}]{HardyWright2008}%
  \BibitemOpen
  \bibfield  {author} {\bibinfo {author} {\bibfnamefont {G.~H.}\ \bibnamefont
  {Hardy}}\ and\ \bibinfo {author} {\bibfnamefont {E.~M.}\ \bibnamefont
  {Wright}},\ }\href@noop {} {\emph {\bibinfo {title} {An Introduction to the
  Theory of Numbers}}},\ \bibinfo {edition} {6th}\ ed.\ (\bibinfo  {publisher}
  {Oxford University Press},\ \bibinfo {year} {2008})\BibitemShut {NoStop}%
\bibitem [{\citenamefont {Cassels}(1957)}]{Cassels1957}%
  \BibitemOpen
  \bibfield  {author} {\bibinfo {author} {\bibfnamefont {J.~W.~S.}\
  \bibnamefont {Cassels}},\ }\href@noop {} {\emph {\bibinfo {title} {An
  Introduction to Diophantine Approximation}}}\ (\bibinfo  {publisher}
  {Cambridge University Press},\ \bibinfo {year} {1957})\BibitemShut {NoStop}%
\bibitem [{\citenamefont {Blais}\ \emph {et~al.}(2004)\citenamefont {Blais},
  \citenamefont {Huang}, \citenamefont {Wallraff}, \citenamefont {Girvin},\
  and\ \citenamefont {Schoelkopf}}]{Blais2004}%
  \BibitemOpen
  \bibfield  {author} {\bibinfo {author} {\bibfnamefont {A.}~\bibnamefont
  {Blais}}, \bibinfo {author} {\bibfnamefont {R.-S.}\ \bibnamefont {Huang}},
  \bibinfo {author} {\bibfnamefont {A.}~\bibnamefont {Wallraff}}, \bibinfo
  {author} {\bibfnamefont {S.~M.}\ \bibnamefont {Girvin}},\ and\ \bibinfo
  {author} {\bibfnamefont {R.~J.}\ \bibnamefont {Schoelkopf}},\ }\bibfield
  {title} {\bibinfo {title} {Cavity quantum electrodynamics for superconducting
  electrical circuits: An architecture for quantum computation},\ }\href
  {https://doi.org/10.1103/PhysRevA.69.062320} {\bibfield  {journal} {\bibinfo
  {journal} {Phys. Rev. A}\ }\textbf {\bibinfo {volume} {69}},\ \bibinfo
  {pages} {062320} (\bibinfo {year} {2004})}\BibitemShut {NoStop}%
\bibitem [{\citenamefont {Schuster}\ \emph {et~al.}(2007)\citenamefont
  {Schuster}, \citenamefont {Houck}, \citenamefont {Schreier}, \citenamefont
  {Wallraff}, \citenamefont {Gambetta}, \citenamefont {Blais}, \citenamefont
  {Frunzio}, \citenamefont {Majer}, \citenamefont {Johnson}, \citenamefont
  {Devoret}, \citenamefont {Girvin},\ and\ \citenamefont
  {Schoelkopf}}]{Schuster2007}%
  \BibitemOpen
  \bibfield  {author} {\bibinfo {author} {\bibfnamefont {D.~I.}\ \bibnamefont
  {Schuster}}, \bibinfo {author} {\bibfnamefont {A.~A.}\ \bibnamefont {Houck}},
  \bibinfo {author} {\bibfnamefont {J.~A.}\ \bibnamefont {Schreier}}, \bibinfo
  {author} {\bibfnamefont {A.}~\bibnamefont {Wallraff}}, \bibinfo {author}
  {\bibfnamefont {J.~M.}\ \bibnamefont {Gambetta}}, \bibinfo {author}
  {\bibfnamefont {A.}~\bibnamefont {Blais}}, \bibinfo {author} {\bibfnamefont
  {L.}~\bibnamefont {Frunzio}}, \bibinfo {author} {\bibfnamefont
  {J.}~\bibnamefont {Majer}}, \bibinfo {author} {\bibfnamefont
  {B.}~\bibnamefont {Johnson}}, \bibinfo {author} {\bibfnamefont {M.~H.}\
  \bibnamefont {Devoret}}, \bibinfo {author} {\bibfnamefont {S.~M.}\
  \bibnamefont {Girvin}},\ and\ \bibinfo {author} {\bibfnamefont {R.~J.}\
  \bibnamefont {Schoelkopf}},\ }\bibfield  {title} {\bibinfo {title} {Resolving
  photon number states in a superconducting circuit},\ }\href
  {https://doi.org/10.1038/nature05461} {\bibfield  {journal} {\bibinfo
  {journal} {Nature}\ }\textbf {\bibinfo {volume} {445}},\ \bibinfo {pages}
  {515} (\bibinfo {year} {2007})}\BibitemShut {NoStop}%
\end{thebibliography}%

\end{document}